\documentclass[
 reprint,
 superscriptaddress,
 amsmath,amssymb,
 aps,
 pra,
floatfix,
]{revtex4-2}

\usepackage[utf8]{inputenc}
\usepackage{commath}
\usepackage{amssymb}
\usepackage{dcolumn}
\usepackage{bm}
\usepackage{booktabs}
\usepackage{accents}
\usepackage{mathrsfs}
\usepackage{amsmath}
\usepackage{soul}
\usepackage{amssymb}
\usepackage{braket}
\usepackage{mathtools}
\usepackage[switch]{lineno}
\usepackage{soul}
\usepackage[dvipsnames]{xcolor}
\usepackage{aligned-overset}
\usepackage{multirow}
\usepackage{cuted}
\usepackage{lipsum}
\usepackage{tabularray}
\usepackage{tablefootnote}
\usepackage{mdframed}

\newcommand{\varphie}{\varphi_{\text{ext}}}
\newcommand{\Phie}{\Phi_{\text{ext}}}
\newcommand{\dt}{\mathrm{d}t}
\newcommand{\dell}{\mathrm{d}\boldsymbol{\ell}}

\newcommand{\A}{\mathbf{A}}
\newcommand{\B}{\mathbf{B}}
\newcommand{\E}{\mathbf{E}}
\newcommand{\EQ}{\mathbf{E}_{Q}}

\newcommand{\EBd}{\mathbf{E}_{\dot{\boldsymbol{B}}}}
\newcommand{\Eng}{\mathbf{E}_{n_{g}}}

\newcommand{\elementdepsymbol}[3]{#1_{\mathrm{#2}_{#3}}}

\DeclareMathOperator{\Real}{\mathrm{Re}}
\DeclareMathOperator{\Imag}{\mathrm{Im}}

\DeclareSymbolFont{sfletters}{OML}{cmbrm}{m}{it}
\DeclareMathAlphabet\mathbfcal{OMS}{cmsy}{b}{n}

\DeclareMathSymbol{\sphi}{\mathord}{sfletters}{"1E}
\DeclareMathSymbol{\spsi}{\mathord}{sfletters}{"20}
\usepackage[dvipsnames]{xcolor}

\usepackage{hyperref}
\hypersetup{
    colorlinks=true,
    linkcolor=blue,
    citecolor=blue,
    urlcolor=blue
}
\usepackage{cleveref}
\crefname{fig}{Fig.}{Figs.}
\crefname{extendedfig}{Extended Data Fig.}{Extended Data Figs.}
\crefname{eq}{Eq.}{Eqs.}
\crefname{Methods}{Methods }{Methods Sections}

\begin{document}

\title{Systematic Construction of Time-Dependent Hamiltonians for Microwave-Driven Josephson Circuits}

\author{Yao Lu}
\email{yaolu@fnal.gov}
\affiliation{Departments of Applied Physics and Physics, Yale University, New Haven, Connecticut 06511, USA}
\affiliation{Yale Quantum Institute, Yale University, New Haven, Connecticut 06511, USA}
\affiliation{Superconducting Quantum Materials and Systems Division, Fermi National Accelerator Laboratory (FNAL), Batavia, Illinois 60510, USA}

\author{Tianpu Zhao}
\affiliation{Graduate Program in Applied Physics, Northwestern University, Evanston, Illinois 60208, USA}

\author{Andr\'e Valli\`eres}
\affiliation{Superconducting Quantum Materials and Systems Division, Fermi National Accelerator Laboratory (FNAL), Batavia, Illinois 60510, USA}
\affiliation{Graduate Program in Applied Physics, Northwestern University, Evanston, Illinois 60208, USA}

\author{Kevin C. Smith}
\affiliation{Departments of Applied Physics and Physics, Yale University, New Haven, Connecticut 06511, USA}
\affiliation{Yale Quantum Institute, Yale University, New Haven, Connecticut 06511, USA}
\affiliation{Brookhaven National Laboratory, Upton, New York 11973, USA}

\author{Daniel Weiss}
\affiliation{Departments of Applied Physics and Physics, Yale University, New Haven, Connecticut 06511, USA}
\affiliation{Yale Quantum Institute, Yale University, New Haven, Connecticut 06511, USA}

\author{Xinyuan You}
\affiliation{Superconducting Quantum Materials and Systems Division, Fermi National Accelerator Laboratory (FNAL), Batavia, Illinois 60510, USA}

\author{Yaxing Zhang}
\affiliation{Departments of Applied Physics and Physics, Yale University, New Haven, Connecticut 06511, USA}
\affiliation{Yale Quantum Institute, Yale University, New Haven, Connecticut 06511, USA}

\author{Suhas Ganjam}
\affiliation{Departments of Applied Physics and Physics, Yale University, New Haven, Connecticut 06511, USA}
\affiliation{Yale Quantum Institute, Yale University, New Haven, Connecticut 06511, USA}

\author{Aniket Maiti}
\affiliation{Departments of Applied Physics and Physics, Yale University, New Haven, Connecticut 06511, USA}
\affiliation{Yale Quantum Institute, Yale University, New Haven, Connecticut 06511, USA}

\author{John W. O. Garmon}
\affiliation{Departments of Applied Physics and Physics, Yale University, New Haven, Connecticut 06511, USA}
\affiliation{Yale Quantum Institute, Yale University, New Haven, Connecticut 06511, USA}

\author{Shantanu Mundhada}
\affiliation{Quantum Circuits, Inc., 25 Science Park, New Haven, Connecticut 06511, USA}

\author{Ziwen Huang}
\affiliation{Superconducting Quantum Materials and Systems Division, Fermi National Accelerator Laboratory (FNAL), Batavia, Illinois 60510, USA}

\author{Ian Mondragon-Shem}
\affiliation{Department of Physics and Astronomy, Northwestern University, Evanston, Illinois 60208, USA}

\author{Steven M. Girvin}
\affiliation{Departments of Applied Physics and Physics, Yale University, New Haven, Connecticut 06511, USA}
\affiliation{Yale Quantum Institute, Yale University, New Haven, Connecticut 06511, USA}

\author{Jens Koch}
\affiliation{Department of Physics and Astronomy, Northwestern University, Evanston, Illinois 60208, USA}
\affiliation{Center for Applied Physics and Superconducting Technologies, Northwestern University, Evanston, Illinois 60208, USA}

\author{Robert J. Schoelkopf}
\email{robert.schoelkopf@yale.edu}
\affiliation{Departments of Applied Physics and Physics, Yale University, New Haven, Connecticut 06511, USA}
\affiliation{Yale Quantum Institute, Yale University, New Haven, Connecticut 06511, USA}
\affiliation{Quantum Circuits, Inc., 25 Science Park, New Haven, Connecticut 06511, USA}

\date{\today}

\begin{abstract}
Time-dependent electromagnetic drives are fundamental for controlling complex quantum systems, including superconducting Josephson circuits. In these devices, accurate time-dependent Hamiltonian models are imperative for predicting their dynamics and designing high-fidelity quantum operations. Existing numerical methods, such as black-box quantization (BBQ) and energy-participation ratio (EPR), excel at modeling the static Hamiltonians of Josephson circuits. However, these techniques do not fully capture the behavior of driven circuits stimulated by external microwave drives, nor do they include a generalized approach to account for the inevitable noise and dissipation that enter through microwave ports. Here, we introduce numerical techniques that leverage classical microwave simulations, efficiently executable in finite-element solvers, to obtain the time-dependent Hamiltonian of microwave-driven superconducting circuits with arbitrary geometries under charge, flux, or mixed electromagnetic modulation. Importantly, our techniques do not rely on a lumped-element description of the superconducting circuit, in contrast to previous approaches to tackling this problem. We demonstrate the versatility of our approach by characterizing the driven properties of realistic circuit devices in complex electromagnetic environments, including coherent dynamics due to charge and flux modulation, as well as drive-induced relaxation and dephasing. Our techniques offer a powerful toolbox for optimizing circuit designs and advancing practical applications in superconducting quantum computing.
\end{abstract}

\maketitle

\section{INTRODUCTION}\label{intro}

\begin{figure*}[!t]
    \centering
    \includegraphics[width=\textwidth
    ]{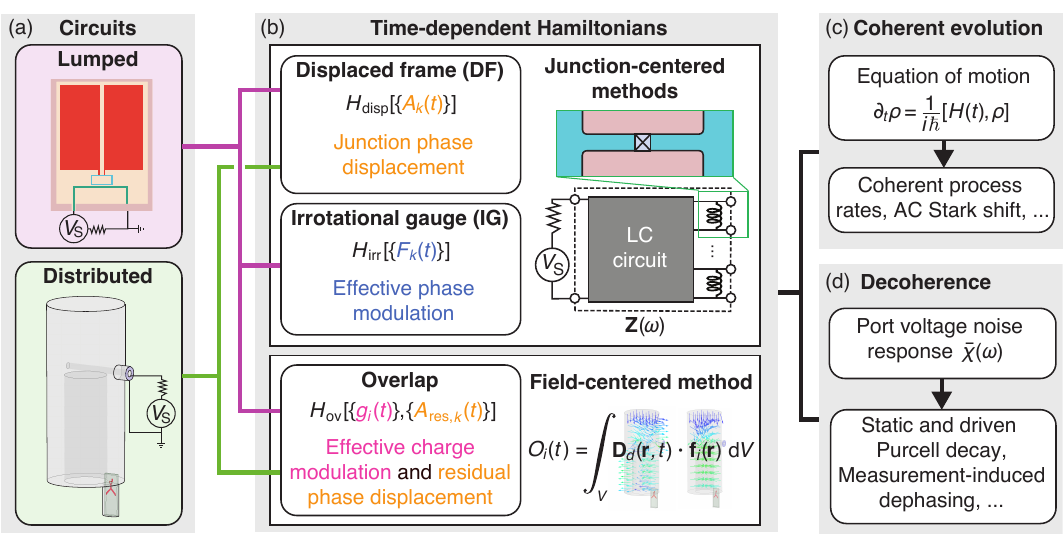}
    \caption{\textbf{Flow diagram for numerically modeling a general Josephson circuit.} 
    (a) Two types of driven Josephson circuits: (i) lumped circuits with negligible distributed linear inductance, and (ii) distributed circuits. Each type is amenable to different methods developed in this work.
    (b) The time-dependent Hamiltonians in different reference frames are derived using three methods: (i) the displaced frame (DF) method, (ii) the irrotational-gauge (IG) method, and (iii) the overlap method. The first two methods are junction-centered approaches, focusing on simulating flux across junctions under excitation (or equivalently the impedance matrix $\mathbf{Z}(\omega)$) to extract the junction phase displacements $A_k(t)$ or effective phase modulations $\elementdepsymbol{F}{J}{k}(t)$ for each junction $k$. Conversely, the overlap method focuses on three-dimensional electric fields of the driven system.
    By computing the overlap between the displacement field of the driven circuit $\mathbf{D}_d(\mathbf{r}, t)$ and the electric field mode profile of each eigenmode $\mathbf{f}_i(\mathbf{r})$,  we can extract the effective charge and residual phase displacement, denoted as $g_i(t)$ and $A_{\mathrm{res}, k}(t)$, respectively. While all three methods apply to lumped circuits, only the DF and overlap methods are suitable for distributed circuits.
    (c) Hamiltonians obtained with these methods describe the coherent dynamics of a driven circuit, enabling the extraction of relevant properties such as coherent process rates and AC Stark shifts.
    (d) The drive parameters are also useful for obtaining the circuit susceptibility function with respect to the drive port. This function allows for Fermi's golden rule or Floquet-Markov calculations of the circuit decoherence due to the thermal bath or signal fluctuations at the drive port.}
    \label{fig:table-of-contents}
\end{figure*}

Externally driven, open quantum systems with multiple interacting degrees of freedom arise across many platforms, including atoms or ions with discrete internal states coupled to motional~\cite{HAFFNER2008} and cavity~\cite{Raimond2001} modes, optomechanical devices~\cite{Aspelmeyer2014}, and superconducting circuits~\cite{Devoret2013}. In all such settings, time-dependent control fields are used to engineer useful dynamics, while the same control channels can also mediate dissipation and noise through an electromagnetic environment with nontrivial frequency structure~\cite{Clerk2010}. Circuit quantum electrodynamics is a particularly versatile platform for quantum information processing, offering a wide variety of tunable Hamiltonians enabled by the Josephson nonlinearity~\cite{Krantz2019,Blais2021,Yvonne2021}. Static or biased Josephson nonlinear elements can be combined in diverse topologies to engineer distinct nonlinear potentials for superconducting circuits~\cite{Gyenis2021}. A defining feature of circuit QED is the interplay between such programmable Josephson nonlinearities and a surrounding linear electromagnetic environment whose mode structure and couplings are often complex and strongly geometry-dependent. Under microwave drive, this interplay can generate many important parametric processes and applications, 
including beamsplitting and photon-exchange interactions~\cite{ZakkaBajjani2011, Strand2013, Sirois2015, Mckay2016, Caldwell2018, Gao2018, Chapman2023, Lu2023, Kim2025}, single- and two-mode squeezing~\cite{CastellanosBeltran2008, Bergeal2010, Puri2017, Sivak2019, Grimm2020, zhou2024}, drive-activated frequency shifts and Kerr nonlinearities~\cite{Sung2021, Sete2021,Noh2023,Eriksson2024, Huang2024, You2025}, subharmonic Rabi driving~\cite{Xia2023, Sah2024, Fillip2024}, conditional interaction for control and readout~\cite{Chow2011,Didier2015,Touzard2019, Heya2021, Eickbusch2022, Diringer2024, You2024}, and engineered driven-dissipation and autonomous stabilization~\cite{Murch2013, Leghtas2015, Yao2017, Ma2019, Lescanne2020, Gertler2021, Berdou2023}, to name a few. The same drive and environment that enable these capabilities can also activate spurious interactions and open additional port-mediated decoherence channels~\cite{Gambetta2006,Houck2008,Boissonneault2009,Boissonneault2012,Sete2014,Sunada2022,Khezri2023,Dumas2024,Frattini2024,Dai2025,Othmane2025, Xia2025}. Precisely modeling the driven Hamiltonian and dissipation of a superconducting Josephson circuit is therefore a prerequisite for making accurate predictions about device performance and enabling circuit optimization for high-quality quantum operations.

Many useful simulation techniques have been developed in recent years for deriving Hamiltonians of closed- or open-system superconducting circuits using electromagnetic simulations~\cite{Solgun2019,Minev2021quasilumped,Peng2022,Chitta2022,Pham2023,Rajabzadeh2023,Moon2024,Roth2024,Khan2024,Smitham2024,Labarca2024,Pham2025,LevensonFalk2025,Parra-Rodriguez2025,Bakr2026}. Among them, a representative class of approaches includes the black box quantization (BBQ) and the energy participation ratio (EPR) methods~\cite{Nigg2012,Solgun2014,Minev2021,IEPR}.
Related impedance- or admittance-based approaches also incorporate microwave drives in important dispersive circuit-QED settings~\cite{Solgun2019,Labarca2024}.
However, existing techniques generally fall short of a direct layout-level workflow for translating signals and noise applied at physical drive ports into effective drive parameters and noise couplings in the Hamiltonian. In particular, circuits driven with time-dependent external magnetic fields experience induced electromotive forces (EMF), which are not captured by static BBQ and EPR constructions alone, and are not the focus of existing impedance- or admittance-based driven-circuit treatments.
The subtleties surrounding time-dependent fluxes in the context of circuit quantization were first pointed out by You {\it et al.} \cite{You2019}, who derived the Hamiltonian of a general lumped-element superconducting circuit subject to time-dependent external flux. Riwar and DiVincenzo recently generalized this work to the case of lumped-element circuits with realistic geometries \cite{Riwar2022},
which requires knowledge of the vector potential induced by an external magnetic field. Despite these previous works, how to model realistic Josephson circuits interacting with arbitrary electromagnetic fields induced by microwave signals at the drive ports, using layouts of 2D or 3D devices as the direct input information, remains challenging.

In this work, we extend previous approaches by introducing three complementary numerical recipes that extract the driven Hamiltonian of arbitrary Josephson circuits: (i) the displaced-frame (DF) method, (ii) the irrotational-gauge (IG) method, and (iii) the overlap method. These methods (summarized in Fig.~\ref{fig:table-of-contents}) lead to Hamiltonians defined in different reference frames, applicable in different contexts. They are compatible with well-established microwave-engineering simulation techniques, and can be readily applied to a wide range of superconducting circuits with complicated geometries. Additionally, they enable a more complete understanding of the static and dynamic behavior of realistic circuits. Drives inevitably introduce noise to the circuit through lossy ports; our framework incorporates the corresponding noise spectral densities and, combined with the driven Hamiltonian, yields drive-induced decay and dephasing rates for modes embedded in such circuits. Taken together, these techniques form a practical toolbox for numerical modeling of driven superconducting circuits, enabling fast design iteration and optimization, and thus accelerating quantum-information-processing research based on superconducting circuits. Although our discussion is specialized to Josephson circuits, the underlying principle, mapping port-level electromagnetic response into effective drive and noise couplings for a multimode quantum system, is broadly applicable to driven open quantum devices.

Below, we briefly summarize the advantages and constraints of each technique, before detailing them in the subsequent sections.

The displaced frame method (Section~\ref{sec:displaced-frame}) focuses on the displacement of the (linearized) Josephson junction phase due to the microwave drives. This allows for the calculation of many important parametric processes such as the beamsplitter interaction and AC-Stark shift~\cite{Chapman2023, Lu2023} for circuits with arbitrary geometry. However, since the displacement is defined with respect to the linearized circuit, this method only serves as an approximation for compact-variable circuits as it does not preserve the periodicity boundary condition of the wavefunction. Since the linear drive terms are eliminated in the displaced frame, this method does not capture the direct interactions between the external fields and the circuit, but only the induced parametric interactions from the stimulated nonlinearity.

The irrotational-gauge method (Section~\ref{sec:irrotational-gauge}), by contrast, yields the driven lab-frame Hamiltonian, with the drives modeled as effective phase modulations of the Josephson energy. From this Hamiltonian, we can immediately extract coherent transition rates such as the Rabi rate, as well as incoherent transition rates such as Purcell decay due to coupling to the lossy ports. The drive parameter in this technique arises from the direct field-circuit interaction and, unlike the linear displacement, applies to circuits with either compact or extended variables. Because this method works in the irrotational gauge \cite{You2019, Riwar2022} where external drives are explicitly coupled to lumped-element inductances, it is difficult to extend the method to inductively driven circuits with distributed geometry (such as 3D cavities). 

The overlap method (Section~\ref{sec:overlap-method}) is the most general of the three, providing the driven lab-frame Hamiltonian for circuits in both the lumped-element and distributed regimes. The drives are represented as effective charge modulations and residual phase displacements coupled to the nonlinear Josephson energy. This generality comes with increased computational cost. In a typical workflow, the overlap method may be used to cross-check the Hamiltonian obtained from the other two methods, while also providing valuable information on the contribution of the individual modes to the junction phase displacements.

Our paper is structured as follows. We first motivate the general Hamiltonian structure for a microwave-driven Josephson circuit and define the charge- and phase-modulation parameters (Section~\ref{sec:general_Hamiltonian_Josephson_circuit}). We then introduce the displaced-frame (Section~\ref{sec:displaced-frame}), the irrotational-gauge (Section~\ref{sec:irrotational-gauge}), and the overlap methods (Section~\ref{sec:overlap-method}). For each technique, we derive the Hamiltonian and explain how the modulation parameters can be determined numerically through classical microwave simulations. We then explore how these methods, combined with the linear response approach, facilitate the calculation of decoherence rates in practical circuits, such as fixed-frequency and flux-tunable transmons coupled to charge and flux drive lines (Section~\ref{sec:decoherence-rates}). We demonstrate the effectiveness of our methods by applying them to compute key processes, including the Purcell effect (Section~\ref{sec:Purcell_effect}) and drive-induced decoherence (Section~\ref{sec:drive-induced-decay-dephasing}), and conclude with a discussion of the main findings and outlook.

\section{MODELING THE HAMILTONIAN OF A MODULATED JOSEPHSON CIRCUIT}
\label{sec:general_Hamiltonian_Josephson_circuit}
\begin{figure}[t]
    \centering\includegraphics[width=\columnwidth]{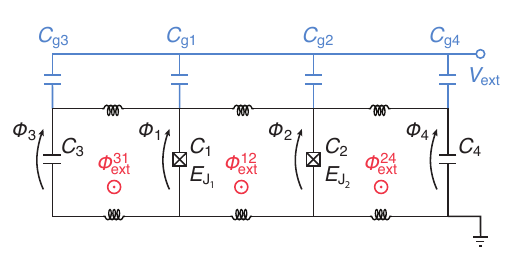}
    \caption{\textbf{A lumped-element flux-tunable circuit under voltage and flux modulation.} 
    Circuit quantization that properly accounts for the EMF effect can produce a gauge-dependent circuit Hamiltonian, Eq.~(\ref{eq:H_SQUID}). However, translating a physical circuit layout into such a lumped-element model is challenging due to the complexities of the circuit's geometry and the external field profile.}
    \label{fig:SQUID-circuit}
\end{figure}

In this section, we describe the circuit quantization for Josephson circuits driven by an external electromagnetic field, and highlight the necessity of new techniques for modeling complicated circuit geometries not amenable to standard lumped-element circuit analysis. We start by considering the minimal nontrivial example of a flux-tunable transmon based on a dc-SQUID and additional shunting capacitance, shown in Fig.~\hyperref[fig:SQUID-circuit]{\ref*{fig:SQUID-circuit}}. Besides the closed SQUID loop formed by the two Josephson junctions and the loop inductor, the circuit also contains loops formed by the capacitors and linear inductors, representing shunting capacitor pads with realistic, continuous geometry. The external drive source generates both voltage and flux modulation of the circuit, via coupling capacitance $\elementdepsymbol{C}{g}{i}$ and mutual inductance (not explicitly drawn). Due to the effect of electromotive force, both the magnetic field within and outside of the SQUID loop need to be included for the correct description of the driven circuit.  Considering only the low-energy dynamics where the high-frequency modes due to the linear geometric inductance can be neglected~\cite{Riwar2022, Lu2023}, we can write down the Lagrangian of the driven circuit as
\begin{align}
\label{eq:SQUID_Lagrangian} 
\mathcal{L}&=\frac{1}{2}\sum_i\left[C_i \dot{\Phi}_i^2+\elementdepsymbol{C}{g}{i}\left(\dot{\Phi}_i-V_{\mathrm{ext}}(t)\right)^2\right]\nonumber\\ 
&\quad +\elementdepsymbol{E}{J}{1}\cos\left(\frac{\Phi_1}{\phi_0}\right)+\elementdepsymbol{E}{J}{2}\cos\left(\frac{\Phi_2}{\phi_0}\right),
\end{align}
where summation is taken over all branches $i\in \{1,2,3,4\}$. Here, $\elementdepsymbol{E}{J}{1}$ and $\elementdepsymbol{E}{J}{2}$ are the Josephson energies of the junctions, $C_i$ and $\Phi_i$ are the capacitance and the branch flux on the $i$-th branch, $V_{\mathrm{ext}}(t)$ is the time-dependent external voltage and $\phi_0=\hbar/2e$ is the reduced flux quantum. Importantly, the branch fluxes are not independent, as they are constrained by Faraday’s law
and the flux quantization relationship,
\begin{equation}
    \Phi_i-\Phi_j=\Phi_{\mathrm{ext}}^{ij}(t)+n_{ij}\Phi_0,
\end{equation}
where $\Phi_{\mathrm{ext}}^{ij}(t)$ stands for the external flux between the $i$-th and the $j$-th branch, $\Phi_0 = 2\pi\phi_0$ is the flux quantum, and $n_{ij}$ is an arbitrary integer.
By choosing to write the Lagrangian using the $j$-th branch flux variable, we obtain
\begin{align}
\label{eq:SQUID_Lagrangian_continuous_2}
\mathcal{L}_j&=\frac{1}{2}C_{\Sigma}\left(\dot{\Phi}_j-\mathcal{V}_{j}(t)\right)^2\nonumber \\ &+\sum_{k=1,2}\elementdepsymbol{E}{J}{k}\cos\left(\frac{\Phi_j}{\phi_0}+\mathcal{F}_{kj}(t)\right), 
\end{align}
where $C_{\Sigma}=\sum_i\left(C_i+\elementdepsymbol{C}{g}{i}\right)$ is the total capacitance of the device, and the modulation parameters $\mathcal{V}_j(t)$ and $\mathcal{F}_{kj}(t)$ are related to $V_{\mathrm{ext}}(t)$ and $\Phi_{\mathrm{ext}}(t)$ through 
\begin{align}
\mathcal{V}_j(t) &= \sum_{i}\frac{\elementdepsymbol{C}{g}{i} V_{\text{ext}}(t)}{C_\Sigma}-\sum_{i \neq j} \frac{( C_i + \elementdepsymbol{C}{g}{i})\dot{\Phi}_{\text{ext}}^{ij}(t)}{C_\Sigma},\label{eq:app-SQUID-modulation-params1}\\
\mathcal{F}_{kj}(t)&= \frac{\Phi_{\text{ext}}^{kj}(t) }{\phi_0} = \varphi_{\text{ext}}^{kj}.\label{eq:app-SQUID-modulation-params2}
\end{align}
A possible integer fluxoid contribution would shift the Josephson phase by an integer multiple of $2\pi$ and therefore leaves the Josephson Hamiltonian unchanged; we absorb this choice into the definition of the phase branch and do not include it in $\mathcal{F}_{kj}(t)$. The second term in Eq.~\eqref{eq:app-SQUID-modulation-params1} manifests the effect of the time-dependent flux as an electromotive force driving the circuit. The dependence of the modulation parameters on the choice of the branch flux reflects the gauge freedom discussed in \cite{You2019, Riwar2022}. More generally, one may choose different linear combinations of the branch fluxes as flux variables, with resulting Hamiltonians equivalent up to a gauge transformation. Per the gauge choice made in Eq.~\eqref{eq:SQUID_Lagrangian_continuous_2}, the Hamiltonian is given by 
\begin{align}
\label{eq:H_SQUID}
\mathcal{H}_j &= \dot{\Phi}_j Q_j - \mathcal{L}_j \nonumber \\ 
&= 4 E_{C_\Sigma} n_j^2 + 2e\mathcal{V}_j(t) n_j -\sum_{k=1,2} \elementdepsymbol{E}{J}{k}\cos\left(\varphi_j+\mathcal{F}_{kj}(t)\right),
\end{align}
where $Q_j = \frac{\partial \mathcal{L}_j}{\partial \dot{\Phi}_j} = C_\Sigma (\dot{\Phi}_j - \mathcal{V}_j )$ is the conjugate momentum to $\Phi_j$ and $E_{C_\Sigma} = e^2/2C_\Sigma$ is the charging energy. Upon quantization, the reduced charge $n_j = Q_j/2e$ and the canonical phase operator $\varphi_j = \Phi_j/\phi_0$ satisfy $[n_j, e^{ i \varphi_j}] =  e^{ i \varphi_j}$ (equivalently $[n_j, \varphi_j] =  -i$ in the small-phase limit)~\cite{Koch2007}. For notational simplicity, we omit operator hats throughout. Since Eq.~\eqref{eq:H_SQUID} retains the same form for any gauge choice, in the following discussions we will suppress the explicit branch index, with the gauge dependence carried by the modulation parameters.

As revealed by Eqs.~\eqref{eq:app-SQUID-modulation-params1} and \eqref{eq:app-SQUID-modulation-params2}, the modulation parameters depend on the applied external voltage, capacitances, and loop magnetic fluxes of the lumped-element circuit, which are intimately related to the external field profile and the circuit geometry. Such a lumped-element model, mapping the complicated geometry of a realistic circuit into loops and nodes interacting with the external electromagnetic field to accurately capture the EMF effect, is generally non-trivial to obtain. In this work, we show how this challenge can be bypassed, by presenting three numerical methods for extracting the driven circuit Hamiltonian without requiring such an effective lumped-element circuit description. 

Through a similar derivation (refer to Appendix \ref{sec:app_describing_circuits}), 
we can write down the general form of the Hamiltonian for a driven $N$-mode circuit of $M$ Josephson junctions and $W$ linear inductors as
\begin{align}
\label{eq:multimode_Hamiltonian_start}
\mathcal{H} &= \sum_{i,j = 1}^N 4 \elementdepsymbol{E}{C}{ij} n_i n_j + \sum_{i=1}^N \epsilon_i(t) n_i \nonumber \\* \nonumber 
&\quad + \frac{1}{2} \sum\limits_{\ell=1}^W \elementdepsymbol{E}{L}{\ell} {{{\left( {\sum\limits_{i=1}^N {{\elementdepsymbol{b}{L}{\ell i}}{{\varphi }_i}} } + \elementdepsymbol{\mathcal{F}}{L}{\ell}(t) \right)}^2}}\\* 
&\quad - \sum\limits_{k=1}^{M} {E_{\mathrm{J}_k}\cos {{\left( {{\sum\limits_{i=1}^N {{b_{\mathrm{J}_{ki}}}{\varphi_i}} + \mathcal{F}_{\mathrm{J}_k}(t) }} \right)}}},
\end{align}
where $E_{C_{ij}}$ is the charging energy between the $i$-th and $j$-th modes, $\epsilon_i(t)$ is the time-dependent charge modulation on the $i$-th mode, including both ac and dc drives, $\elementdepsymbol{E}{L}{\ell} = \phi_0^2/L_\ell$ is the energy of the $\ell$-th inductor, $\elementdepsymbol{b}{L}{\ell i}$, $\elementdepsymbol{b}{J}{ki}$ are the participation factors of the $i$-th mode's phase across the $\ell$-th inductor and $k$-th junction, and $\elementdepsymbol{\mathcal{F}}{L}{\ell}(t)$, $\elementdepsymbol{\mathcal{F}}{J}{k}(t)$ represent the (combined ac and dc) phase modulations on the $\ell$-th inductor and $k$-th junction, respectively.

The lumped-element derivation leading to Eq.~\eqref{eq:multimode_Hamiltonian_start} is used here to establish the general Hamiltonian structure and to identify the charge- and phase-modulation parameters that enter a driven Josephson circuit. For a realistic device, the numerical methods developed below do not require the construction of this effective lumped-element model explicitly. Instead, they determine the required modulation parameters directly from layout-level electromagnetic response. In the remainder of the paper, we demonstrate the three numerical methods, using Eq.~\eqref{eq:multimode_Hamiltonian_start} as a starting point. The general workflow will be to apply a unitary transformation to the Hamiltonian, yielding a new Hamiltonian in a specific frame where the modulation parameters can be conveniently extracted from microwave simulations. We also discuss how these methods enable useful applications such as coherent evolution or decoherence rate calculations.

\section{NUMERICALLY OBTAINING THE HAMILTONIAN PARAMETERS}
\subsection{Junction-centered approaches}

In this section, we present two junction-centered methods for obtaining the effective Hamiltonian and modulation parameters of a driven Josephson circuit from the layout-level microwave response. In both cases, we first write the Hamiltonian in an appropriate frame or gauge through time-dependent unitary transformations~\cite{Gambetta2006,Boissonneault2009, Verney_2019,You2019,Riwar2022}, and then show how the corresponding modulation parameters are determined by the induced junction response from signals applied at the drive ports. This response can be obtained either directly from finite-element simulations or from the impedance matrix of the linearized circuit.

\subsubsection{The displaced-frame (DF) method}
\label{sec:displaced-frame}
The displaced-frame (DF) Hamiltonian is a useful representation of the driven superconducting circuit, providing important information regarding the parametric processes arising from the stimulated nonlinearities. Here, we first obtain the general form of the displaced-frame Hamiltonian for driven circuits, and then discuss how to extract Hamiltonian parameters from classical microwave simulations, using a concrete example for clarity.

A convenient choice for establishing the displaced-frame Hamiltonian is to work with the ladder operator basis, by diagonalizing the linear and quadratic parts of the Hamiltonian in Eq.~(\ref{eq:multimode_Hamiltonian_start}), followed by canonical quantization. Doing so results in the following Hamiltonian
\begin{align}
\begin{split}
\mathcal{H} &= \mathcal{H}_\textrm{lin}+\mathcal{H}_\textrm{nl},\\
\mathcal{H}_{\text{lin}}&=\sum_{i=1}^N \hbar \omega_i a_i^\dagger a_i + i \sum_{i=1}^N \tilde{\epsilon}_i (t) (a_i^\dagger-a_i)\\
&+ \sum_{\ell=1}^W \elementdepsymbol{E}{L}{\ell} \elementdepsymbol{\mathcal{F}}{L}{\ell}(t) \sum_{i=1}^N \elementdepsymbol{\beta}{L}{\ell i} (a_i + a_i^\dagger)\\
&+ \sum_{k=1}^M \elementdepsymbol{E}{J}{k} \elementdepsymbol{\mathcal{F}}{J}{k}(t) \sum_{i=1}^N \elementdepsymbol{\beta}{J}{ki} (a_i + a_i^\dagger),\\
\mathcal{H}_{\mathrm{nl}}&=-\sum_{k=1}^M \elementdepsymbol{E}{J}{k} \cos_{\mathrm{nl}} \left( \sum_{i=1}^N \elementdepsymbol{\beta}{J}{ki} (a_i + a_i^\dagger) + \elementdepsymbol{\mathcal{F}}{J}{k}(t) \right).
\end{split}
\label{eq:dfh-Hamiltonian-lin-nl}
\end{align}
Here $\mathcal{H}_\textrm{lin}$ is the Hamiltonian of the linearized circuit, describing normal modes with angular frequencies $\omega_i$, coupled to the external charge and phase modulations. The nonlinear part, $\mathcal{H}_\textrm{nl}$, represents the phase-modulated Josephson energies beyond the quadratic order, with $\cos_{\mathrm{nl}}(x) = \cos(x) + x^2/2$. The normal mode participation ratios in the junctions $\elementdepsymbol{\beta}{J}{ki}$ can be readily obtained from finite-element simulations using methods described in \cite{Nigg2012, Minev2021}. The inductor participation ratios $\elementdepsymbol{\beta}{L}{\ell i}$ do not need to be computed for the final displaced-frame Hamiltonian, since they are eliminated by the displacement transformation and do not appear in Eq.~\eqref{eq:dfh-Hamiltonian-displaced}. They are therefore not required for the DF dynamics and rates considered in this work. Additional details are required to perform the explicit displacement transformation when evaluating certain lab-frame dynamics; see Appendix~\ref{app:df_vs_lab}. In the presence of non-zero dc-flux bias, the normal modes are defined by expanding the potential around the minimum (or minima), see Appendix \ref{app:dc_flux} for more information. It is worth noting that since Eq.~\eqref{eq:dfh-Hamiltonian-lin-nl} treats the circuit as linearized harmonic oscillators with Josephson nonlinearity as an addition, it is only exact for circuits with only extended phase variables, such as the fluxonium. For circuits with periodic boundary conditions, such as the transmon, this Hamiltonian serves as an approximation because it does not satisfy the required periodicity (see Appendix~\ref{app:boundary_conditions}). Despite this limitation, it effectively describes many practical scenarios, such as the low-lying states of the transmon, though it does not account for charge dispersion~\cite{Koch2007}.

Instead of focusing on finding out the modulation parameters $\tilde{\epsilon}_i (t) $, $\elementdepsymbol{\mathcal{F}}{L}{\ell}(t)$, and $\elementdepsymbol{\mathcal{F}}{J}{k}(t)$ in Eq.~
\eqref{eq:dfh-Hamiltonian-lin-nl}, we eliminate them by performing the following displacement transformation:
\begin{align}
    &U_{\text{disp}} = \exp\left({\sum_{i=1}^N \left(\xi_i(t) a_i^\dagger - \xi_i(t)^* a_i\right)}\right),
    \label{eq:dfh-transform}
\end{align}
where $\xi_i(t)$ represents the coherent displacement (classical response) of mode $i$ induced by the applied modulations, obtained from the damped Heisenberg equation of motion (refer to Appendix \ref{sec:app_dfh}), 
\begin{align}
\dot{a}_i(t) = \frac{i}{\hbar}\left[\mathcal{H}_{\text{lin}},a_i(t)\right] 
- \frac{\kappa_i}{2} a_i(t),\quad\xi_i(t)=\langle a_i(t) \rangle.
\label{eq:dfh-HEOM}
\end{align}
Here, $\kappa_i$ is the total damping rate of mode $i$, which can include both external loading through the drive ports and internal material loss modeled in the simulation. The displacement transformation effectively shifts away the coherent classical component in the ladder operators, leaving them with only the quantum fluctuations around the classical trajectories. 
Using the solution to Eq.~(\ref{eq:dfh-HEOM}), the full displaced-frame Hamiltonian becomes 
\begin{align}
\begin{split}
\mathcal{H}_{\text{disp}} &= \sum_{i=1}^N \hbar \omega_i a_i^\dagger a_i \\
&-\sum_{k=1}^M \elementdepsymbol{E}{J}{k} \cos_{\mathrm{nl}}\left(\sum_{i=1}^N\elementdepsymbol{\beta}{J}{ki} (a_i +a_i^\dagger)+A_k(t) \right), 
\end{split}
\label{eq:dfh-Hamiltonian-displaced}
\end{align}
where $A_k(t) = \sum_{i=1}^N\elementdepsymbol{\beta}{J}{ki}\left(\xi_i(t)+\xi_i^*(t)\right)+\elementdepsymbol{\mathcal{F}}{J}{k}(t)$ is the phase displacement on the $k$-th junction. Considering the linearized circuit, where the Josephson junctions are taken as linear inductors of inductance $\elementdepsymbol{L}{J}{k}$, a key observation is that the value of $\langle a_i + a_i^\dagger\rangle$ is zero in this displaced frame. Hence, $A_k(t)$ can be obtained from microwave simulations as the phase across the $k$-th linearized junction. It is also important to note that, while the beginning Hamiltonian in Eq.~\eqref{eq:dfh-Hamiltonian-lin-nl} is specific to the gauge choice, $A_k(t)$ is gauge-invariant as it represents the physical observable of the (linearized) Josephson phase under drive (see discussion in Appendix~\ref{app:disp_invariance}). 

The form of Eq.~\eqref{eq:dfh-Hamiltonian-displaced} is widely used for analyzing parametrically driven superconducting circuits and the resulting frequency-conversion processes. In many applications, the displacement $A_k(t)$ is obtained from a reduced circuit model or calibrated from experiment. Here, we instead determine $A_k(t)$ directly from the microwave response of the physical layout. In practice, it is convenient to obtain the phase displacements from simulated voltages across the linearized junctions. Although the same procedure applies in principle to arbitrary waveforms (see Appendix~\ref{app:general_waveforms}), for simplicity, we assume monochromatic sinusoidal modulation at frequency $\omega_d$ such that
\begin{equation}
    A_k(t) = 
    \Real[\bar{A}_k e^{i\omega_d t}]
    = |\bar{A}_k| \cos [\omega_d t + \arg(\bar{A}_k)],
\end{equation}
where $\bar{A}_k = |\bar{A}_k| e^{i \arg(\bar{A}_k)}$ is the phasor containing the magnitude and phase information of $A_k(t)$. The time derivative of the flux across the $k$-th junction, $\phi_0\dot{A}_k(t)$, is then simply equal to the voltage across the same junction,
\begin{align}
    \frac{d}{dt} \phi_0 \bar{A}_k e^{i \omega_d t} 
    = i \phi_0\omega_d \bar{A}_k e^{i \omega_d t} 
    = \elementdepsymbol{\bar{V}}{J}{k} e^{i\omega_d t},
\end{align}
where $\elementdepsymbol{\bar{V}}{J}{k} $ represents the voltage phasor across the $k$-th junction. In vector form, we may then write
\begin{align}
    \mathbf{\bar{A}} = \frac{\elementdepsymbol{\mathbf{\bar{V}}}{\mathbf{J}}{}}{i \phi_0 \omega_d},
    \label{eq:A-vs-V}
\end{align}
where $\mathbf{\bar{A}}=\left(\bar A_{1},\bar A_{2},\cdots,\bar A_{M}\right)^\textrm{T}$ and $\elementdepsymbol{\mathbf{\bar{V}}}{\mathbf{J}}{}=\left(\elementdepsymbol{\bar{V}}{J}{1},\elementdepsymbol{\bar{V}}{J}{2},\cdots,\elementdepsymbol{\bar{V}}{J}{M}\right)^\textrm{T}$. The phase displacements $\bar{\mathbf{A}}$ can then be obtained from the voltages $\mathbf{\bar{V}_J}$, calculated by integrating the electric field phasor $\bar{\mathbf{E}}$ along the directed line DL$_k$ of the $k$-th junction,
\begin{align}
    \bar{A}_k = \frac{1}{i \phi_0 \omega_d}\int_{\text{DL}_k}\bar{\mathbf{E}}\cdot \mathbf{dl}.
    \label{eq:A_line_int}
\end{align}
Crucially, we use the same directed line for calculating the mode participation ratios in the $k$-th junctions, $\elementdepsymbol{\beta}{J}{ki}$~\cite{Minev2021}, ensuring the relative phase between the phase displacement and the mode participation is correctly captured for each junction.

\begin{figure*}
    \centering
    \includegraphics[width=\textwidth,
    ]{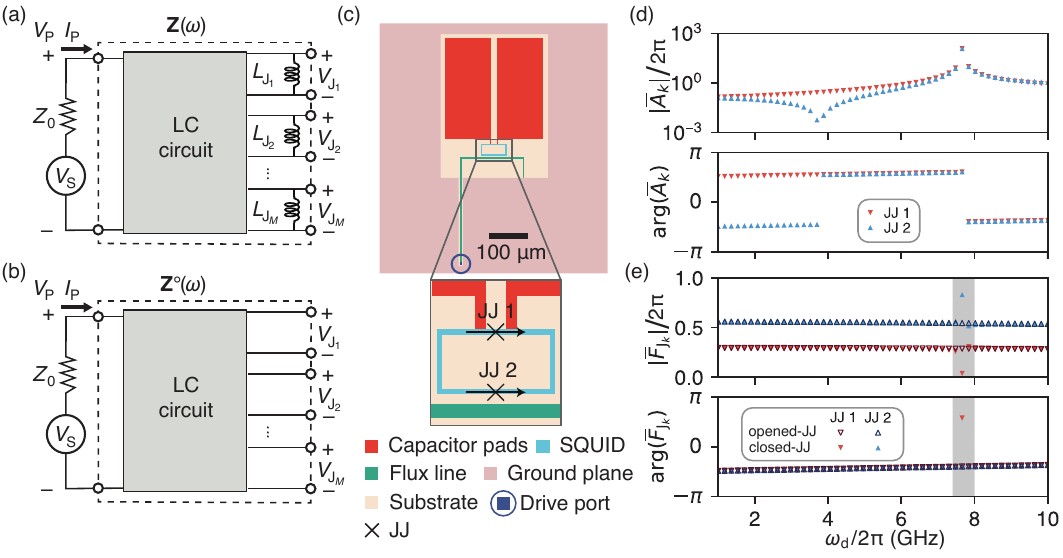}
    \caption{\textbf{Illustration of using the junction-centered approaches.}
    (a) Black-box representation of Josephson circuit with $M$ linearized Josephson junctions driven by a voltage source $V_{\text{S}}$ with internal impedance $Z_0$. The impedance matrix $\mathbf{Z}(\omega)$ is used to obtain the modulation parameters in the displaced frame (junction phase displacements) or the irrotational gauge Hamiltonians (effective phase modulations) using Eqs.~(\ref{eq:A-from-Z}) and (\ref{eq:f-from-Z}), respectively. (b) The irrotational gauge effective phase modulations can also be obtained using the impedance matrix of the same circuit, but with the junctions opened, using Eq.~(\ref{eq:f-from-Z-open}). (c) Illustration of a SQUID-based flux-tunable transmon circuit that is modulated by an on-chip flux line (green). The two arrows indicate the directed lines $\mathrm{DL}_1$ and $\mathrm{DL}_2$, along which the phase displacements of the two junctions are evaluated in Eq.~\eqref{eq:A_line_int}. The inductances of the two junctions are $\elementdepsymbol{L}{J}{1} = \elementdepsymbol{L}{J}{2} = 15$ nH, and the (linearized) mode frequency is 7.673 GHz. The input power is set to be -50\,dBm.
    (d) Phase displacements $\bar{A}_1$ and $\bar{A}_2$ of the two junctions in circuit (c) obtained with the displaced-frame method. Magnitudes and phases are shown separately. (e) Effective phase modulation parameters $\elementdepsymbol{\bar{F}}{J}{1}$ and $\elementdepsymbol{\bar{F}}{J}{2}$ of the same circuit obtained with the irrotational-gauge method, where filled markers refer to the circuit with closed junctions (a) and empty markers to opened junctions (b). Magnitudes and phases are shown separately. The shaded region indicates the frequency range where numerical convergence is difficult to reach for the closed junction approach due to the mode resonance, whereas the opened junction approach mitigates this issue and provides correct results over the entire frequency range.
}
    \label{fig:junction-centered}
\end{figure*} 

A simpler alternative for obtaining the phase displacements, motivated by microwave engineering, is to relate the modulation parameters to the network response functions (such as the impedance matrix) of the linearized circuit~\cite{Solgun2019,Naaman2022,Chapman2023,Labarca2024}. Here, to keep the notation simple, we focus on the case where the circuit is driven by a single drive port; when multiple drive ports are present, the resulting modulation parameters are simply the sum of the contributions of each individual drive port. Consider Fig.~\hyperref[fig:junction-centered]{\ref*{fig:junction-centered}a}, where the black box represents the circuit excluding the drive port and the junctions. The system is driven by applying a source voltage $V_{\mathrm{S}}$ through the drive port of impedance $Z_0$. From the definition of the impedance
\begin{align}
    Z_{mn} = \frac{\bar{V}_m}{\bar{I}_n} \Bigg|_{I_\ell = 0 \text{ for } \ell \neq n},
\end{align}
where $\bar{V}_m$ represents the (complex) voltage across the port $m$ and $\bar{I}_n$ the current from port $n$, we find that
\begin{align}
\elementdepsymbol{\bar{V}}{J}{k} &= \bar{I}_{\mathrm{P}} Z_{\textrm{J}_k \textrm{P}},\label{eq:Zjp}\\
\bar{V}_{\mathrm{P}} &= \bar{I}_{\mathrm{P}} Z_{\mathrm{P}\mathrm{P}}.\label{eq:Zpp}
\end{align}
Here, $Z_{\textrm{J}_k \textrm{P}}$ is the transfer impedance between the drive port $\mathrm{P}$ and the $k$-th junction, while $Z_{\mathrm{P}\mathrm{P}}$ is the input impedance as seen from port $\mathrm{P}$. Similar to the direct integral approach in Eq.~\eqref{eq:A_line_int}, the sign of the junction voltage in the impedance calculation needs to be in accordance with the directed line specified for each junction. The port and directed-line conventions used in these simulations are summarized in Appendix~\ref{app:simulation_port_conventions}. Defining $\mathbf{Z_J} = \left(Z_{\mathrm{J}_1\mathrm{P}},Z_{\mathrm{J}_2\mathrm{P}},\cdots,Z_{\mathrm{J}_M\mathrm{P}}\right)^\textrm{T}$, Eq.~(\ref{eq:A-vs-V}) can be rewritten as
\begin{align}
\mathbf{\bar{A}} = \frac{\bar{I}_\mathrm{P}}{i\phi_0 \omega_d} \mathbf{Z_J} = \frac{\bar{V}_\mathrm{P}}{i\phi_0 \omega_d Z_{\mathrm{P}\mathrm{P}}} \mathbf{Z_J}.
\label{eq:A_from_Vp}
\end{align}
As this expression is to be used in electromagnetic simulations, it can be more convenient to re-express $\bar{V}_\mathrm{P}$ in terms of the source voltage $\bar{V}_{\mathrm{S}}$ and the maximum available power from the source $P$ \cite{Pozar, power_comment}. From the expression for $P$ and voltage division at the input port of the black box
\begin{align}
P = \frac{|\bar{V}_\mathrm{S}|^2}{8Z_0},\quad 
\bar{V}_\mathrm{P} = \frac{Z_{\mathrm{P}\mathrm{P}}}{Z_0+Z_{\mathrm{P}\mathrm{P}}}
\bar{V}_{\mathrm{S}},
\label{eq:Vs-vs-P-and-Vp}
\end{align}
we obtain
\begin{align}
\mathbf{\bar{A}} = \frac{2\sqrt{2 P Z_0}}{i\phi_0 \omega_d \left( Z_0 + Z_{\mathrm{P}\mathrm{P}} \right)} \mathbf{Z_J}.
\label{eq:A-from-Z}
\end{align}
This expression shows how the impedance matrix is used in the present extraction procedure: the transfer impedance from the physical drive port to the directed junction terminal pairs determines the junction phase displacements $\mathbf{\bar{A}}$ that enter the DF Hamiltonian. In reduced-model impedance or admittance treatments, the relevant qubit ports or circuit degrees of freedom are typically assumed to have already been identified. Here, by contrast, the impedance matrix supplies the layout-level map from a physical drive port to the drive-induced junction response itself. The required circuit impedances can be extracted from electromagnetic simulations or analytically calculated if the lumped-element circuit representation is known. In Fig.~\hyperref[fig:junction-centered]{\ref*{fig:junction-centered}d}, we present the results of applying Eq.~\eqref{eq:A-from-Z} to the circuit shown in Fig. \hyperref[fig:junction-centered]{\ref*{fig:junction-centered}c}, using Ansys HFSS finite-element simulations.

The displaced-frame Hamiltonian reveals how the combination of the drive and the Josephson nonlinearity generates parametric frequency-conversion processes, making it a useful tool for analyzing parametrically-driven circuits. From the displaced-frame Hamiltonian [Eq.~\eqref{eq:dfh-Hamiltonian-displaced}], we can numerically compute experimentally-relevant process rates such as for the beamsplitter interaction~\cite{Chapman2023,Lu2023}, two-mode squeezing \cite{zhou2024}, subharmonic qubit drive~\cite{Xia2023, Sah2024, Fillip2024}, driven-dissipative reset and stabilization~\cite {Yao2017, Pfaff2017,Nie2024, Li2024}, as well as drive-induced frequency shifts due to the ac Stark effect~\cite{ACStarkshift}. 
On the other hand, while the DF method captures the junction phase displacement $A_k(t)$ entering Eq.~\eqref{eq:dfh-Hamiltonian-displaced}, it does not by itself reconstruct the full time-dependent displacement transformation needed to evaluate arbitrary lab-frame dynamics at intermediate times. This distinction matters most when interpreting dynamics during a finite pulse: without knowing
the displacement transformation, agreement between displaced-frame and lab-frame evolutions typically relies on ramp segments that adiabatically prepare and undo the relevant displacement~\cite{Zhang2019}. Appendix~\ref{app:df_vs_lab} gives the corresponding ramp conditions and a numerical example illustrating both agreement and failure of the steady-state DF description. In contrast, the lab-frame Hamiltonian offers a more direct representation for the dynamics of the system, thereby allowing for a broader application that goes beyond these limitations. 

\subsubsection{The irrotational-gauge (IG) method}
\label{sec:irrotational-gauge}
As an alternative to the displaced-frame method, we introduce the irrotational-gauge method in this section, which provides the lab-frame Hamiltonian of the driven circuit in a form where the external drives explicitly couple to the Josephson inductive elements. Compared to earlier discussions on establishing the IG Hamiltonian for driven circuits~\cite{You2019,Riwar2022}, our approach can handle multimodal, multi-junction circuits with realistic geometry in the presence of both electrical and magnetic drive fields using well-established microwave simulation techniques. In what follows, we only consider circuits whose inductive elements are Josephson junctions. The method can also be generalized to include lumped-element inductors; however, its applicability to systems with distributed inductance remains limited (see Appendix \ref{app:lin_inductance} for more details). 

To arrive at the irrotational gauge Hamiltonian, we first perform the following time-dependent unitary transformation to Eq.~\eqref{eq:multimode_Hamiltonian_start} (refer to Appendix \ref{sec:app_irg}), 
\begin{align}
U_{\text{irr}} = \exp \left\{ -\frac{i}{\hbar} \int^t_{0}{dt' \sum_{i=1}^N \epsilon_i(t') n_i} \right\},
\label{eq:irg-transform}
\end{align}
where the drive begins at $t'=0$. Under this gauge transformation, the circuit Hamiltonian becomes
\begin{align}
\label{eq:multimode_Hamiltonian_IG}
\mathcal{H_{\text{irr}}} &= \sum_{i,j=1}^N 4 \elementdepsymbol{E}{C}{ij}n_i n_j  - \sum\limits_{k=1}^M {\elementdepsymbol{E}{J}{k}\cos {{\left( {{\sum\limits_{i=1}^N {{\elementdepsymbol{b}{J}{ki}}{\varphi_i}} + \elementdepsymbol{F}{J}{k}(t) }} \right)}}},
\end{align}
where $\elementdepsymbol{F}{J}{k}(t) = \elementdepsymbol{\mathcal{F}}{J}{k}(t) + \frac{1}{\hbar} \int^t_0 dt' \sum_{i=1}^N \elementdepsymbol{b}{J}{ki}  \epsilon_i(t')$ is the effective phase modulation in the irrotational gauge. The second term in $\elementdepsymbol{F}{J}{k}(t)$ absorbs the charge-drive contribution into the Josephson phase; related phase translations have been used for pure charge drives~\cite{Verney_2019}. Here, $\epsilon_i(t)$ may include both explicit capacitive drives and EMF-induced terms associated with time-dependent flux, whose distribution among the coordinates depends on the chosen gauge, as illustrated in Eq.~\eqref{eq:app-SQUID-modulation-params1}.

For convenience in the numerical implementation, we express this Hamiltonian in the normal-mode basis of the linearized circuit,
\begin{align}
\begin{split}
\mathcal{H}_{\text{irr}} &= \sum_{i=1}^N \hbar \omega_i a_i^\dagger a_i + \sum_{k=1}^M \elementdepsymbol{E}{J}{k}\elementdepsymbol{F}{J}{k}(t)  \sum_{i=1}^N \elementdepsymbol{\beta}{J}{ki} (a_i + a_i^\dagger)\\
&- \sum_{k=1}^M \elementdepsymbol{E}{J}{k} \cos_\textrm{nl} \left( \sum_{i=1}^N \elementdepsymbol{\beta}{J}{ki} (a_i + a_i^\dagger) + \elementdepsymbol{F}{J}{k}(t) \right).
\end{split}
\label{eq:irg-Hamiltonian-ladder}
\end{align} 
It is worth noting that while the IG Hamiltonian in Eq.~\eqref{eq:irg-Hamiltonian-ladder} and the DF Hamiltonian in Eq.~\eqref{eq:dfh-Hamiltonian-displaced} share a similar algebraic form, the modulation terms have different meanings in the two representations. The IG Hamiltonian captures the classical interaction between the circuit and the external fields (the second term in Eq.~\eqref{eq:irg-Hamiltonian-ladder}), whereas the DF Hamiltonian only contains the classical response of the linearized circuit as a result of the field-circuit interaction. 

Next, we explain how the drive parameters $\elementdepsymbol{F}{J}{k}(t)$ can be obtained from microwave simulations. To see this, we first note that the classical response of the ladder operator $a_i$ can be obtained, in a similar fashion to Eq.~\eqref{eq:dfh-HEOM}, as the solution to the equation of motion
\begin{align}
\dot{a}_i &= \frac{i}{\hbar}\left[ \mathcal{H}_\text{irr}^\text{lin}, a_i \right]
-\frac{\kappa_i}{2}a_i\nonumber\\
&= -i \omega_i a_i - \frac{\kappa_i}{2}a_i - \frac{i}{\hbar} \sum_{k=1}^M \elementdepsymbol{E}{J}{k} \elementdepsymbol{\beta}{J}{ki} \elementdepsymbol{F}{J}{k}(t).
\label{eq:irg-HEOM}
\end{align}
Here, $\mathcal{H}_\text{irr}^\text{lin}$ is the linearized circuit part of the IG Hamiltonian (i.e., the first line in Eq.~\eqref{eq:irg-Hamiltonian-ladder}). 
For simplicity, assuming monochromatic sinusoidal modulation at frequency $\omega_d$ (see Appendix~\ref{app:general_waveforms} for the arbitrary-waveform case), we represent the effective phase modulation as a phasor
\begin{align}
\elementdepsymbol{F}{J}{k}(t) = \Real [\elementdepsymbol{\bar{F}}{J}{k} e^{i \omega_d t}],
\label{eq:irg-sinusoidal-mod}
\end{align}
containing both magnitude and phase information. The steady-state solution to Eq.~(\ref{eq:irg-HEOM}) is then
\begin{align}
    \left\langle a_i + a_i^\dagger \right\rangle 
    &= 
    \Real\Bigg\{
    \frac{2 \omega_i}{\hbar \left(\omega_d^2 - \omega_i^2 -\kappa_i^2/4 - i\kappa_i \omega_d \right)} \nonumber \\
    &\times \sum_{k=1}^M \elementdepsymbol{E}{J}{k}\elementdepsymbol{\beta}{J}{ki} \elementdepsymbol{\bar{F}}{J}{k} e^{i \omega_d t}
    \Bigg\},
    \label{eq:irg-ai-steady-state}
\end{align}
where $\kappa_i$ is the damping rate of mode $i$. 
As introduced in the displaced-frame method section, we denote the junction phase displacement of the $j$-th junction as
\begin{equation}
    A_j(t) = \left\langle\sum_{i=1}^N \elementdepsymbol{\beta}{J}{ji} (a_i + a_i^\dagger) + \elementdepsymbol{F}{J}{j}(t)\right\rangle= \Real[\bar{A}_j e^{i \omega_d t}].\label{eq:irg-A(t)}
\end{equation}
From Eqs.~\eqref{eq:irg-ai-steady-state} and \eqref{eq:irg-A(t)}, we establish the relation between the effective phase modulation parameters $\elementdepsymbol{\bar{F}}{J}{k}$ and the junction phase displacements $\bar{A}_j$ as a compact matrix equation:
 \begin{align}
     \elementdepsymbol{\bar{\mathbf{F}}}{\mathbf{J}}{} =  \mathbf{R}^{-1} \mathbf{\bar{A}}.\label{eq:irg-matrix-eq}
 \end{align}
 Here, $\elementdepsymbol{\bar{\mathbf{F}}}{\mathbf{J}}{}$ and $\mathbf{\bar{A}}$ are vectors $\mathbf{\elementdepsymbol{\bar{F}}{J}{}}
=\left(\elementdepsymbol{\bar{F}}{J}{1}, \elementdepsymbol{\bar{F}}{J}{2},\cdots,\elementdepsymbol{\bar{F}}{J}{M}\right)^\textrm{T}$ and $\mathbf{\bar{A}}=\left(\bar{A}_1, \bar{A}_2,\cdots,\bar{A}_M\right)^\textrm{T}$, and $\mathbf{R}$ is the $M\times M$ matrix connecting them,
\begin{align}
(\mathbf{R})_{jk}
=\sum_{i=1}^N\frac{2\omega_i\elementdepsymbol{E}{J}{k}\elementdepsymbol{\beta}{J}{ji}\elementdepsymbol{\beta}{J}{ki}}{\hbar \left(\omega_d^2 - \omega_i^2 -\kappa_i^2/4 - i\kappa_i \omega_d \right)}
+\delta_{jk},  \label{eq:irg-r-matrix}
\end{align}
with $\delta_{ij}$ the Kronecker delta. We note that Eq.~\eqref{eq:irg-matrix-eq} assumes the matrix $\mathbf{R}$ is invertible, which we have consistently observed in our simulations. However, a formal investigation of the invertibility of $\mathbf{R}$ is warranted and left for future work.

The effective phase modulation parameters $\elementdepsymbol{\bar{\mathbf{F}}}{\mathbf{J}}{}
$ can be calculated from Eqs.~\eqref{eq:irg-matrix-eq} and \eqref{eq:irg-r-matrix}, knowing the junction phase displacement $\mathbf{\bar{A}}$ from the DF method simulation, and the $\boldsymbol{\mathbf{R}}$ matrix from the eigenmode EPR simulation.
However, when the drives are nearly resonant with a circuit eigenmode, the calculation of $\elementdepsymbol{\bar{\mathbf{F}}}{\mathbf{J}}{}
$ may become particularly sensitive 
to the values of $\mathbf{\bar{A}}$ (see Appendix \ref{app:IG_sensitivity} 
for detailed discussions). This sensitivity implies that uncertainties in $\mathbf{\bar{A}}$, for instance from finite-element discretization, can be strongly amplified in the extraction of $\elementdepsymbol{\bar{\mathbf{F}}}{\mathbf{J}}{}
$, leading to significant uncertainties.
In addition, the $\mathbf{R}$ matrix must contain all the modes with significant contribution to the junction phase displacement (see Appendix~\ref{app:mode_truncation} for detailed discussions). 

Remarkably, we show that there exists a simpler approach for obtaining $\elementdepsymbol{\bar{\mathbf{F}}}{\mathbf{J}}{}
$ without requiring the computation of $\boldsymbol{\mathbf{R}}$. Recalling that the junction phase displacement can be expressed in terms of the impedance parameters and the drive power [Eq.~\eqref{eq:A-from-Z}], we rewrite Eq.~\eqref{eq:irg-matrix-eq} as
\begin{align}
\elementdepsymbol{\bar{\mathbf{F}}}{\mathbf{J}}{}
= \frac{2 \sqrt{2 P Z_0} }{i\phi_0 \omega_d \left( Z_0 + Z_{\mathrm{P}\mathrm{P}} \right)} \boldsymbol{\mathbf{R}}^{-1} \mathbf{Z_J}.
\label{eq:f-from-Z}
\end{align}
Because the modulation parameters $\elementdepsymbol{\bar{\mathbf{F}}}{\mathbf{J}}{}$ are set by the external field-circuit coupling and gauge allocation, rather than by the normalization of the linearized modes, they are independent of the junction kinetic inductances; see Appendix~\ref{sec:app_irg}. This is in contrast to the junction phase displacement $\bar{\mathbf A}$ in Eq.~\eqref{eq:A-from-Z}, which depends on the modes of the linearized circuit and therefore on the linearized junction inductances. We can therefore simplify Eq.~(\ref{eq:f-from-Z}) by taking the opened-junction limit $\elementdepsymbol{L}{J}{k}\rightarrow \infty$ ($\elementdepsymbol{E}{J}{k} \rightarrow 0$) in Eq.~(\ref{eq:irg-r-matrix}). In this case, $\boldsymbol{\mathbf{R}}$ reduces to the identity matrix (see Appendix \ref{app:R_to_identity} for formal derivation),
and $\elementdepsymbol{\bar{\mathbf{F}}}{\mathbf{J}}{}$ simplifies to
\begin{align}
\elementdepsymbol{\bar{\mathbf{F}}}{\mathbf{J}}{} = \frac{2 \sqrt{2 P Z_0}}{i\phi_0 \omega_d \left( Z_0 + Z^\circ_{\mathrm{P}\mathrm{P}} \right)}\mathbf{Z^\circ_J}.
\label{eq:f-from-Z-open}
\end{align}
Here, the circle symbol refers to impedance parameters obtained for the transformed circuit where all the junctions are removed, corresponding to the opened-junction limit (see Fig. \hyperref[fig:junction-centered]{\ref*{fig:junction-centered}b}). 

For illustration, Fig.~\hyperref[fig:junction-centered]{\ref*{fig:junction-centered}e} compares two calculations of the parameters $\elementdepsymbol{\bar{F}}{J}{k}$ for the circuit in Fig. \hyperref[fig:junction-centered]{\ref*{fig:junction-centered}c}: one obtained from Eq.~\eqref{eq:irg-matrix-eq} using driven simulations of circuits with linearized Josephson inductance (labeled ``closed-JJ"), and the other from Eq.~\eqref{eq:f-from-Z-open} using driven simulations of circuits with opened junctions (labeled ``opened-JJ"). The two results show excellent agreement, except for the region around 7.8\,GHz close to the mode resonance. There, the modulation parameters $\elementdepsymbol{\bar{F}}{J}{k}$ obtained from Eq.~\eqref{eq:irg-matrix-eq} become more sensitive to errors in $\bar{\mathbf{A}}$ than at off-resonant frequencies, whereas the opened-JJ calculation circumvents this issue and yields continuous and accurate results across the resonance.  

Even though we adopted the linearized normal-mode basis for the IG Hamiltonian in this section, this choice is a practical representation that allows us to use the static Hamiltonian parameters obtained from EPR/BBQ-style linearized-mode analysis, such as mode frequencies and junction participation factors. It is not a restriction of the IG formulation to a perturbative Kerr-expanded regime. The established method can in principle be applied to lumped-element circuits with known bare-mode parameters (such that $\elementdepsymbol{E}{C}{ij}$ and $\elementdepsymbol{b}{J}{ki}$ in Eq.~\eqref{eq:multimode_Hamiltonian_IG} are readily available). In such cases, the IG Hamiltonian can be written in the charge-phase basis, useful for preserving the periodic boundary condition for certain kinds of circuits. On the other hand, the IG method requires that the effective phase modulations produced by the external drive can be assigned to a finite set of localized inductive branches with well-defined terminal pairs, such as Josephson junctions or lumped linear inductors. As a result, the IG method has limited applicability when the external drive produces non-negligible effective phase modulation of distributed inductive elements that are not associated with well-defined branch ports; see Appendix~\ref{app:lin_inductance} for further discussion.

\subsection{Field-centered approach: The \textit{overlap} method}
\label{sec:overlap-method}
The junction-centered methods allow for the extraction of useful circuit information by simulating the phase response of the junctions. However, they have limitations that hinder their applicability to broader classes of driven circuits. For example, if the modes of interest have significant energy participation in the distributed linear inductive elements that are strongly driven, the IG method is incapable of obtaining the complete modulation information. While the DF method is applicable to general circuits for obtaining the junction phase displacement, it is unable to predict the complete dynamics of the circuit where the lab-frame Hamiltonian is needed. In this section, we introduce a new approach that enables establishing the Hamiltonian in domains where DF or IG breaks down, which we dub the \textit{overlap} method. 

\begin{figure*}[t!]
  \centering 
  \includegraphics[width=\textwidth]{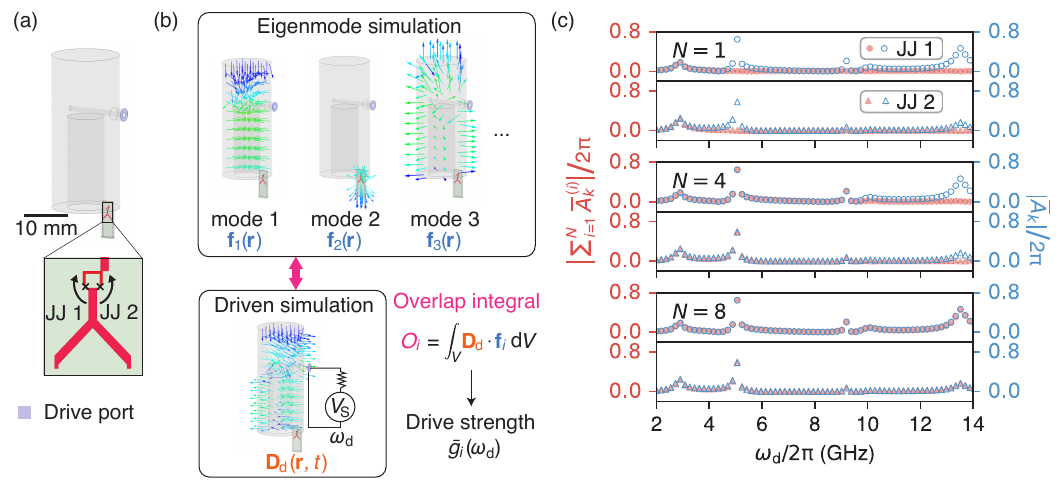}
  \caption{\textbf{Illustration of the overlap method.} 
  (a) An example circuit where a flux-tunable transmon device is inductively coupled to a stub cavity~\cite{Lu2023}. 
  (b) The overlap method involves two finite-element simulations: the eigenmode simulation of the electric field mode profile $\mathbf{f}_i$ of the $i$-th normal mode, and the driven simulation of the displacement field $\mathbf{D}_d$ at frequency $\omega_d$. By computing the overlap integral $O_i$ [Eq.~(\ref{eq:electric_overlap})] between $\mathbf{f}_i$ and $\mathbf{D}_d$, we can obtain the effective charge modulation $\bar g_i$ for mode $i$ through Eq.~(\ref{eq:gi_overlap}).
  (c) The amplitude of the $k$-th junction phase displacement contributed by the first $N$ modes $\lvert\sum_{i=1}^N\bar{A}_k^{(i)}\rvert$ (empty markers), compared against the amplitude of the total junction phase displacement $\lvert A_k\rvert$ (filled markers). The former is calculated from $\bar g_i$ obtained via the overlap method, while the latter is extracted using the displaced-frame method. Both simulations are performed under a drive power of -50\,dBm and a drive frequency range of $2-14$\,GHz. The contribution from the lowest $N$ modes progressively reproduces the total junction phase displacement as $N$ increases.
  }
\label{fig:overlap}
\end{figure*}

Figure~\ref{fig:overlap} provides a representative example of the overlap method using a hybrid lumped-distributed circuit. Starting from a model of the physical circuit, we decompose the electric field of the driven system into the contributions of each normal mode. In practice, this is achieved by computing the overlap integral between (i) the displacement field $\mathbf{D}_d$ of the system when driven at frequency $\omega_d$ and (ii) the electric field mode profile for each normal mode of the undriven circuit $(\mathbf{f}_i)$. Compared with junction- or branch-centered extraction, this field-centered projection captures drive coupling to distributed modes without assigning the driven response entirely to a finite set of localized circuit branches. Since the projection includes the port-driven response, the extracted quantities are port- and frequency-dependent drive amplitudes rather than static participation parameters. This enables us to extract the effective charge modulation $g_i(t)$ for each normal mode $i$ and, from this, predict important quantities such as coherent transition and decoherence rates due to the drive ports.

\subsubsection{Obtaining the Hamiltonian}
We start from Eq.~(\ref{eq:multimode_Hamiltonian_start}) and perform a time-dependent unitary transformation to obtain the Hamiltonian useful for the overlap method (refer to Appendix \ref{sec:app-overlap-Hamiltonian} for derivation). This transformation effectively brings the circuit Hamiltonian into an ``opposite" gauge to the irrotational gauge, where the drive terms manifest explicitly as linear charge drives, rather than as linear flux drives. The resulting Hamiltonian is
\begin{align}
\label{eq:overlap-ov-Hamiltonian}
    \mathcal{H}_{\text{ov}} &= \sum_{i=1}^N \hbar \omega_i a_i^\dagger a_i + i \sum_{i=1}^N g_i(t) (a_i^\dagger - a_i) \nonumber \\ 
    &- \sum_{k=1}^M \elementdepsymbol{E}{J}{k} \cos_\mathrm{nl} \left( \sum_{i=1}^N \elementdepsymbol{\beta}{J}{ki} (a_i + a_i^\dagger) +  A_{\mathrm{res}, k}(t) \right),
\end{align}
where $g_i(t)$ is the effective charge modulation parameter of the $i$-th mode, and $A_{\mathrm{res}, k}(t)$ is the ``residual" phase displacement across the $k$-th junction after elimination of the linear flux drive terms, which only couples to the nonlinear Josephson energy.

\subsubsection{Obtaining the modulation parameters}
Under sinusoidal drive (see Appendix~\ref{app:general_waveforms} for the extension to arbitrary waveforms), the effective charge modulation parameter of the $i$-th mode can be expressed in phasor representation as
\begin{align}
g_i(t)=\Real[\bar g_i(\omega_d)e^{i\omega_d t}].\label{eq:g(t)}
\end{align}
The phasor of the effective charge modulation parameter, $\bar g_i$, is then connected to the ``electric overlap" phasor, $\bar O_i$, by (see Appendix \ref{sec:app-overlap-derivation} and \ref{app:overlap_EM} for two alternative derivations)
\begin{equation}
    \bar g_i(\omega_d) =\left[\frac{\omega_d^2 - \omega_i^2 -\kappa_i^2/4- i\omega_d \kappa_i}{\omega_i^2 + \kappa_i^2/4}\right]
    \sqrt{\frac{\hbar\omega_i}{2}}\bar O_i(\omega_d),
    \label{eq:gi_overlap}
\end{equation}
where the effects of weak dissipation are included via the damping parameter $\kappa_i$. For simplicity, Eq.~\eqref{eq:gi_overlap} assumes that the dissipation is solely due to the port load, so $\kappa_i$ denotes the port-induced linewidth. The more general case with nonzero intrinsic loss, together with practical ways to determine the linewidth contributions, is given in Appendix~\ref{sec:app-overlap-derivation}. The electric overlap $O_i(t)$ and its phasor $\bar O_i(\omega_d)$ at frequency $\omega_d$ are defined as
\begin{align}
O_i(t) = \Real[\bar O_i(\omega_d) e^{i\omega_d t}] =\int_V \mathbf{D}_d (\mathbf{r}, t) \cdot \mathbf{f}_i(\mathbf{r})\,dV, \label{eq:electric_overlap}
\end{align}
where the integration is over the volume $V$ of the driven circuit, excluding the external voltage source and its source impedance. Here, $\mathbf{D}_d (\mathbf{r},t)$ is the displacement field of the driven system and $\mathbf{f}_i(\mathbf{r})$ is the electric field mode function of the $i$-th normal mode of the undriven circuit following the normalization condition (see Appendix \ref{app:overlap_EM})
\begin{equation}
\int_V \varepsilon(\mathbf{r})\mathbf{f}_i(\mathbf{r})\cdot\mathbf{f}_j(\mathbf{r})\,dV = \delta_{ij},
\label{eq:normalization_normal_mode}
\end{equation} 
where $\varepsilon(\mathbf{r})$ is the dielectric function. 

In practice, a finite truncation of the mode number $N$ has to be imposed, typically keeping only modes of interest (such as the high-Q modes that participate in the desired parametric process) and their drive terms in the Hamiltonian (see discussion in Appendix~\ref{app:mode_truncation}). Even though the rest of the modes are not explicitly kept, their contribution to the junction phase displacement is captured by the residual phase displacement $A_{\mathrm{res}, k}(t)$, which may play an important role in activating nonlinear parametric processes. The residual phase displacements $A_{\mathrm{res}, k}(t)$ are obtained numerically through subtracting the contributions of the $N$ modes retained explicitly in the Hamiltonian model, from the total junction phase $A_k(t)$,
\begin{align}
    A_{\mathrm{res}, k}(t) = A_k(t) - \sum_{i=1}^N A_k^{(i)}(t),
    \label{eq:overlap-Fk}
\end{align}
where 
\begin{align}
    A_k^{(i)}(t) = \elementdepsymbol{\beta}{J}{ki} \left<a_i + a_i^\dagger\right>=\frac{\elementdepsymbol{\beta}{J}{ki}}{\omega_i^2 + \kappa_i^2/4} \sqrt{\frac{2\omega_i}{\hbar}}  \dot{O}_i(t)\label{eq:A_k^i}
\end{align}
is the $i$-th mode contribution to the phase of the $k$-th junction.
The last equation is obtained by solving the Heisenberg equation of motion for the $i$-th mode, 
\begin{equation}
    \dot{a}_i = -i\omega_i a_i - \frac{\kappa_i}{2}a_i + \frac{g_i(t)}{\hbar},
    \label{eq:overlap-HEOM-flux}
\end{equation}
using $g_i(t)$ from Eq.~\eqref{eq:g(t)} and \eqref{eq:gi_overlap}.
Thus, the dynamic parameters of the overlap method Hamiltonian in  Eq.~\eqref{eq:overlap-ov-Hamiltonian} are provided by Eqs.~\eqref{eq:gi_overlap} and \eqref{eq:overlap-Fk}, which are obtained from the electric overlap $O_i(t)$, the normal mode frequencies $\omega_i$, their participation ratios $\elementdepsymbol{\beta}{J}{ki}$, and the total junction flux $A_k(t)$ extracted from microwave simulations. Importantly, the participation ratios $\elementdepsymbol{\beta}{J}{ki}$ are calculated using the same eigenmode field functions as for the overlap calculation, so that the phase correlations between $g_i(t)$ and $A_k(t)$ are preserved.

In stark contrast with the previous junction-centered methods, the overlap method provides the linear drive strength $g_i$ of each retained mode, whether lumped-element or distributed, and also resolves how the driven junction phase response is built up from these modes. This additional information is illustrated in Fig.~\hyperref[fig:overlap]{\ref*{fig:overlap}}c, where we compare the total phase displacement of each SQUID junction, obtained from the DF extraction, with the partial modal reconstruction obtained from the lowest $N$ modes using the overlap-derived contributions $A_k^{(i)}$. As $N$ is increased, the partial sums progressively reproduce the total junction response over a broader frequency range. This modal decomposition identifies which eigenmodes dominate the driven junction response at a given drive frequency, providing a useful framework for circuit optimization, for example by suppressing $A_k$ at noise-sensitive frequencies to mitigate dephasing and relaxation, or by enhancing $A_k$ to increase the strength of a desired parametric modulation. This mode-resolved information comes at the cost of greater computational complexity and memory usage, since the method requires storing electric-field profiles from both driven and eigenmode simulations and evaluating the overlap integral in Eq.~\eqref{eq:electric_overlap}.

The overlap construction represents the driven Hamiltonian in the normal-mode basis of the linearized circuit, as in the DF method and in the normal-mode implementation of the IG method used above. The accuracy of this representation is therefore controlled by the suitability of the retained linear-mode basis and Hilbert-space truncation for the nonlinear circuit under study. Compact-variable effects such as charge dispersion are not naturally captured by this representation, while extended-phase circuits can in principle be treated with sufficiently large mode and Fock-space truncations~\cite{yilmaz2024}.

\section{NUMERICALLY CALCULATING DECOHERENCE RATES}
\label{sec:decoherence-rates}
In the previous sections, we demonstrated how to model the driven Josephson circuit using classical microwave simulations, which allows us to predict the coherent evolution of the system. Crucially, the same electromagnetic transfer functions that set coherent drive strengths also determine how environmental voltage fluctuations at external ports stochastically drive the circuit. In the weak-coupling (Born) and near-Markovian limits, these fluctuations act as small, stationary perturbations that induce incoherent transitions and dephasing. 

Building on this, we develop numerical procedures, using the microwave simulation workflow established in the previous sections, to compute decoherence rates induced by drive port voltage noise. We first establish the linear susceptibility that connects the drive port voltage to the effective drive parameters, and then apply Fermi's golden rule (FGR) to evaluate transition rates in the static or Floquet basis. We refer to this approach as the port voltage noise response (PVNR) method. Critically, the susceptibility extracted from the microwave simulation carries the full device geometry and field distributions into the rate formulas. As a result, the matrix elements and their weighting by the specified voltage noise spectrum reflect the actual electromagnetic coupling between the source and the circuit. This automatically includes mode mixing, parasitic paths, and impedance transformations, without relying on a simplified lumped-element model. Another powerful aspect of PVNR is that it accepts a realistic voltage noise spectral density with full temperature and frequency dependence, including thermal noise at the port load resistance and other noise present in the input signal. The PVNR method provides a quantitative assessment of the system's performance across environmental conditions, guiding circuit design to minimize port-induced loss while maintaining sufficient drive strength for desired operations.

\begin{figure}[t]
  \centering 
  \includegraphics[width=\columnwidth]{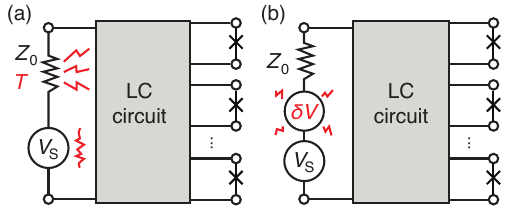}
  \caption{
  \textbf{Equivalent representations of drive-port voltage noise.}
  (a) Josephson circuit driven by a sinusoidal voltage source $\elementdepsymbol{V}{S}{}$ with fluctuations, loaded by a resistance $Z_0$ at temperature $T$. (b) The voltage noise at the lossy resistor and the voltage source can be modeled as an ideal resistor in series with a voltage noise source $\delta V$ that incoherently drives the Josephson circuit, in addition to the coherent voltage drive.}
\label{fig:decoherence-sketch}
\end{figure}

As a concrete case, we consider a drive port loaded by a lumped-element resistor of resistance $Z_0$ at temperature $T$, as shown in Fig. \hyperref[fig:decoherence-sketch]{\ref*{fig:decoherence-sketch}a}. The voltage source at the drive port generally results in a time-dependent Hamiltonian, 
\begin{align}
    \mathcal{H} = \mathcal{H} \left[\left\{s_i(t)\right\}\right],
    \label{eq:H-functional-si}
\end{align}
where $\left\{s_i(t)\right\}$ is a set of effective modulation parameters (such as $\elementdepsymbol{F}{J}{k}(t)$ or $g_i(t)$) dependent on the frame and gauge choices. Within the framework of linear response theory, these modulation parameters are related to the source voltage $\elementdepsymbol{V}{S}{}(t)$ through a linear response function $\chi_{s_i}(t)$
\begin{align}
s_i(t)=\int_{-\infty}^{t} \chi_{s_i}(t-t')\elementdepsymbol{V}{S}{}(t')dt'. 
\label{eq:si_convolution_chi}
\end{align}
Under a sinusoidal source voltage with fixed amplitude, $\elementdepsymbol{V}{S}{}(t) = \Real[\elementdepsymbol{\bar{V}}{S}{} e^{i\omega t}]$, the drive parameters can be written as
\begin{align}
s_i(t)=\Real[\bar{s}_i (\omega) e^{i \omega t}]=
\Real [\bar{\chi}_{s_i}(\omega) \bar{V}_{\mathrm{S}}e^{i \omega t}]
\label{eq:susceptibility_freq}
\end{align}
where ${\bar{\chi}_{s_i}(\omega) \equiv \mathcal{F}\left[\chi_{s_i}(t)\right] = \left\lvert\bar{\chi}_{s_i}(\omega)\right\rvert \exp\left[i \arg \bar{\chi}_{s_i}(\omega) \right]}$ is the Fourier transform of the linear response function. 
The susceptibility function, $\bar{\chi}_{s_i}(\omega)$, can then be numerically obtained as 
\begin{equation}
    \bar{\chi}_{s_i}(\omega) = \frac{\bar{s}_i(\omega)}{\elementdepsymbol{\bar{V}}{S}{}}.
    \label{eq:susceptibility}
\end{equation}

In the spirit of the fluctuation-dissipation theorem, the dissipation of the system due to the bath can be understood as the result of the voltage fluctuations $\delta V(t)$. They may arise from the noisy resistor or from the noisy input signal (e.g., noise from the control electronics), incoherently driving the system in addition to the coherent source, as depicted in Fig. \hyperref[fig:decoherence-sketch]{\ref*{fig:decoherence-sketch}b}. The power spectral density (PSD) of such voltage noise is given by

\begin{align}
S_{\delta V}(\omega)&=\int_{-\infty}^{\infty}\left<\delta V(t)\delta V(0)\right>e^{i\omega t}dt\nonumber\\
&=2\hbar\omega Z_0 \left[1+n_\mathrm{B}(\omega)\right] + S_\textrm{in}(\omega), 
\label{eq:Johnson-Nyquist-spectral-density}
\end{align}
where the first term represents the quantum extension of Johnson-Nyquist noise~\cite{Callen1951} with Bose occupation factor $n_\mathrm{B}(\omega) = 1/(e^{\hbar\omega/k_\mathrm{B} T} -1)$, and $S_\textrm{in}(\omega)$ represents noise in the input signal (such as thermal noise from higher temperature stages or control electronics). The voltage noise interacts with the circuit through the same drive port as the coherent voltage drive, creating fluctuations in the drive parameters through the same linear response functions,  
\begin{equation}
    \delta s_i(t) = \int_{-\infty}^{t} \chi_{s_i}(t-t')\delta V\left(t'\right)dt',
    \label{eq:del_i_convolution}
\end{equation}
such that the noisy drive parameters become $s_i(t) \rightarrow s_i(t) + \delta s_i(t)$. The PSD of $\delta s_i(t)$ relates to that of $\delta V(t)$ through (see Appendix \ref{sec:app-Sxy-to-SVN})
\begin{align}
S_{\delta s_i}(\omega)&=\int_{-\infty}^{\infty}\left<\delta s_i(t)\delta s_i(0)\right>e^{i\omega t}dt=\lvert\bar{\chi}_{s_i}(\omega)\rvert^2 S_{\delta V}(\omega). 
\label{eq:SVV_to_SFF}
\end{align}
Although written here with classical noise, the same relations hold for quantum noise with unsymmetrized quantum spectrum~\cite{Schoelkopf2003}, in which case the linear mapping follows directly from the quantum Langevin/input–output treatment of a linearized system (see Appendix \ref{sec:app-quantum-noise} for more discussions).

Under the assumption that the susceptibility $\bar{\chi}_{s_i}(\omega)$ is broadband near the frequencies of interest, such that the correlation time of $\delta s_i(t)$ is still short compared with the system timescale, we can calculate the decoherence rate induced by the linear system-bath coupling via the FGR approach. For the noise-perturbed Hamiltonian $\mathcal{H} \left[\left\{\delta s_i(t)\right\}\right]$ in the absence of all coherent modulations, the transition rate from state $\ket{m}$ to $\ket{n}$ is given by 
\begin{equation}
    \Gamma_{m \rightarrow n} = \frac{1}{\hbar^2}\left\lvert\sum_iD_{\delta s_i,nm}\bar{\chi}_{s_i}\left(\omega_{nm}\right)\right\rvert^2S_{\delta V}\left(\omega_{nm}\right).
    \label{eq:Gamma_m_to_n}
\end{equation}
Here, $\omega_{nm} = \omega_n - \omega_m$ is the frequency difference between these states, and $D_{\delta s_i,nm}$ is
\begin{align}
D_{\delta s_i,nm}=\left\langle n \middle|
\frac{\partial \mathcal{H}[\{\delta s_i\}]}{\partial \delta s_i}
\middle| m \right\rangle\Bigg\rvert_{\{\delta s_i\}=0}. \label{eq:linear_noise_coupling}
\end{align}
Note that, in the displaced-frame formulation, the leading-order static Purcell channel does not appear as a noise term; see Appendix~\ref{sec:app-gauge} for the gauge and displaced-frame subtleties. 

Similarly, the pure dephasing rate in the long-time, Markovian limit can be obtained by considering the transition between the superposition states $\ket{m} + \ket{n}\leftrightarrow\ket{m} - \ket{n}$,
\begin{equation}
\begin{split}
\Gamma^{\phi}_{mn}&=\frac{1}{2\hbar^2}\left\lvert\sum_i\left(D_{\delta s_i,mm}-D_{\delta s_i,nn}\right)\bar{\chi}_{s_i}(0)\right\rvert^2S_{\delta V}(0). 
    \label{eq:Gamma_phi_mn}
\end{split}
\end{equation}
The derivation of Eqs.~(\ref{eq:Gamma_m_to_n}) and (\ref{eq:Gamma_phi_mn}) is given in Appendix~\ref{sec:app-decoherence-rates}. As is well known, for $1/f$-type spectra that diverge at zero frequency, Eq.~\eqref{eq:Gamma_phi_mn} is ill-defined and FGR generally fails; in this case the non-exponential dephasing can be estimated by Monte Carlo sampling~\cite{Timmer1995, Kerman2008,Kolesnikow2024} \(\delta V\) trajectories from $S_{\delta V}(\omega)$, which yields a more accurate result. 
\begin{figure*}[t]
  \centering \includegraphics[width=1\textwidth]{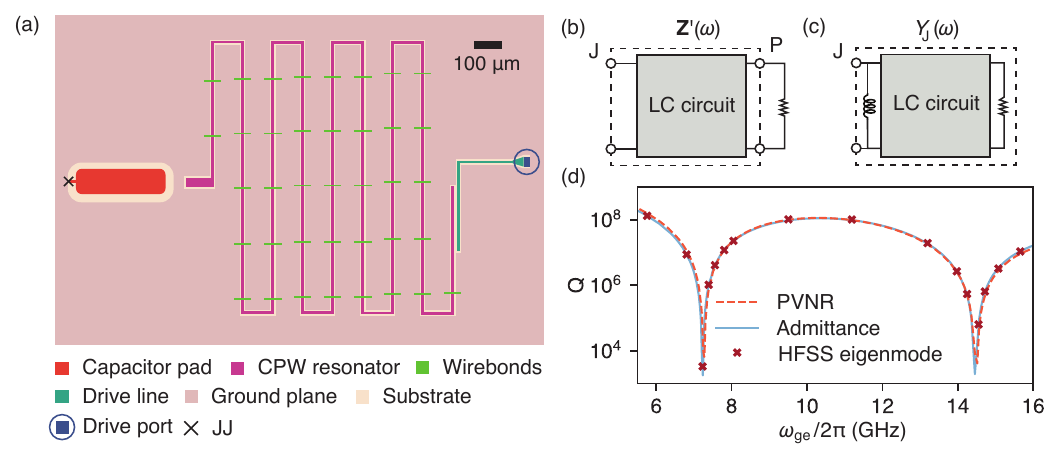}
  \caption{\textbf{Calculation of the transmon qubit decay rate.} 
  (a) Circuit layout of the single-junction transmon qubit capacitively coupled to a 50~$\Omega$ source.
  (b) Black-box model for extracting the effective phase modulation of the transmon circuit using the opened-junction IG method. The effective phase modulation is then used in the PVNR method to compute the qubit decay rate.
  (c) Black-box model for determining the qubit decay rate based on the junction port admittance. The real and imaginary parts of the admittance are used in Eq.~\eqref{eq:transmon_decay_BBQ} \cite{Nigg2012}.
  (d) 
  Quality factors at different qubit frequencies $\omega_{\mathrm{ge}}$ obtained from Eqs.~\eqref{eq:transmon_decay_LR} and \eqref{eq:transmon_decay_BBQ}, compared with results from direct HFSS eigenmode simulations. The junction inductance is varied to sweep the qubit frequency.
  In the admittance method, the qubit frequency is determined using the standard black-box quantization approach, which identifies the zeros of the admittance \cite{Nigg2012}. In the PVNR method, the qubit frequency is given by $1/\sqrt{L_{\mathrm{J}}C_{\Sigma}}$, where the total capacitance $C_\Sigma = 92.64$ fF is obtained by fitting to qubit frequencies from eigenmode calculations.
  }
\label{fig:transmon-decay-rate}
\end{figure*} 

Importantly, Eqs.~(\ref{eq:Gamma_m_to_n}) and (\ref{eq:Gamma_phi_mn}) fully capture the interference effects between correlated noise terms, which physically stem from the coupling to the same lossy drive port. This is reflected in these expressions as the sum of the complex susceptibility functions with varying amplitudes and phases inside the squared modulus, which can lead to constructive or destructive interference that can significantly affect the coherence of the system. This calculation can be extended to circuits in the presence of coherent, periodic modulations using the Floquet-Markov formalism \cite{Grifoni_1998}. Considering the general Hamiltonian
\begin{equation}
    \mathcal{H} \left[\left\{s_i + \delta s_i\right\}\right] \approx  \mathcal{H} \left[\left\{s_i\right\}\right] + \sum_i{\delta s_i \mathcal{V}_i 
    },
    \label{eq:app-H-first-order-pert}
\end{equation}
expanded here into the coherent part, $\mathcal{H}\left[\left\{s_i\right\}\right]$, and the first-order interaction between the noise $\delta s_i$ and the driven system, $\mathcal{V}_i(t) = \partial_{\delta s_i}\mathcal{H}\left[\left\{s_i+\delta s_i\right\}\right]\Big\rvert_{\{\delta s_i\}=0}$. Note that this interaction is generally time-dependent when it contains the drive terms, $s_i(t)$. The coherent part of the Hamiltonian defines the Floquet modes ($\ket{\Phi_\alpha(t)}$, $\ket{\Phi_\beta(t)}$, ...) and their corresponding quasi-energies ($\epsilon_{\alpha}$, $\epsilon_{\beta}$, ...), while the noise interaction gives rise to stochastic transitions between the Floquet modes. In a manner similar to the standard FGR approach, we can obtain the transition rate from the Floquet mode $\ket{\Phi_\alpha(t)}$ to $\ket{\Phi_\beta(t)}$, and the pure dephasing rate between them, as ~\cite{Verney_2019,Zhang2019,You2025, Othmane2025,Carde2025}
\begin{equation}
    \Gamma_{\alpha\rightarrow\beta} = \frac{1}{\hbar^2}\sum_{n=-\infty}^{+\infty}\Big\lvert \sum_{i} P^{i}_{\alpha\beta n} \Big\rvert^2S_{\delta V}\left(\Delta_{\alpha\beta n}\right),
\label{eq:Floquet_decay}
\end{equation}
\begin{equation}
    \Gamma^{\phi}_{\alpha\beta} = \frac{1}{2\hbar^2}\sum_{n=-\infty}^{+\infty}\Big\lvert \sum_{i}\left(P^{i}_{\alpha\alpha n}-P^{i}_{\beta\beta n}\right) \Big\rvert^2S_{\delta V}\left(n\omega_d\right).
\label{eq:Floquet_dephasing}
\end{equation}
Here, 
\begin{equation}
P^j_{\alpha\beta n}=\frac{i\bar{\chi}_{s_j}\left(\Delta_{\alpha\beta n}\right)}{T}\int^T_0 e^{-in\omega_d t} \bra{\Phi_\alpha} \mathcal{V}_j(t) \ket{\Phi_\beta}dt 
\label{eq:FM_elements}
\end{equation}
is the transition matrix element in the Floquet mode basis coupled to the $j$-th noise source, $T$ is the time-period of the drive, and $\hbar \Delta_{\alpha\beta n}=\epsilon_\beta-\epsilon_\alpha+n\hbar\omega_d$ is the quasi-energy difference at the drive frequency $\omega_d$ in the $n$-th Brillouin zone. Using the Floquet modes and quasi-energies defined above, together with the rates in Eqs.~\eqref{eq:Floquet_decay} and \eqref{eq:Floquet_dephasing}, a master equation can be constructed for numerically calculating the incoherent dynamics of the driven circuit. For a finite drive pulse, the Floquet modes defined by the periodic part of the drive can often be connected to the static eigenstates before and after the pulse through the adiabatic drive envelope~\cite{Weinberg2017}. Once this mapping is established, the Floquet-Markov master equation can be used to describe the driven dynamics and decoherence of an initial state prepared before the pulse.

In devices with multiple drive ports, voltage fluctuations at different ports can be correlated, for example, through shared upstream electronics or imperfect isolation between control lines. In such cases, the preceding analysis carries over verbatim by promoting the scalar susceptibility to a port-indexed vector, and the single-port PSD to the port-port cross–PSD matrix. We provide the detailed derivations and closed-form expressions in Appendix~\ref{sec:app-decoherence-rates-multiports}.

\subsection{Case studies}
\subsubsection{Purcell effects in static circuits}
\label{sec:Purcell_effect}
As a validation of our numerical method, we use Eq.~(\ref{eq:Gamma_m_to_n}) to calculate the Purcell decay rates due to the lossy ports~\cite{Esteve1986, Houck2008, Sunada2022,Wassaf2024,Bakr2025} for two weakly nonlinear circuits: a single-junction transmon capacitively coupled to a coplanar waveguide (CPW) resonator (Fig.~\hyperref[fig:transmon-decay-rate]{\ref*{fig:transmon-decay-rate}a}), and a SQUID-based flux-tunable transmon at zero external flux (Fig.~\hyperref[fig:junction-centered]{\ref*{fig:junction-centered}c}). By using an EM-extracted response function instead of a mode-by-mode decomposition, PVNR avoids the apparent divergences associated with incomplete modeling of the multimode circuit~\cite{Gely2017,Bakr2026,Malekakhlagh2025}. The results are compared with decay rates obtained from using admittance matrices \cite{Houck2008, Nigg2012} and HFSS eigenmode simulations. 

For the coupled transmon-CPW resonator circuit, we apply the IG method using the two-port impedance (Fig.~\hyperref[fig:transmon-decay-rate]{\ref*{fig:transmon-decay-rate}b}), and compute the effective phase modulation of the transmon with Eq.~(\ref{eq:f-from-Z-open}). The decay rate of the transmon mode is obtained by employing the PVNR approach,
\begin{align}
\begin{split}
\Gamma_{\downarrow}&= \elementdepsymbol{L}{J}{}^{-1}Z_0 \left[1+n_\mathrm{B}(\omega_{ge})\right]\left\lvert\frac{Z^\circ_{\mathrm{J}\mathrm{P}}(\omega_{ge})}{Z_0+Z^\circ_{\mathrm{P}\mathrm{P}}(\omega_{ge})}\right\rvert^2. \label{eq:transmon_decay_LR} 
\end{split}
\end{align}
The detailed derivation of the expression is presented in Appendix \ref{sec:app-transmon-SQUID-decay}. Here,
$\elementdepsymbol{L}{J}{}$ is the junction inductance, $Z^\circ_{\mathrm{J}\mathrm{P}}$ and $Z^\circ_{\mathrm{P}\mathrm{P}}$ are impedance parameters of the opened-junction circuit, and $\omega_{ge}$ is the transmon transition frequency between the ground and excited states. We set $n_{\mathrm{B}} = 0$ in the calculation.

As a comparison to the PVNR method, the decay rate can also be obtained using the admittance extracted from the junction port at the qubit frequency $Y_{\mathrm{J}}(\omega_{ge})$ (Fig. \hyperref[fig:transmon-decay-rate]{\ref*{fig:transmon-decay-rate}c}) \cite{Houck2008, Nigg2012}
\begin{align}
\Gamma_{\downarrow}=2\frac{\Real Y_{\mathrm{J}}(\omega_{ge})}{\Imag Y_{\mathrm{J}}'(\omega_{ge})},
\label{eq:transmon_decay_BBQ}
\end{align} 
where the prime denotes the derivative with respect to frequency and $\omega_{ge}$ is determined by the real part of the zeros of $\elementdepsymbol{Y}{J}{}(\omega)$. In addition, we directly extract the quality factors of eigenmodes from HFSS. The quality factors $Q = \omega_{ge}/\Gamma_\downarrow$ calculated using the three methods are presented in Fig.~\hyperref[fig:transmon-decay-rate]{\ref*{fig:transmon-decay-rate}d}, with excellent agreement with each other. For this comparison, we evaluate $\omega_{ge}$ at the linearized resonance so that PVNR can be compared directly to the admittance and eigenmode results computed for the linearized circuit; however, PVNR also supports evaluation at the Lamb-shifted transition frequency as the result of the Josephson nonlinearity. This correction is negligible for circuits with weak anharmonicity, but may have a significant impact in circuits with stronger nonlinearity. We emphasize that the eigenmode simulation and the admittance method give only the decay rate at the resonance frequency, whereas PVNR uses the specified noise PSD to yield frequency-resolved contributions to relaxation and dephasing. This is essential for determining off-resonance nonlinear transition rates and for quantifying drive-induced decoherence.

Next, we study the decay of the flux-tunable transmon shown in Fig.~\hyperref[fig:junction-centered]{\ref*{fig:junction-centered}c}, with zero external flux. Unlike the single-junction transmon, the SQUID-based transmon circuit consists of two junctions, 
leading to interference effects that influence its decoherence.
Following a derivation similar to that of Eq.~\eqref{eq:transmon_decay_LR} for the single-junction transmon, we use the PVNR method to obtain the decay rate of the SQUID transmon as
\begin{align}
\begin{split}
\Gamma_{\downarrow}&= L_{\Sigma}Z_0\left[1+n_\mathrm{B}(\omega_{ge})\right]\left\lvert\sum_{k=1}^2\frac{1}{\elementdepsymbol{L}{J}{k}}\frac{Z^\circ_{\mathrm{J}_k\mathrm{P}}(\omega_{ge})}{Z_0+Z^\circ_{\mathrm{P}\mathrm{P}}(\omega_{ge})}\right\rvert^2,
\label{eq:decay_SQUID}
\end{split}
\end{align}
where $k =1,2$ labels the junctions and ${L_\Sigma} = ( \elementdepsymbol{L}{J}{1}^{-1}+ \elementdepsymbol{L}{J}{2}^{-1})^{-1}$ is the total inductance of the two junctions. 

The key point is that the two junctions couple to the same noisy port, so the noise is correlated, and the rate depends on the coherent sum inside the modulus. For a purely differential coupling corresponding to ideal flux modulation~\cite{Lu2023}, the response currents flow in the opposite directions, hence the two impedance functions are out of phase and the decay is suppressed. For a common-mode coupling corresponding to charge modulation, the impedance functions constructively interfere, resulting in a finite decay rate.     

\begin{figure}[t!]
  \centering 
  \includegraphics[width=\columnwidth]{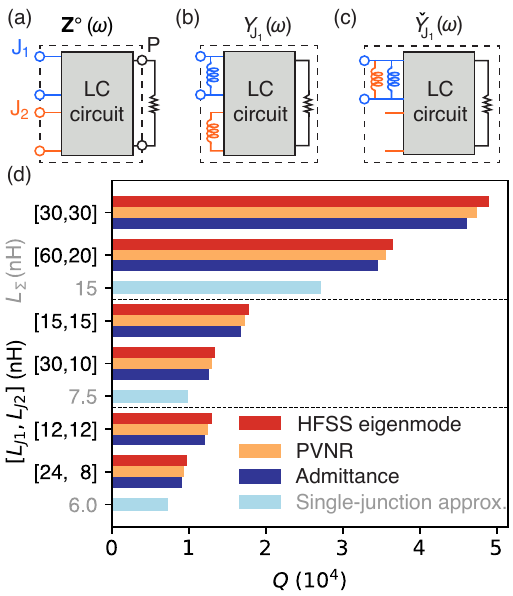}
  \caption{\textbf{Calculation of the SQUID circuit decay rate.}
  (a) 
  Black-box model for extracting the effective phase modulation of the SQUID circuit in Fig.~\hyperref[fig:junction-centered]{\ref{fig:junction-centered}c} using the opened-junction IG method. Colors distinguish junction ports $\mathrm{J}_1$ and $\mathrm{J}_2$. 
  The extracted phase modulations are used in the PVNR method to compute the decay rate.
  (b) 
  Black-box model for calculating the decay rate using the one-port admittance of junction port $\mathrm{J}_1$.
  (c)
  Black-box model under the single-junction approximation, where the two junctions are replaced by an effective inductance $L_{\Sigma} = (\elementdepsymbol{L}{J}{1}^{-1} + \elementdepsymbol{L}{J}{2}^{-1})^{-1}$ at junction port $\mathrm{J}_1$. The admittance at $\mathrm{J}_1$ is used to compute the decay rate.
  (d)
  Comparison of SQUID mode quality factors obtained using the PVNR method [Eq.~\eqref{eq:decay_SQUID}], the admittance method, and HFSS eigenmode simulations for different junction inductance pairs $(\elementdepsymbol{L}{J}{1}, \elementdepsymbol{L}{J}{2})$. Dashed lines group results by total inductance $L_{\Sigma}$ with varying $\elementdepsymbol{L}{J}{1}$ and $\elementdepsymbol{L}{J}{2}$.
  Quantitative agreements are achieved among the quality factors obtained from PVNR method, the admittance method, and HFSS eigenmode simulations. In contrast, the single-junction approximation leads to significant deviation (shown in light blue), indicating that such an approximation fails to capture the interference effect.
  }
\label{fig:SQUID_Q}
\end{figure}

The decay rate for the SQUID transmon can also be obtained using admittances, using a slight generalization of the previous single-junction transmon admittance approach \cite{Nigg2012}: we select the junction $\mathrm{J}_1$, obtain its one-port admittance $Y_{\mathrm{J}_1}$, treating the other junction $\mathrm{J}_2$ as part of the black box, and apply Eq.~\eqref{eq:transmon_decay_BBQ} to obtain the decay rate (see Fig.~\hyperref[fig:SQUID_Q]{\ref*{fig:SQUID_Q}b}). The same decay rate can be obtained using the admittance of the other junction port, $Y_{\mathrm{J}_2}$, even though it generally differs from $Y_{\mathrm{J}_1}$. Intuitively, this follows from the fact that the eigenmode frequency and decay rate are intrinsic properties of the mode and thus independent of the chosen reference port for evaluating admittance.

In Fig.~\hyperref[fig:SQUID_Q]{\ref*{fig:SQUID_Q}d}, we compute the quality factors of the SQUID circuit for different pairs of junction inductances $(\elementdepsymbol{L}{J}{1}, \elementdepsymbol{L}{J}{2})$
using both the admittance and the PVNR methods (with $n_{\mathrm{B}} = 0$). The quality factors from both methods agree with HFSS eigenmode simulations. In contrast, reducing the SQUID to a single effective junction of inductance $L_{\Sigma}$ and computing Eq.~\eqref{eq:transmon_decay_BBQ} for this reduced circuit (Fig.~\hyperref[fig:SQUID_Q]{\ref*{fig:SQUID_Q}c}) yields an incorrect $Q$. This collapses the two correlated drive pathways into one and removes the interference captured by Eq.~\eqref{eq:decay_SQUID}. Consequently, circuits with the same total inductance $L_{\Sigma}$ can have different quality factors because the two junctions have different effective flux modulation susceptibilities.

\subsubsection{Drive-induced decay and dephasing}
\label{sec:drive-induced-decay-dephasing}
\begin{figure*}[t!]
  \centering \includegraphics[width=\textwidth]{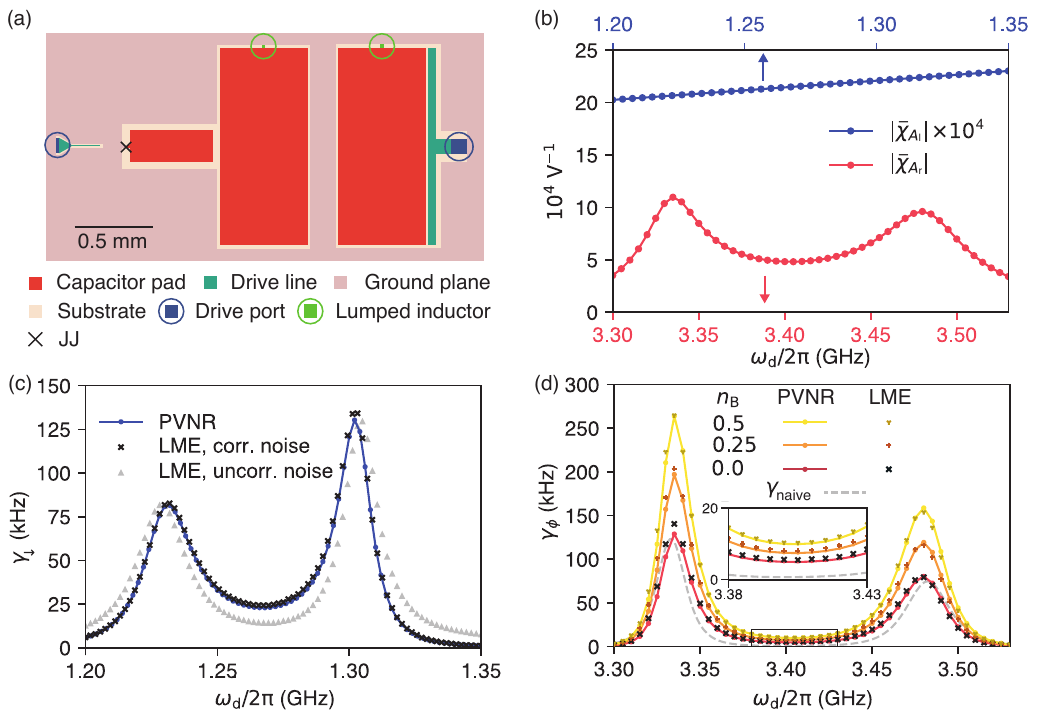}
  \caption{\label{fig:two_filter_modes}
  \textbf{Calculation of the transmon drive-induced Purcell decay and dephasing rates due to a lossy filter network.}
  (a)
  Circuit layout of a single-junction transmon capacitively coupled to a filter network formed by two strongly coupled, nearly degenerate modes. The $g-e$ transition frequency of the transmon is 5.959\,GHz, and the two normal modes are at 3.331\,GHz and 3.485\,GHz. The filter network is strongly coupled to a drive port on the right, while the transmon is weakly coupled to a drive port on the left. Both ports are loaded with 50\,$\Omega$ impedance.
  (b) Amplitudes of susceptibilities of the transmon junction phase displacement with respect to voltages at the two ports. 
  (c) 
  Drive-induced decay rates of the transmon mode at zero photon number. Decay rates are computed using the PVNR method and Floquet theory, and direct Lindblad master equation simulations, respectively. 
  A naive computation using a Lindblad master with an uncorrelated noise model is also shown for comparison, which yields incorrect results.
  (d)
  Drive-induced dephasing rates of the transmon mode at different thermal photon numbers $n_{\mathrm{B}}$. Dephasing rates computed using the PVNR and the direct LME simulation show good agreement, clearly deviating from the naive sum [Eq.~\eqref{eq:single_mode_dephasing}] of the single-mode contribution that ignores the noise correlation and coherent interference between the two filter modes.
  }
\end{figure*}

Finally, we discuss how our method applies to the calculation of the drive-induced decay and dephasing rates of a driven transmon circuit. For concreteness, we consider the circuit in Fig.~\hyperref[fig:two_filter_modes]{\ref*{fig:two_filter_modes}a} where a fixed-frequency transmon is weakly coupled to a filter network (see Appendix~\ref{app:master} for circuit parameters). The latter is composed of two near-degenerate resonators, made of planar capacitor pads and lumped-element linear inductors. This lumped-element choice is solely for RF-simulation convenience, and the method applies equally to distributed circuits such as coplanar waveguide or stripline resonators. The coupling between the two resonators gives rise to two normal modes ($a$ and $b$), which are weakly coupled to the transmon mode ($q$) with similar strengths. Two 50\,$\Omega$-loaded ports, with the left one weakly coupled to the transmon and the right one strongly coupled to the filter network, allow driving the circuit to create various parametric processes. For instance, when the drive frequency is around half of the detuning between the transmon and the resonator modes, a transmon-resonator sideband interaction is activated through a four-wave-mixing process. In the regime where this interaction strength is much weaker than the resonator decay rate, an exponential decay of the transmon population occurs as a result of the resistive bath at the drive port. Dephasing can also happen due to fluctuations of the transmon frequency induced by the coherent-photon noise in the driven filter network, sometimes known as the measurement-induced dephasing in the context of readout back-action. These effects are well-known~\cite{ACStarkshift,Gambetta2006,Clerk2010,Sete2014,Sete2015,Malekakhlagh2020,petrescu2020,Hanai2021,Blais2021} and they are often simulated with Lindblad master equation (LME) models. Here, we discuss how the PVNR method allows for the efficient calculation of both processes using susceptibilities obtained directly from microwave simulations that automatically capture the multipath interference in both the coherent drive response and the port-noise response, without introducing an explicit lumped-element circuit model.  

While our analysis can also be made with the lab-frame Hamiltonian obtained via the overlap method or the IG method, here, we choose to work with the displaced-frame (DF) method that provides an intuitive view of these decoherence mechanisms, as well as higher numerical efficiency. We first write down the DF Hamiltonian of the driven circuit, according to Eq.~\eqref{eq:dfh-Hamiltonian-displaced}, as
\begin{align}
\label{eq:qubit-cavity-dephasing}
&\mathcal{H}_{\rm disp}  = \sum_{k=q,a,b}\hbar\omega_k k^\dag k \nonumber \\ 
&- E_\mathrm{J}\cos_\textrm{nl}\left[\sum_{k=q,a,b} \beta_k \left(k + {k^\dag }\right)+\sum_{i=l,r} A_i(t) \right].
\end{align}
Here, $A_l(t)$ and $A_r(t)$ are the displacements generated by the coherent drives from the left and right ports. Since our focus here is on the transmon decoherence processes, we don't need to consider the explicit quantum dynamics of the resonator modes. This is further validated by the regime of interest where the resonator modes are strongly damped, with linewidths far exceeding their parametric coupling strengths to the transmon and the dispersive shifts~\cite{Sete2014,Boissonneault2012}. Moreover, their participation in the Josephson element is small, so their Kerr nonlinearities are negligible, and their driven response remains well described by linear susceptibility. Therefore, we can adiabatically eliminate the resonator modes, approximating them as a linear filter network (and thus ignoring the drive-induced corrections to their susceptibility function), retaining only their linear coherent response and noise response in our model. This yields the following effective Hamiltonian 
\begin{align}
\label{eq:PVNR-qubit-dephasing}
&\mathcal{H}'_{\rm disp}  = \hbar\omega_q q^\dag q \\ \nonumber 
&- E_\mathrm{J}\cos_\textrm{nl}\left\{\beta_q \left(q + {q^\dag }\right)+\sum_{i=l,r}\left[ A_i(t)+\delta A_i(t)\right] \right\}.
\end{align}

This Hamiltonian provides an intuitive perspective on the drive-induced transmon decoherence processes. To the leading order, the drive-induced Purcell effect can be understood as the result of $\beta_q\left(q+q^\dagger\right)\left[\sum_{i=l,r}A_i(t)\right]^2\left[\sum_{i=l,r}\delta A_i(t)\right]$ from the expansion of $\cos_\textrm{nl}$, whereas the coherent-photon-induced dephasing originates from $\beta_q^2 q^\dagger q\left[\sum_{i=l,r}A_i(t)\right]\left[\sum_{i=l,r}\delta A_i(t)\right]$ (see discussion in Appendix~\ref{app:dephasing_susceptibility}). For a more accurate quantitative analysis, we make use of Eqs.~\eqref{eq:app-H-first-order-pert} to \eqref{eq:FM_elements} based on the Floquet-Markov approach. Expanding the noises $\delta A_{l,r}(t)$ to the first order, we rewrite Eq.~\eqref{eq:PVNR-qubit-dephasing} as
\begin{align}
    &\mathcal{H}'_{\rm disp} \approx \mathcal{H}_0+\mathcal{V}\sum_{i=l,r}\delta A_i(t),\nonumber\\
    &\mathcal{H}_0 = \mathcal{H}'_{\rm disp}\Big|_{\delta A_{l,r}(t)=0},\nonumber\\
    &\mathcal{V} = E_\mathrm{J}\left\{\sin\left[\beta_q\left(q+q^\dagger\right)+\sum_{i=l,r}A_i(t)\right]-\beta_q\left(q+q^\dagger\right)\right\}.\label{eq:noise_int}
\end{align}
We calculate the Floquet modes $\lvert\Phi_g\rangle$ and $\lvert\Phi_e\rangle$, adiabatically connected to the ground and excited states of the transmon with respect to $\mathcal{H}_0$. Then, the drive-induced decay and dephasing rates are obtained respectively from Eqs.~\eqref{eq:Floquet_decay} and \eqref{eq:Floquet_dephasing}, using the noise interaction $\mathcal{V}$ [Eq.~\eqref{eq:noise_int}], the voltage noise PSD at the port, and the susceptibility connecting the source voltage of the $i$-th port to the junction displacement $\bar{\chi}_{A_i}$. The susceptibility functions are depicted in Fig.~\hyperref[fig:two_filter_modes]{\ref*{fig:two_filter_modes}b}, extracted from the microwave simulation using Eqs.~\eqref{eq:A-from-Z} and \eqref{eq:susceptibility}, as
\begin{align}
    \bar{\chi}_{A_i}(\omega) = \frac{Z_{\mathrm{JP}_i}}{i\phi_0 \omega \left( Z_0 + Z_{\textrm{P}_i \textrm{P}_i} \right)}. \label{eq:disp_susceptibility}
\end{align}

While the complete treatment includes noise from both ports, here, for clarity, we focus on calculating the drive-induced decoherence rates due to the filtered 50\,$\Omega$ noise from the strongly-coupled right port, ignoring noise from the weakly-coupled left port. The drive-induced decay and dephasing rates are illustrated in Fig.~\hyperref[fig:two_filter_modes]{\ref*{fig:two_filter_modes}c} and Fig.~\hyperref[fig:two_filter_modes]{\ref*{fig:two_filter_modes}d}, respectively, obtained from both Floquet-Markov (FM) simulations with the PVNR method and Lindblad master equation (LME) simulations with explicit mode dissipators (see Appendix~\ref{app:master}). To guarantee reasonable rates and enable a fair, tractable comparison between PVNR and LME, we intentionally use two distinct drive configurations differing in drive port, power, and frequency, which we detail in the next two paragraphs.

For the simulation of drive-induced Purcell effect, we choose to apply a drive tone through the left port at a drive power of -39.4\,dBm 
(corresponding to a source voltage of 6.77\,mV), sweeping the drive frequency from 1.20 GHz to 1.35 GHz. This range spans across half of the detunings between the transmon and both filter modes, where their sideband interactions are activated.
The susceptibilities $\tilde\chi_{A_l}$ and $\tilde\chi_{A_r}$ shown in Fig.~\hyperref[fig:two_filter_modes]{\ref*{fig:two_filter_modes}b} are then used to compute the coherent displacement \(A_l(t)\) generated by the left port, and the right-port coupling entering the transition matrix element, respectively. The resulting decay rate $\gamma_\downarrow$ (Fig.~\hyperref[fig:two_filter_modes]{\ref*{fig:two_filter_modes}c}) shows two peaks, corresponding to the sideband resonances between the transmon and the two filter modes. Our PVNR calculation agrees with the LME result only when the dissipator correctly includes the correlated noise of the two resonators induced by the same right port. Treating the resonator normal-mode losses as independent Lindblad channels misses this interference effect, causing the result to deviate significantly between the peaks and away from the resonances.

For simulating drive-induced dephasing, the drive is applied through the right port near the frequencies of the filter modes at a drive power of 
-128.1\,dBm (0.250\,$\mu$V source voltage). Dephasing rates are evaluated for port temperatures corresponding to bath occupations of $n_\textrm{B} = 0, 0.25, 0.5$. The PVNR method captures the temperature dependence of the dephasing rate by including $n_\textrm{B}$ in the Johnson-Nyquist noise PSD, showing consistent agreement with the LME results where detailed balance is enforced. For contrast, the dashed gray curve shows a “naive” prediction obtained by incoherently summing single-mode contributions~\cite{Gambetta2006, Yan_dephasing, Clerk_Utami},
\begin{align}
    \gamma _\textrm{naive}(\omega_d)= \sum_{k=a,b}\frac{\bar n_k^\text{coh}}{2}\frac{\chi^2_{qk}\kappa_k }{\left( {{\omega _d} - {\omega _k}} \right)^2 + \left( {\frac{\kappa_k}{2}} \right)^2},\label{eq:single_mode_dephasing}
\end{align}
where $\bar n_k^\text{coh}$ is the coherent photon population in the $k$-th mode, and $\chi_{qk}$ is the cross-Kerr between the transmon and that mode. The clear discrepancy between this naive sum and the PVNR/LME results arises because the naive approach treats the qubit as coupled to each filter mode independently, summing their contributions as if the two filter modes couple to two uncorrelated baths. Such an approach also neglects the correlations in the coherent dynamics of the two modes. In contrast, PVNR and LME retain the correlations in the coherent modulations and the noise originating from the same physical port, which is essential for obtaining accurate dephasing rates for circuits with complicated filter networks where multiple drive pathways interfere.

Finally, we remark that while PVNR and LME yield consistent results in our examples, PVNR offers broader practical advantages. PVNR is typically more computationally efficient, directly turning susceptibilities and noise spectra into transition rates using Floquet-Markov evaluation in the frequency domain. Under proper conditions, adiabatic elimination of the strongly damped filter modes further reduces the system dimension and speeds up the calculation. In contrast, LME must retain the filter modes and time-step through many drive periods for an accurate fitting of the decoherence timescale. 
For example, in our two-filter-mode circuit, observing a noticeable drive-induced decay rate requires large junction phase displacements. Achieving this with a drive on the right port would populate many photons in the filter modes, making the Hilbert space prohibitively large for LME calculations. To keep the calculation tractable, we instead used a left-port drive, which produces the required phase modulations primarily by displacing the qubit mode and thus with far fewer photons. However, PVNR handles drive from either port at essentially the same cost. Furthermore, PVNR inherits the complex port-to-mode and mode-to-mode couplings from the microwave simulations, where common-port noise correlations and interference are automatically embedded in the susceptibility functions. In contrast, LME reproduces these effects only if one specifies an explicit correlated dissipator with the correct relative amplitudes and phases. The lumped-element filter circuit in Fig.~\ref{fig:two_filter_modes} was deliberately chosen because this correlated dissipator can be constructed from a reduced circuit model and used as a controlled LME benchmark. For more distributed geometries or larger multimode networks, constructing such a dissipator by hand can become ambiguous or unavailable, since a single port may address many modes through a superposition of capacitive and inductive pathways~\cite{Sank2025}, and the modes themselves may also be coupled. In summary, PVNR offers a compact, geometry-aware route to accurately estimate decoherence rates induced by drive ports directly from electromagnetic susceptibilities and realistic noise spectra, and at lower computational cost. This makes it a practical tool for circuit design, allowing one to directly assess the tradeoff between desired parametric coupling and the decoherence caused by the drive.

\section{CONCLUSIONS} 
In conclusion, we present a geometry-aware workflow that turns standard electromagnetic simulations into quantitative Hamiltonians and drive-port-induced decoherence rates for driven Josephson circuits. We introduce and contrast three complementary constructions: the displaced-frame (DF), the irrotational-gauge (IG), and the overlap methods. They map impedances and fields of the circuit, obtained from microwave simulation, to effective drive parameters and Hamiltonians, directly applicable to
a broad class of superconducting Josephson circuits. In addition to modeling the coherent modulation of the driven circuit, the same workflow provides a port-voltage noise response (PVNR) that maps port voltage noise to effective perturbations, useful for the decoherence rate estimation via Fermi's golden rule or Floquet-Markov theory, automatically taking into account multi-path interference and common port correlations with complex circuit geometries. These coherent and incoherent descriptions together provide a comprehensive picture of a driven circuit: they predict circuit dynamics under realistic drives, identify which modes and ports are responsible for the effective modulation and loss of the circuit, and expose the trade-offs between achievable drive strength and back-action. This enables practical design tasks such as optimizing filter networks, setting control line specifications incorporating temperature gradients and transfer functions of attenuators and filters at each stage, and benchmarking operating points against performance metrics.

Looking forward, several important extensions remain. On the coherent drive side, an immediate generalization is to move from monochromatic steady-state drives to arbitrarily shaped pulses by extracting time-domain drive parameters, either from transient electromagnetic simulations or from the circuit susceptibility convolved with the experimental waveform, both of which fit within the same linear-response mapping used here. The resulting Hamiltonians also provide natural inputs to recent comprehensive formalisms for evaluating and optimizing parametric processes in strongly driven Josephson circuits~\cite{Venkatraman2022,xiao2022,Petrescu2023,Baskov2025,Xia2025}. On the noise side, our decoherence estimates use weak-coupling Markovian treatments (Fermi’s golden rule and Floquet-Markov), which allow frequency-dependent spectra but neglect long memory in the bath correlations. Going beyond this regime, non-Markovian classical fluctuations specified by a power spectral density (such as $1/f$) can be incorporated by generating stochastic time traces and averaging Schr\"odinger evolution over Monte Carlo realizations~\cite{Timmer1995, Kerman2008,Kolesnikow2024}, which also accommodates explicitly non-periodic control waveforms. Achieving comparable realism for colored quantum noise generally requires non-Markovian open-system methods~\cite{Breuer2016, deVega2017}. Moreover, in PVNR calculations, it is often efficient to represent filter modes through effective linear susceptibilities, but when the filter-system dressing becomes state dependent (even without bath memory), a single state-independent susceptibility can misestimate driven coherent and dissipative dynamics, motivating either explicit inclusion of the relevant filter modes or dressed, state-dependent response functions.

More broadly, our geometry-aware mapping should also interface naturally with hybrid electromechanical platforms~\cite{Teufel2011,Chu2017,Satzinger2018}. For example, one could extend an overlap-based construction to include mechanical generalized coordinates~\cite{Banderier2023}. This would provide a unified way to compute not only electromechanical coupling strengths, but also how external ports drive and damp the resulting hybrid modes using a layout-resolved description. Another direction is to generalize the workflow to include nonreciprocal circuit elements~\cite{Labarca2024,Parra-Rodriguez2025}, where reciprocity-based simplifications break down and quantum-noise constraints must be treated consistently. Finally, because the workflow fundamentally combines linear simulations with localized nonlinear constitutive elements, it should be adaptable to lumped nonlinearities beyond Josephson junctions, including dual elements such as quantum phase-slip devices~\cite{osborne2023,Koliofoti2023,Parra-Rodriguez2024}, provided an appropriate compact-variable quantization and participation analysis is incorporated. 

\section*{DATA AVAILABILITY}
The data and source files supporting the findings of this study are available at \href{https://dx.doi.org/10.5281/zenodo.18028722}{10.5281/zenodo.18028722}.

\section*{ACKNOWLEDGMENTS}
We thank Benjamin Chapman, Stijn de Graaf, Douglas Stone, Rohan Narayan Rajmohan, Danyang Chen, Changqing Wang, Basil Smitham, Patrick Winkel, Luigi Frunzio, and Walter E. Lawrence for helpful discussions. This work was supported by the Army Research Office (ARO) under Grant No. W911NF-23-1-0051, the Air Force Office of Scientific Research (AFOSR) under Grant No. FA9550-21-1-0209, and the U.S. Department of Energy (DOE), Office of Science, National Quantum Information Science Research Centers, including the Co-design Center for Quantum Advantage (C2QA) under Contract No. DE-SC0012704. Y.L., A.V., X.Y., and Z.H. also acknowledge support from the DOE Office of Science National Quantum Information Science Research Centers, Superconducting Quantum Materials and Systems Center (SQMS) under Contract No. DE-AC02-07CH11359. Y.L. acknowledges support from the DOE Early Career Research Program. The views and conclusions contained in this document are those of the authors and should not be interpreted as representing the official policies, either expressed or implied, of the ARO, DOE, AFOSR, or the U.S. Government. The U.S. Government is authorized to reproduce and distribute reprints for Government purposes notwithstanding any copyright notation herein.

Y.L. conceived the project and carried out the primary development and implementation of the theory, numerical methods, simulations, and manuscript. T.Z. and A.V. made the largest contributions after Y.L., spanning theory, simulations, and manuscript preparation. K.C.S. developed key aspects of the overlap method using a field-based formulation. D.W. contributed to derivations for the irrotational-gauge method and to numerical simulations. X.Y. contributed to interpreting time-dependent flux quantization and connecting it to the present framework. Y.Z. provided guidance on the Floquet–Markov and linear-response approaches. S.G. supported microwave finite-element simulations. Z.H. contributed to PVNR-related analysis. A.M., J.W.O.G., S.M., and I.M.-S. contributed through discussions, early-stage exploration, and application-driven feedback. S.M.G., J.K., and R.J.S. provided supervision and critical feedback. All authors discussed the results and commented on the manuscript.

R.J.S. is a founder and equity holder, and S.M.G. is a consultant and equity holder of Quantum Circuits, Inc. (QCI).

\appendix

\section*{Appendix overview}

The appendices provide supporting derivations, implementation details, and numerical examples for the methods discussed in the main text. Appendix~\ref{sec:app_describing_circuits} derives the general driven-circuit Hamiltonian used in Section~\ref{sec:general_Hamiltonian_Josephson_circuit}. Appendix~\ref{app:dc_flux} discusses dc-flux bias and the corresponding linearized normal-mode basis. Appendix~\ref{app:junction_methods} gives details of the junction-centered methods, including the displaced-frame construction, arbitrary drive waveforms, finite-pulse effects, the irrotational-gauge construction, opened-junction extraction, applicability to inductive branches, and simulation port conventions. Appendix~\ref{app:overlap_method} presents the overlap-method Hamiltonian, circuit- and field-based derivations, and simulation workflow. Appendix~\ref{app:mode_truncation} discusses the effect of finite mode truncation. Appendix~\ref{app:pvnr_details} provides details of the port-voltage-noise-response calculation and related FGR estimates. Appendix~\ref{app:master} describes the master-equation benchmark used for the drive-induced decay and dephasing example.

\section{Describing general Josephson circuits}\label{sec:app_describing_circuits}

\begin{figure}[b]
  \centering \includegraphics[width=\columnwidth]{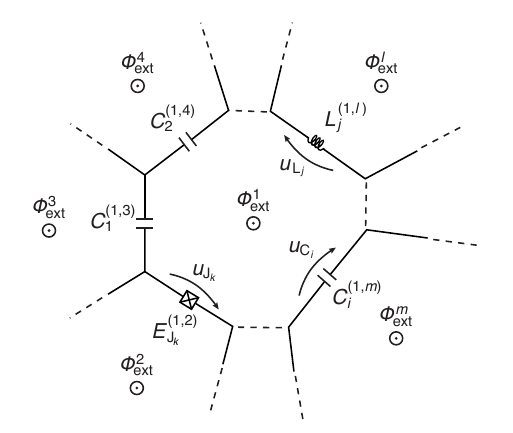}
  \caption{\textbf{Lumped-element model of a multimode multi-junction circuit in a continuous geometry with both voltage and flux modulation.} A circuit with multiple modes and junctions, including geometric inductors. The capacitors, inductors, and junctions form loops that allow external flux to penetrate through. The subscript of each element represents the index of the element in the circuit, while the superscript represents the indices of the loops the element belongs to. The Lagrangian of the circuit is defined with respect to the directions assigned to each element, labeled by $\elementdepsymbol{u}{C}{i}$, $\elementdepsymbol{u}{L}{j}$, and $\elementdepsymbol{u}{J}{k}$, referring to currents crossing the capacitor $\mathrm{C}_i$, inductor $\mathrm{L}_j$, and junction $\mathrm{J}_k$, respectively.}
\label{fig:describing_JJ}
\end{figure}

In this section, we derive the time-dependent Hamiltonian for the general Josephson circuit [Eq.~\eqref{eq:multimode_Hamiltonian_start}]. We consider the abstract circuit model shown in  Fig. \hyperref[fig:describing_JJ]{\ref*{fig:describing_JJ}}, where there are $N$ modes, $M$ junctions, and $W$ linear inductors. For a specified lumped-element network, such coordinate choices and circuit Lagrangians can be constructed systematically using standard graph-based circuit quantization methods~\cite{Burkard2004}. Here, however, we do not focus on mapping a realistic electromagnetic layout to a particular lumped model, since our numerical methods do not rely on obtaining the specific values of these lumped elements. The static bare-mode Lagrangian of the circuit can be compactly written as
\begin{align}
\mathcal{L} &= \frac{1}{2}\sum\limits_i C_i^{{R_i}}\left( \sum\limits_{n=1}^N a_{in}\dot \Phi_n \right)^2- \frac{1}{2}\sum\limits_j \frac{1}{L_j^{S_j}}\left( \sum\limits_{n=1}^N b_{jn}\Phi _n \right)^2\nonumber\\  
&+ \sum\limits_{k=1}^M \elementdepsymbol{E}{J}{k}^{T_k}\cos {\left( \frac{\sum\limits_{n=1}^N {\elementdepsymbol{b}{J}{kn}}\Phi _n }{\phi _0} \right)},
\label{eq:app-multimode-Lagrangian} 
\end{align}
where $R_i$, $S_j$ and $T_k$ are the sets of loop indices that the $i$-th capacitor, $j$-th inductor and $k$-th junction belong to, respectively. For the $n$-th bare circuit coordinate, $\Phi_n$ is the corresponding generalized flux coordinate before diagonalizing the linearized circuit, and $a_{in}$ ($b_{in}$) specifies its contribution to the voltage (flux) drop across the $i$-th capacitor (inductor). Such a coordinate basis can be obtained, in principle, from a standard graph-based circuit quantization by choosing a complete set of independent circuit coordinates and eliminating the flux constraints. Note that $a_{in}$ and $b_{in}$ can be 1, -1, or 0, depending only on the circuit geometry but not on the circuit parameters~\cite{Devoret1997}, including the Josephson inductance.

Next, we include the external flux modulation $\Phi_{\mathrm{ext}}^i(t)$ in the $i$-th loop, which we define to be positive if pointing out of the page. Now, the Lagrangian becomes
\begin{align}
\mathcal{L} &= \frac{1}{2}\sum\limits_i {C_i^{{R_i}}{{\left( {\sum\limits_{n=1}^N {{a_{in}}{{\dot \Phi }_n}} } + \mathcal{V}_i(t) \right)}^2}}\nonumber\\ 
&- \frac{1}{2}\sum\limits_j \frac{\phi_0^2}{L_j^{{S_j}}}{{{\left( {\frac{\sum\limits_{n=1}^N{b_{jn}{\Phi }_n}}{\phi_0} } + \mathcal{F}_j(t) \right)}^2}}\nonumber\\ 
&+ \sum\limits_{k=1}^M {\elementdepsymbol{E}{J}{k}^{{T_k}}\cos {{\left( {\frac{{\sum\limits_{n=1}^N {{\elementdepsymbol{b}{J}{kn}}{\Phi _n}} }}{{{\phi _0}}}} + \elementdepsymbol{\mathcal{F}}{J}{k}(t) \right)}}}.
\label{eq:app-multimode-Lagrangian-modulations} 
\end{align}
Here, $\mathcal{V}_i(t)$ is the drive-induced voltage modulation on the $i$-th capacitive branch, while $\mathcal{F}_{j}(t)$ and $\elementdepsymbol{\mathcal{F}}{J}{k}(t)$ are the drive-induced phase modulations on the $j$-th linear inductive branch and the $k$-th Josephson branch, respectively. Faraday's law constrains these modulations through
\begin{align}
   \frac{\dot \Phi _{\mathrm{ext}}^\ell(t)}{\phi_0} 
   &= \sum\limits_{i\lvert \ell \in {R_i}} 
   \elementdepsymbol{w}{C}{i\ell} \mathcal{V}_i(t)/\phi_0 
   + \sum\limits_{j\lvert \ell \in {S_j}} 
   \elementdepsymbol{w}{L}{j\ell}\dot{\mathcal{F}}_j(t) \nonumber\\
   &+ \sum\limits_{k\lvert \ell \in {T_k}} 
   \elementdepsymbol{w}{J}{k\ell}\elementdepsymbol{\dot{\mathcal{F}}}{J}{k}(t)
\label{eq:V_j_F_i}
\end{align}
for the $\ell$-th loop. The summations are done over all elements belonging to the $\ell$-th loop, with the signs $w_{\mathcal{E}_{i\ell}}$ defined as
\begin{align}
\label{eq:sign}
w_{\mathcal{E}_{i\ell}} = \begin{cases}
+1, & \text{if } u_{\mathcal{E}_i} \text{ is clockwise in loop } \ell \\
-1, & \text{if } u_{\mathcal{E}_i} \text{ is counter-clockwise in loop } \ell \end{cases}
\end{align}
for each element $\mathcal{E} = \mathrm{C, L, J}$, where $\elementdepsymbol{u}{C}{i}$, $\elementdepsymbol{u}{L}{j}$, and $\elementdepsymbol{u}{J}{k}$ represent the direction of the voltage or flux for the $i$-th capacitor, $j$-th inductor and $k$-th junction, respectively. Importantly, the drive–modulation parameters $\mathcal{V}_{i}(t)$, $\mathcal{F}_{j}(t)$ and $\elementdepsymbol{\mathcal{F}}{J}{k}(t)$ are kinematic quantities fixed by Faraday’s law [Eq.~\eqref{eq:V_j_F_i}] and the gauge allocation. They are determined by the overlap of the external drive field with the circuit’s geometric capacitances and inductances, and do not depend on the Josephson nonlinearity. In other words, replacing Josephson elements with linear lumped inductors in the microwave simulation does not alter the drive parameters. This independence is already explicit in Eqs.~\eqref{eq:app-SQUID-modulation-params1} and \eqref{eq:app-SQUID-modulation-params2} of the main text. One may therefore extract the drive parameters from a linearized circuit model and insert them into the driven nonlinear Hamiltonian in Eq.~\eqref{eq:app-general-Hamiltonian}.

The general Hamiltonian can be obtained by performing a Legendre transformation to the Lagrangian in Eq.~(\ref{eq:app-multimode-Lagrangian-modulations}) $\mathcal{H} = \sum_{n=1}^N \dot{\Phi}_n Q_n - \mathcal{L}$, where $Q_n = \frac{\partial \mathcal{L}}{\partial \dot{\Phi}_n}$ are the conjugate momenta, but is more easily obtained using matrix notation. Dropping loop indices and explicit time dependence, we can rewrite the kinetic part of Eq.~(\ref{eq:app-multimode-Lagrangian-modulations}) as
\begin{align}
\begin{split}
    \mathcal{L}_{kin} = \frac{1}{2} \left( \mathbf{a}\dot{\mathbf{\Phi}} + \mathbfcal{V} \right)^\textrm{T} \mathbf{C} \left( \mathbf{a}\dot{\mathbf{\Phi}} + \mathbfcal{V} \right),
\end{split}
\label{eq:app-matrix-Lagrangian}
\end{align}
where $(\mathbf{C})_{ij}$ represents the capacitance between branch $i$ and $j$, $\mathbf{a}$ is the $J \times N$ voltage participation factor matrix, with $J$ and $N$ the number of capacitors and modes, respectively, and $\mathbfcal{V}$ is the voltage modulation parameter column-vector. With this, the conjugate momentum column-vector is
\begin{align}
\mathbf{Q} &= \frac{\partial L}{\partial \dot{\mathbf{\Phi}}} = \mathbf{a}^\textrm{T} \mathbf{C} (\mathbf{a}\dot{\mathbf{\Phi}} + \mathbfcal{V}) \\
&\Rightarrow \mathbf{a} \dot{\mathbf{\Phi}} = \mathbf{C}^{-1} (\mathbf{a}^\textrm{T})^{-1} \mathbf{Q} - \mathbfcal{V}, \label{eq:app-charge-matrix}
\end{align}
where the inverse should be understood as the pseudoinverse in the case of $J \neq N$. We may now perform a Legendre transformation to obtain the general Hamiltonian, focusing here on the kinetic energy terms
\begin{align}
\begin{split}
\mathcal{H}_{kin} &= \mathbf{Q}^\textrm{T} \dot{\mathbf{\Phi}} - \mathcal{L}_{kin}\\
&= \frac{1}{2} (\mathbf{a}\dot{\mathbf{\Phi}})^\textrm{T} \mathbf{C} (\mathbf{a}\dot{\mathbf{\Phi}}) - \frac{1}{2} \mathbfcal{V}^\textrm{T} \mathbf{C} \mathbfcal{V}.
\end{split}
\label{eq:app-Hkin-matrix}
\end{align}
We can now recover the full Hamiltonian, Eq.~(\ref{eq:multimode_Hamiltonian_start}), by converting back to summation notation
\begin{align}
\mathcal{H} &= \sum_{i,j=1}^N 4 \elementdepsymbol{E}{C}{ij}n_i n_j + \sum_{i=1}^N \epsilon_i(t) n_i \nonumber\\
&+ \frac{1}{2} \sum\limits_{\ell=1}^W \elementdepsymbol{E}{L}{\ell} {{{\left( {\sum\limits_{i=1}^N {{\elementdepsymbol{b}{L}{\ell i}}{{\varphi }_i}} } + \elementdepsymbol{\mathcal{F}}{L}{\ell}(t) \right)}^2}}\nonumber\\
&- \sum\limits_{k=1}^M {\elementdepsymbol{E}{J}{k}\cos {{\left( {{\sum\limits_{i=1}^N {{\elementdepsymbol{b}{J}{ki}}{\varphi _i}} + \elementdepsymbol{\mathcal{F}}{J}{k}(t) }} \right)}}},
\label{eq:app-general-Hamiltonian}
\end{align}
where $\elementdepsymbol{E}{L}{\ell} = \phi_0^2/L_\ell$ is the energy of the $\ell$-th inductor, $\varphi_i = \Phi_i/\phi_0$ is the gauge-invariant phase across the $i$-th element, and
\begin{align}
\elementdepsymbol{E}{C}{ij}&= \frac{e^2}{2} \sum_{m,n=1}^J (\mathbf{a}^{-1})_{im} (\mathbf{a}^{-1})_{nj} (\mathbf{C}^{-1})_{mn},\\
\epsilon_i(t) &= -e \sum_{n=1}^J \left[ (\mathbf{a}^{-1})_{in} + (\mathbf{a}^{-1})_{ni} \right] \mathcal{V}_i(t)
\label{eq:app-eps_i}
\end{align}
are the charging energy between modes $i$ and $j$, and the time-dependent charge modulation on the $i$-th mode, respectively.

\section{Including dc-flux biases}
\label{app:dc_flux}
When deriving the Hamiltonians in the linearized normal-mode basis, one must first expand the potential about its minimum location (or one of the minimum locations, if there are multiple minima), which depends in general on dc-flux biases for circuits containing loops \cite{Weiss2021}. Here, the linearized normal-mode basis refers to the ladder-operator basis obtained by diagonalizing the quadratic Hamiltonian of the linearized circuit, as in black-box/EPR-style quantization~\cite{Nigg2012,Minev2021}. In this appendix, we show how the effect of finite dc-flux biases can be captured in the Hamiltonian without explicitly modeling the dc flux in the microwave simulation.

In the linearized normal-mode basis, the general Hamiltonian of an unbiased Josephson circuit can be written as
\begin{align}
&\mathcal{H}_0= \sum_i \hbar \omega_i a^\dagger_i a_i \nonumber\\&- \sum_k\elementdepsymbol{E}{J}{k}\left[\cos\sum_i \elementdepsymbol{\varphi}{J}{ki}+\frac{1}{2}\left(\sum_i \elementdepsymbol{\varphi}{J}{ki}\right)^2\right] \nonumber \\ & +\sum_l \frac{\elementdepsymbol{E}{L}{l}}{2} \left(\sum_i \elementdepsymbol{\varphi}{L}{li}\right)^2-\sum_l \frac{\elementdepsymbol{E}{L}{l}}{2} \left(\sum_i \elementdepsymbol{\varphi}{L}{li}\right)^2,\label{app:H0}
\end{align}
where $\elementdepsymbol{\varphi}{J}{ki} = \elementdepsymbol{\beta}{J}{ki}\left( a_i+a^\dagger_i\right)$ and $\elementdepsymbol{\varphi}{L}{li} = \elementdepsymbol{\beta}{L}{li}\left( a_i+a^\dagger_i\right)$ are the $i$-th normal mode's contribution to the phase of the $k$-th junction and $l$-th inductor, respectively. Even though the last two terms in Eq.~\eqref{app:H0} exactly cancel out, they are explicitly retained here for introducing the dc flux bias in the next step.  

Under a static magnetic field, the biased Hamiltonian reads 
\begin{align}
&\mathcal{H}_\textrm{bias}= \sum_i \hbar \omega_i a^\dagger_i a_i \nonumber\\&- \sum_k\elementdepsymbol{E}{J}{k}\left[\cos\left(\sum_i \elementdepsymbol{\varphi}{J}{ki}+\elementdepsymbol{\phi}{J}{k}\right)+\frac{1}{2}\left(\sum_i \elementdepsymbol{\varphi}{J}{ki}\right)^2\right] \nonumber \\ & + \sum_l \frac{\elementdepsymbol{E}{L}{l}}{2}\left(\sum_i \elementdepsymbol{\varphi}{L}{li}+\elementdepsymbol{\phi}{L}{l}\right)^2- \sum_l \frac{\elementdepsymbol{E}{L}{l}}{2}\left(\sum_i \elementdepsymbol{\varphi}{L}{li}\right)^2.\label{app:Hbias}
\end{align}
The dc phases $\elementdepsymbol{\phi}{J}{k}$ and $\elementdepsymbol{\phi}{L}{l}$ are chosen at a stable extremum of the biased static potential, so that the first derivative with respect to each independent phase coordinate vanishes~\cite{Miano2023}. After expanding around this operating point and expressing the fluctuations in the linearized normal-mode basis, this condition removes all terms linear in $a_i+a_i^\dagger$. Under this condition, we may rewrite Eq.~\eqref{app:Hbias} as 
\begin{align}
&\mathcal{H}_\textrm{bias}\nonumber\\&= \underbrace{\sum_i \hbar \omega_i a^\dagger_i a_i \nonumber+ \sum_k\frac{\elementdepsymbol{E}{J}{k}}{2}\left(\cos\elementdepsymbol{\phi}{J}{k}-1\right)\left(\sum_i \elementdepsymbol{\varphi}{J}{ki}\right)^2\nonumber}_{\mathcal{H}_\textrm{lin}} \\ & -\sum_k\elementdepsymbol{E}{J}{k}\cos\elementdepsymbol{\phi}{J}{k}\cos_\textrm{nl}\sum_i \elementdepsymbol{\varphi}{J}{ki}\nonumber\\ & +\sum_k\elementdepsymbol{E}{J}{k}\sin\elementdepsymbol{\phi}{J}{k}\sin_\textrm{nl}\sum_i \elementdepsymbol{\varphi}{J}{ki}.\label{app:Hbias_1}
\end{align}
Here, $\cos_\textrm{nl} x = \cos x + x^2/2$ removes the quadratic part already included in the linearized Hamiltonian, and $\sin_\textrm{nl} x = \sin x - x$ removes the linear part that is canceled by the dc-equilibrium condition. Remarkably, this Hamiltonian no longer contains the linear inductive energy terms, eliminating the need to obtain their associated participation ratios and the dc-flux biases from the simulation. 

In practice, the magnetic flux penetrating each superconducting loop is first determined via magnetostatic finite-element simulations, such as Ansys Maxwell or COMSOL. Then, the equilibrium dc phases across the junctions can be obtained using, for example, the method of Miano {\em et al.} \cite{Miano2023}. In principle, due to the completeness of the ladder operator basis, the diagonalization of this Hamiltonian with sufficiently large Fock state truncation should recover the flux-biased spectrum, even when the unbiased basis is far from the true eigenbasis of the biased circuit. This has been demonstrated in the diagonalization of a strongly anharmonic circuit by retaining many levels within the EPR framework~\cite{yilmaz2024}. 

For improved numerical convergence under truncation, however, it is advantageous first to diagonalize the Hamiltonian for the linearized circuit, $\mathcal{H}_\textrm{lin}$, via a bosonic Bogoliubov transformation,
\begin{align}
    \mathcal{S}^\dagger \mathcal{H}_\textrm{lin}\mathcal{S} &= \sum_j \hbar\tilde{\omega}_j b^\dagger_j b_j, \\\sum_i \elementdepsymbol{\beta}{J}{ki}\mathcal{S}^\dagger\left( a_i+a^\dagger_i\right)\mathcal{S} &= \sum_j \elementdepsymbol{\tilde{\beta}}{J}{kj}\left( b_j+b^\dagger_j\right).
\end{align}
Substituting them into Eq.~\eqref{app:Hbias_1}, we obtain
\begin{align}
    \tilde{\mathcal{H}}_\textrm{bias}&= \mathcal{S}^\dagger\mathcal{H}_\textrm{bias}\mathcal{S}\nonumber\\&=\sum_j \hbar\tilde{\omega}_j b^\dagger_j b_j \nonumber-\sum_k\elementdepsymbol{E}{J}{k}\cos\elementdepsymbol{\phi}{J}{k}\cos_\textrm{nl}\sum_j \elementdepsymbol{\tilde\varphi}{J}{kj}\nonumber\\ & +\sum_k\elementdepsymbol{E}{J}{k}\sin\elementdepsymbol{\phi}{J}{k}\sin_\textrm{nl}\sum_j \elementdepsymbol{\tilde\varphi}{J}{kj},\label{app:Hbias_2}
\end{align}
where $\elementdepsymbol{\tilde\varphi}{J}{kj} = \elementdepsymbol{\tilde{\beta}}{J}{kj}\left( b_j+b^\dagger_j\right)$. This diagonalization removes the bilinear mixing of the ladder operators in the original biased Hamiltonian (Eq.~\eqref{app:Hbias_1}), thereby suppressing spurious hybridization among low-lying states in the truncated basis, and practically lowering the Fock-space cutoff required for a given spectral accuracy.

\section{Hamiltonians for junction-based methods}
\label{app:junction_methods}
\subsection{Ladder-operator approximation for circuits with periodic potentials}
\label{app:boundary_conditions}
In the derivation of the displaced-frame and the overlap methods, we rely on representing the Hamiltonian in terms of raising and lowering operators (in contrast to the irrotational-gauge method, where this choice is made purely for convenience, since the derivation can also be formulated in terms of charge–phase operators). For circuits with periodic degrees of freedom, this ladder-operator description should be viewed as an approximation, since it does not, by itself, enforce the compact nature of the coordinate. As an example, consider the transmon/Cooper pair box (CPB) Hamiltonian
\begin{equation}
    \mathcal{H} = 4 E_C \left(n-n_g\right)^2 - \frac{\elementdepsymbol{E}{J}{}}{2} \left(e^{i \phi} + e^{-i\phi}\right)
\end{equation}
where $n_g$ is the offset charge. Here, the phase $\phi$ is a compact variable, which is restricted to the interval $0 \leq \phi < 2 \pi$. As such, the shift operators $e^{\pm i\phi}$ are well-defined, and $\phi$ is not. The exact spectrum exhibits a periodic dependence on $n_g$ (charge dispersion). As we write the Hamiltonian in terms of the ladder operators, we implicitly treat $\phi$ as a non-compact variable $-\infty < \phi < \infty$, which breaks the periodic boundary condition, and modifies the eigenstates and eigenenergies (including the offset-charge dependence)~\cite{Koch2007}. In the CPB regime, it is necessary to retain the compactness of $\phi$ to capture the correct spectrum with significant charge dispersion. In contrast, in the transmon regime, the charge dispersion is exponentially suppressed, so the ladder-operator formalism remains a good approximation for the low-lying states~\cite{Koch2007, Girvin2014}, provided that an appropriate basis truncation keeps the wavefunctions confined within a single potential well.
It also applies naturally to systems without compact $2\pi$ periodicity, such as fluxonium~\cite{yilmaz2024}. Moreover, for lumped-element circuits whose static Hamiltonian can be written directly in charge–phase variables via circuit quantization, the IG method circumvents this limitation, providing the driven Hamiltonian where the exact periodic boundary condition is preserved.

\subsection{The displaced-frame Hamiltonian}\label{sec:app_dfh}
Here, we derive the displacement transformation that eliminates the terms linear in the ladder operators $a_i$ and $a_i^\dagger$ at the level of the master equation. Similar derivations can be found in \cite{Blais2007, petrescu2020}.

Starting from the general Hamiltonian in the normal modes basis as in Eq.~(\ref{eq:dfh-Hamiltonian-lin-nl}), consider for each mode $n$ its damping rate $\kappa_n$, such that the master equation is
\begin{equation}
    \dot{\rho} = 
    \frac{1}{i\hbar}
    [\mathcal{H}, \rho] + \sum_{i=1}^{N}\kappa_i \mathcal{D}[a_i] \rho.
\end{equation}
Here, $\mathcal{D}$ is the dissipation superoperator, with ${\mathcal{D}[a] \rho = a \rho a^\dagger - \{a^\dagger a, \rho\}/2}$. 
By applying the displacement transformation 
[Eq.~\eqref{eq:dfh-transform}], 
\begin{align}
    &U_{\mathrm{disp}} = \exp\left({\sum_{i=1}^N \left(\xi_i(t) a_i^\dagger - \xi_i(t)^* a_i\right)}\right),
\end{align}

the ladder operators transform as
\begin{align}
    U_{\text{disp}}^\dagger a_i U_{\text{disp}} &= a_i + \xi_i(t),\\
    U_{\text{disp}}^\dagger a_i^\dagger U_{\text{disp}} &= a_i^\dagger + \xi_i^*(t).
\end{align}
The evolution of the density matrix in the displaced frame $\rho_{\mathrm{disp}} = U_{\text{disp}}^\dagger \rho U_{\text{disp}}$ is governed by the following master equation \cite{Blais2007}
\begin{align}
    \dot{\rho}_{\mathrm{disp}}
    = 
    \frac{1}{i\hbar}
    [\mathcal{H}_{\mathrm{disp}}
    , \rho_{\mathrm{disp}}] + 
    \sum_{i = 1}^N \kappa_i \mathcal{D}[a_i] \rho_{\mathrm{disp}}.
\label{eq:rho_disp_ME_app}
\end{align}
The displaced-frame Hamiltonian $\mathcal{H}_{\mathrm{disp}}$ is given by 
\begin{align}
\mathcal{H}_{\text{disp}} &= U_{\text{disp}}^\dagger \mathcal{H} U_{\text{disp}} - i \hbar U_{\text{disp}}^\dagger \dot{U}_{\text{disp}}\nonumber\\
&+\sum_{i=1}^N\frac{i\hbar \kappa_i}{2} (a_i\xi_i^*(t) - a_i^\dagger \xi_i(t)).
\label{eq:H_disp_ME_app}
\end{align}
Notice that the second line arises from displacing the ladder operators in the dissipators. For convenience, we repeat the lab-frame Hamiltonian [Eq.~\eqref{eq:dfh-Hamiltonian-lin-nl}], split into the linearized circuit part $\mathcal{H}_{\mathrm{lin}}$ and its nonlinear counterpart $\mathcal{H}_{\mathrm{nl}}$:
\begin{align}
\begin{split}
\mathcal{H} &= \mathcal{H}_\textrm{lin}+\mathcal{H}_\textrm{nl},\\
\mathcal{H}_{\text{lin}}&=\sum_{i=1}^N \hbar \omega_i a_i^\dagger a_i + i \sum_{i=1}^N \tilde{\epsilon}_i (t) (a_i^\dagger-a_i)\\
&+ \sum_{\ell=1}^W \elementdepsymbol{E}{L}{\ell} \elementdepsymbol{\mathcal{F}}{L}{\ell}(t) \sum_{i=1}^N \elementdepsymbol{\beta}{L}{\ell i} (a_i + a_i^\dagger)\\
&+ \sum_{k=1}^M \elementdepsymbol{E}{J}{k} \elementdepsymbol{\mathcal{F}}{J}{k}(t) \sum_{i=1}^N \elementdepsymbol{\beta}{J}{ki} (a_i + a_i^\dagger),\\
\mathcal{H}_{\mathrm{nl}}&=-\sum_{k=1}^M \elementdepsymbol{E}{J}{k} \cos_{\mathrm{nl}} \left( \sum_{i=1}^N \elementdepsymbol{\beta}{J}{ki} (a_i + a_i^\dagger) + \elementdepsymbol{\mathcal{F}}{J}{k}(t) \right).
\end{split}
\end{align}
We can similarly split the displaced-frame Hamiltonian
\begin{align}
\mathcal{H}_{\mathrm{disp}}
&= \mathcal{H}_{\mathrm{disp, lin}}
+ \mathcal{H}_{\mathrm{disp, nl}},\label{eq:app-H_disp}\\
\mathcal{H}_{\mathrm{disp, lin}}
&= \sum_{i=1}^N \hbar \omega_i a_i^\dagger a_i \nonumber\\
&+ \sum_{i=1}^N \Bigg\{ \hbar \omega_i \left( \xi_i(t) a_i^\dagger + \xi_i^*(t) a_i \right) + i \tilde{\epsilon}_i (t) (a_i^\dagger - a_i) \nonumber\\
&+ (a_i + a_i^\dagger) \Bigg( \sum_{\ell=1}^W \elementdepsymbol{E}{L}{\ell} \elementdepsymbol{\mathcal{F}}{L}{\ell}(t)\elementdepsymbol{\beta}{L}{\ell i}\nonumber\\
&+ \sum_{k=1}^M \elementdepsymbol{E}{J}{k} \elementdepsymbol{\mathcal{F}}{J}{k}(t) \elementdepsymbol{\beta}{J}{ki} \Bigg) - i\hbar (\dot{\xi}_i(t) a_i^\dagger - \dot{\xi}_i^*(t) a_i) \nonumber\\
&+\frac{i\hbar \kappa_i}{2} (\xi_i^*(t) a_i - \xi_i(t)a_i^\dagger)
\Bigg\},\\
\mathcal{H}_{\mathrm{disp, nl}}&=
-\sum_{k=1}^M \elementdepsymbol{E}{J}{k} \cos_{\mathrm{nl}}\left(\sum_{i=1}^N\elementdepsymbol{\beta}{J}{ki} (a_i +a_i^\dagger)+A_k(t) \right).
\label{eq:H_disp_app}
\end{align}
The terms in curly braces of $\mathcal{H}_{\mathrm{disp,lin}}$ are canceled by choosing the time-dependent displacements $\xi_i(t)$ as the solution of the linear equation of motion below,
\begin{align}
    &\dot{\xi}_i(t) = \frac{i}{\hbar} \Bigg\{ -\hbar \omega_i \xi_i(t) 
    + i\hbar \frac{\kappa_i}{2} \xi_i(t) 
    - i \tilde{\epsilon}_i(t) \nonumber\\
    &-\sum_{\ell=1}^W \elementdepsymbol{E}{L}{\ell} \elementdepsymbol{\mathcal{F}}{L}{\ell}(t)\elementdepsymbol{\beta}{L}{\ell i} - \sum_{k=1}^M \elementdepsymbol{E}{J}{k} \elementdepsymbol{\mathcal{F}}{J}{k}(t) \elementdepsymbol{\beta}{J}{ki} \Bigg\}\,.\label{eq:app-xi-EOM}
\end{align}
With this choice, $\mathcal{H}_\mathrm{disp}$ in Eq.~\eqref{eq:app-H_disp} reduces to the displaced-frame Hamiltonian in Eq.~\eqref{eq:dfh-Hamiltonian-displaced} of the main text. 

The same equation can be understood from the classical response of the driven, damped linearized circuit. Taking the expectation value of the damped Heisenberg equation for the linearized Hamiltonian and identifying $\xi_i(t)=\langle a_i(t)\rangle$ gives Eq.~(\ref{eq:dfh-HEOM})
\begin{equation}
    \dot{a}_i (t) = \frac{i}{\hbar} [\mathcal{H}_{\text{lin}}, a_i]
    - \frac{\kappa_i}{2} a_i(t).
\end{equation}

\subsection{Uniqueness of the displaced frame}
\label{app:disp_invariance}
In the following, we will prove that the displaced frame as defined is gauge-invariant, consequently confirming that the phase displacements $A_k(t)$ are gauge-invariant.

We start with the Hamiltonian in Eq.~\eqref{eq:dfh-Hamiltonian-lin-nl} for a single mode
\begin{align}
    \mathcal{H}_1 &= \hbar \omega a^\dagger a + i \tilde{\epsilon}(t)(a^\dagger - a) \nonumber\\
    &+ \elementdepsymbol{E}{L}{} \elementdepsymbol{\mathcal{F}}{L}{}(t) \elementdepsymbol{\beta}{L}{} (a + a^\dagger)+ \elementdepsymbol{E}{J}{} \elementdepsymbol{\mathcal{F}}{J}{}(t) \elementdepsymbol{\beta}{J}{} (a + a^\dagger) \nonumber\\
    &- \elementdepsymbol{E}{J}{} \cos_\mathrm{nl}\left( \elementdepsymbol{\beta}{J}{}(a + a^\dagger) + \elementdepsymbol{\mathcal{F}}{J}{}(t) \right).
\end{align}
The corresponding Lindblad master equation is
\begin{equation}
    \dot{\rho_1} = \frac{1}{i\hbar}[\mathcal{H}_1,\rho_1]+\kappa \mathcal{D}[a]\rho_1,
\end{equation}
including single-photon loss rate $\kappa$. Next, we compare two cases: 1) going into the displaced frame with displacement parameter $\xi_1(t)$, or 2) first performing a gauge transform with parameter $\alpha(t)$, then going into the displaced frame with displacement parameter $\xi_2(t)$. We will see that the final displaced frame in both cases is the same.

Following the second option, we model the gauge transformation as a displacement transform with parameter $\alpha(t)$
\begin{equation}
    U_2(t)=\exp[\alpha(t)a^\dagger-\alpha^*(t)a].
\end{equation}
The resulting master equation is
\begin{equation}
    \dot{\rho}_2 = \frac{1}{i\hbar}[\mathcal{H}_2,\rho_2]+\kappa \mathcal{D}[a]\rho_2,
    \label{eq:app-transformed-rho-uniqueness-dfh}
\end{equation}
with 
\begin{align*}
    \rho_2&=U_2^\dagger \rho_1 U_2,\\
    \mathcal{H}_2&=U_2^\dagger \mathcal{H}_1 U_2 - i\hbar U_2^\dagger \dot{U}_2 + \frac{i\hbar\kappa}{2}(a\alpha^*(t)-a^\dagger \alpha(t)).
\end{align*}

In each gauge, we now perform a displacement transform that cancels the linear drive terms, transforming into their respective displaced frame. The corresponding unitaries are
\begin{align}
    U_{\mathrm{disp}_{1(2)}}(t) = \exp[\xi_{1(2)}(t)a^\dagger-\xi_{1(2)}^*(t)a].
\end{align}
The resulting master equations have the same form as Eq.~\eqref{eq:app-transformed-rho-uniqueness-dfh} and
\begin{align*}
    \rho_{\mathrm{disp}_k}&=U_{\mathrm{disp}_k}^\dagger \rho_k U_{\mathrm{disp}_k},\\
    \mathcal{H}_{\mathrm{disp}_k}&=U_{\mathrm{disp}_k}^\dagger \mathcal{H}_k U_{\mathrm{disp}_k} - i\hbar U_{\mathrm{disp}_k}^\dagger \dot{U}_{\mathrm{disp}_k}\\
    &+ \frac{i\hbar\kappa}{2}(a\xi_k^*(t)-a^\dagger \xi_k(t))
\end{align*}
for $k=1,2$. The displaced frame parameters $\xi_{1,2}(t)$ are chosen to cancel the linear drive terms in the Hamiltonians, producing the following differential equations:
\begin{equation}
\begin{split}
\dot{\xi}_1(t)&=\frac{1}{\hbar}\tilde{\epsilon}(t) - \frac{i}{\hbar}\left( \elementdepsymbol{E}{L}{} \elementdepsymbol{\mathcal{F}}{L}{}(t)\elementdepsymbol{\beta}{L}{} + \elementdepsymbol{E}{J}{}\elementdepsymbol{\mathcal{F}}{J}{}(t)\elementdepsymbol{\beta}{J}{} \right)\\
&- i\left(\omega-i\frac{\kappa}{2}\right)\xi_1(t),\\
\dot{\alpha}(t)+\dot{\xi}_2(t)&=\frac{1}{\hbar}\tilde{\epsilon}(t) - \frac{i}{\hbar}\left( \elementdepsymbol{E}{L}{} \elementdepsymbol{\mathcal{F}}{L}{}(t)\elementdepsymbol{\beta}{L}{} + \elementdepsymbol{E}{J}{}\elementdepsymbol{\mathcal{F}}{J}{}(t)\elementdepsymbol{\beta}{J}{} \right)\\
&- i\left(\omega-i\frac{\kappa}{2}\right)\left( \alpha(t)+\xi_2(t) \right).
\end{split}
\end{equation}
Let $\xi_2'(t)=\alpha(t)+\xi_2(t)$ be the total displacement due to $U_2(t)$ and $U_{\mathrm{disp}_2}(t)$, we remark that $\xi_1(t)$ and $\xi_2'(t)$ obey the same differential equation. The two solutions coincide for all $t$ once a common boundary
condition is specified. We fix this boundary condition by defining the displacement as the driven response with
$\xi_1(t_0)=\xi_2'(t_0)=0$ at a reference time $t_0$ when the applied drives are off (equivalently, $\xi_1(t\to -\infty)=\xi_2'(t\to -\infty)=0$ for a pulse that
vanishes in the distant past). With this convention, 
\begin{equation}
\xi_1(t)= \alpha(t)+\xi_2(t)=\xi_2'(t)
\label{eq:app-xi-eqn-uniqueness-dfh}
\end{equation} holds for all $t$, and the displaced frame is unique, or invariant of the gauge choice. Additionally, we can write the displaced frame Hamiltonians explicitly
\begin{align*}
    \mathcal{H}_{\mathrm{disp}_{1(2)}} &= \hbar\omega a^\dagger a- \elementdepsymbol{E}{J}{} \cos_\mathrm{nl}\left( \elementdepsymbol{\beta}{J}{} (a+a^\dagger)+A_{1(2)}(t) \right),
\end{align*}
where the phase displacements in each gauge are
\begin{align*}
    A_1(t)&=\elementdepsymbol{\beta}{J}{} \left(\xi_1(t)+\xi_1^*(t)\right) + \elementdepsymbol{\mathcal{F}}{J}{}(t),\\
    A_2(t)&=\elementdepsymbol{\beta}{J}{} \left(\alpha(t)+\alpha^*(t)+\xi_2(t)+\xi_2^*(t)\right) + \elementdepsymbol{\mathcal{F}}{J}{}(t).
\end{align*}
Using Eq.~\eqref{eq:app-xi-eqn-uniqueness-dfh} with the previous equations, we find that
\begin{align*}
\begin{split}
    A_1(t)&=A_2(t)+\elementdepsymbol{\beta}{J}{} (\xi_1(t)+\xi_1^*(t)-\xi_2'(t)-\xi_2'^*(t))\\
    &=A_2(t),
\end{split}
\end{align*}
confirming the gauge invariance of the phase displacements, without much surprise as they are physical observables. We note that the proof could easily be extended to multiple modes, but the end result would be the same.

\subsection{Response to arbitrary drive waveforms}
\label{app:general_waveforms}

In the main text, we present the extraction of modulation parameters using monochromatic steady-state simulations. This is the natural output of frequency-domain electromagnetic solvers, but it does not restrict the method to monochromatic drives. The same simulations determine the frequency-dependent response function from a drive-port voltage to a Hamiltonian parameter. Once this response function is sampled over the relevant bandwidth, it can be combined with the spectrum of an arbitrary input waveform to construct the corresponding time-domain drive parameter.

Let $s_\mu(t)$ denote any of the drive-induced Hamiltonian parameters extracted in this work, such as a DF junction phase displacement $A_k(t)$, an IG effective phase modulation $F_{\mathrm{J}_k}(t)$, or an overlap-derived drive amplitude $g_i(t)$. For a sinusoidal excitation at port $p$ with positive frequency $\omega$, we write
\begin{equation}
    V_p(t)=\operatorname{Re}\!\left[\bar V_p(\omega)e^{i\omega t}\right],
    \qquad
    s_\mu(t)=\operatorname{Re}\!\left[\bar s_\mu(\omega)e^{i\omega t}\right],
\end{equation}
and define the transfer function
\begin{equation}
    \chi_{\mu p}(\omega)
    =
    \frac{\bar s_\mu(\omega)}{\bar V_p(\omega)},
    \qquad \omega>0.
    \label{eq:app_waveform_chi_def}
\end{equation}
Here $\bar V_p(\omega)$ is the voltage phasor at the chosen port reference plane. If the simulation is normalized by source voltage or available source power, it should first be converted to the same voltage convention before forming $\chi_{\mu p}(\omega)$.

For a real drive waveform $V_p(t)$, define its Fourier transform by
\begin{equation}
    \tilde V_p(\omega)=\int_{-\infty}^{\infty}dt\,V_p(t)e^{-i\omega t}.
\end{equation}
Only the positive-frequency part is independent, since $\tilde V_p(-\omega)=\tilde V_p^*(\omega)$. Using the positive-frequency response extracted from monochromatic simulations, the time-domain modulation parameter is
\begin{equation}
    s_\mu(t)
    =
    \operatorname{Re}
    \int_{0}^{\infty}\frac{d\omega}{\pi}\,
    \chi_{\mu p}(\omega)\tilde V_p(\omega)e^{i\omega t}.
    \label{eq:app_waveform_positive_freq}
\end{equation}

For a drive containing a finite number of tones, $V_p(t)=\operatorname{Re}\sum_m \bar V_{p,m}e^{i\omega_m t}$, this expression reduces to the discrete sum $s_\mu(t)=\operatorname{Re}\sum_m \chi_{\mu p}(\omega_m)\bar V_{p,m}e^{i\omega_m t}$.

For multiple drive ports, the contributions add linearly in frequency space,
\begin{equation}
    s_\mu(t)
    =
    \operatorname{Re}
    \int_{0}^{\infty}\frac{d\omega}{\pi}\,
    \left[
    \sum_p \chi_{\mu p}(\omega)\tilde V_p(\omega)
    \right] e^{i\omega t}.
    \label{eq:app_waveform_multiport}
\end{equation}

In practice, the response function is obtained from monochromatic simulations over a finite frequency window. The integral is therefore evaluated over a bandwidth $0<\omega<\Omega_{\mathrm{max}}$,
\begin{equation}
    s_\mu(t)
    \simeq
    \operatorname{Re}
    \int_{0}^{\Omega_{\mathrm{max}}}\frac{d\omega}{\pi}\,
    \left[
    \sum_p \chi_{\mu p}(\omega)\tilde V_p(\omega)
    \right] e^{i\omega t}.
    \label{eq:app_waveform_cutoff}
\end{equation}
The cutoff $\Omega_{\mathrm{max}}$ and the frequency sampling should be chosen large enough to cover the spectral bandwidth of the applied waveform and the relevant variation of $\chi_{\mu p}(\omega)$. With this choice, a set of monochromatic simulations provides the response functions needed to construct the time-domain drive parameters for arbitrary input pulses.

\subsection{Adiabatic mapping between lab and displaced frames}
\label{app:df_vs_lab}

\begin{figure*}
    \centering
    \includegraphics[width=\textwidth]{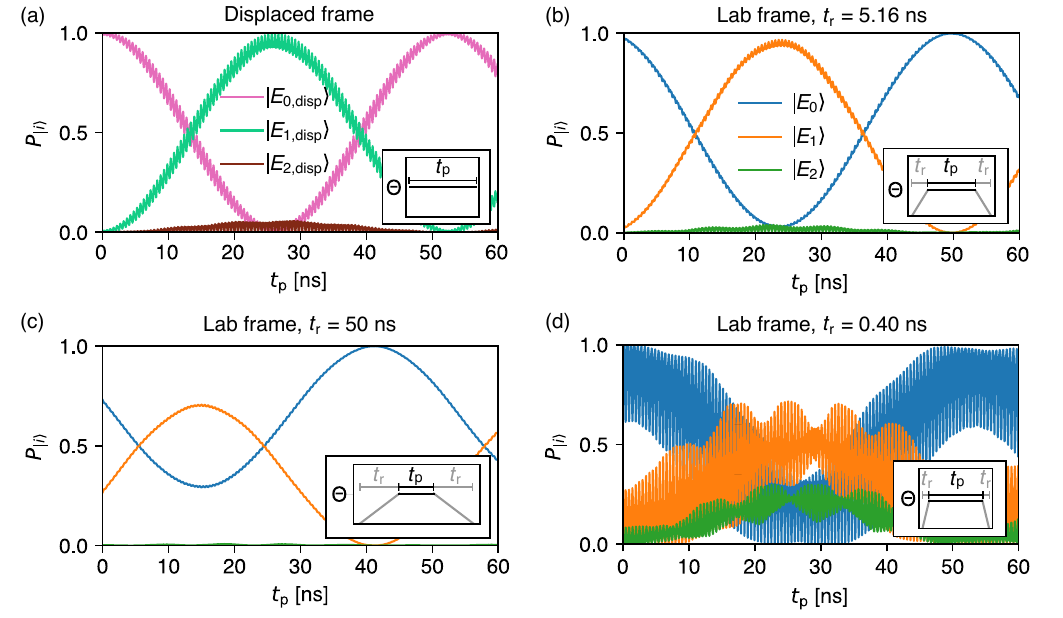}
    \caption{
    \textbf{Dynamics under subharmonic drive in the lab frame and displaced frame.}
    (a) Simulation in the displaced frame of the occupation probabilities $P_n$ of the displaced
    eigenstates $\ket{E_{n,\mathrm{disp}}}$ ($n = 0,1,2$) under a flat pulse of duration $t_p$,
    using the displaced-frame Hamiltonian [Eq.~\eqref{eq:DF_vs_IG_DFH}]. In this frame, the subharmonic drive
    activates a parametric process in which three drive photons induce a one-photon transition,
    leading to clear Rabi oscillations between $\ket{E_{0,\mathrm{disp}}}$ and
    $\ket{E_{1,\mathrm{disp}}}$.
    (b) Simulation in the lab frame of the occupation probabilities of the eigenstates
    $\ket{E_n}$ ($n = 0,1,2$) for the same system, now driven with a pulse of duration $t_p$
    and a piecewise-linear envelope $\Theta(t)$ with ramp time $t_r = 5.16~\mathrm{ns}$,
    using the lab-frame Hamiltonian [Eq.~(C18)]. The ramp segments approximately implement
    the displacement operator, mapping $\ket{E_n} \to \ket{E_{n,\mathrm{disp}}}$, and thereby
    reproduce near-resonant Rabi oscillations between $\ket{E_0}$ and $\ket{E_1}$.
    (c) Same as (b), but with a much longer ramp time $t_r = 50~\mathrm{ns}$.
    (d) Same as (b), but with a much shorter ramp time $t_r = 0.40~\mathrm{ns}$.
    For both overly long and overly short ramps, the lab-frame dynamics differ strongly from
    panel (a), illustrating that the DF Hamiltonian alone does not describe the evolution under
    non-adiabatic ramping.}
    \label{fig:DF_vs_lab}
\end{figure*}

Displaced-frame descriptions are useful both as a compact way to identify and estimate parametric processes and as a practical basis for interpreting driven dynamics in superconducting circuits. In this section, we clarify how the displaced-frame description used in this work relates to the lab frame and what it implies for driven dynamics.

Formally, a displaced-frame Hamiltonian is related to the lab-frame Hamiltonian by a time-dependent displacement transformation $U_{\mathrm{disp}}(t)$. If both the Hamiltonian and the quantum state are transformed by $U_{\mathrm{disp}}(t)$, all physical observables are identical in the two frames. For an arbitrary drive waveform, one can in principle define such a time-dependent displaced frame by choosing the displacement to follow the full classical response of the linearized circuit to the applied waveform. In that construction, the corresponding displaced-frame Hamiltonian cancels the linear drive terms exactly (including during ramps).

In our workflow, however, we do not construct $U_{\mathrm{disp}}(t)$ explicitly. This distinction matters differently depending on what is being compared. If the pulse is arranged so that the displacement vanishes at the initial and final times $t_i$ and $t_f$ (by ringdown and/or waveform design), then $U_{\mathrm{disp}}(t_i)=U_{\mathrm{disp}}(t_f)=\mathbb I$. In this case, transition probabilities between eigenstates of the static Hamiltonian evaluated before and after the pulse are identical whether they are computed in the lab frame or in the displaced frame, even if $U_{\mathrm{disp}}(t)$ is never written down explicitly. By contrast, reproducing lab-frame populations or coherences \emph{during} the pulse from a displaced-frame simulation would require applying $U_{\mathrm{disp}}(t)$ at intermediate times, or, in its absence, invoking an additional adiabatic connection between lab and displaced eigenstates.

While working with the full time-dependent displaced frame is possible in principle, it requires obtaining the classical phase trajectory, for example, from a transient linear simulation or from the circuit susceptibility convolved with the known waveform, as described in Appendix~\ref{app:general_waveforms}. In many practical settings, it is simpler to construct a DF Hamiltonian from the steady-state displacement associated with a sinusoidal drive of fixed frequency and amplitude. For long, flat pulses, this steady-state DF description coincides with the periodic steady-state response of the driven linear circuit. For pulses with finite ramps, it becomes an approximation whose validity is controlled by the ramp conditions derived below. When these conditions are satisfied, the ramps adiabatically map prepared lab-frame eigenstates to their displaced counterparts and back, so that the steady-state DF Hamiltonian reproduces the lab-frame transition probabilities evaluated before and after the pulse and provides a reliable interpretation of the driven dynamics. When the ramps violate these conditions, additional non-adiabatic features appear in the lab-frame evolution that are not captured by the steady-state DF treatment.

We illustrate these points by considering a transmon subject to a ramped drive pulse, described by the full lab-frame Hamiltonian
\begin{equation}
    \mathcal{H}(t) = \hbar \omega a^\dagger a 
    - \elementdepsymbol{E}{J}{} \cos_{\mathrm{nl}}\!\left[\beta(a^\dagger + a)\right] 
    + \mathcal{H}_{\mathrm{d,r}}(t),
    \label{eq:DF_vs_IG_lab_frame}
\end{equation}
with drive term (the subscript r denotes quantities associated with the ramp.)
\begin{align}
    \mathcal{H}_{\mathrm{d, r}}(t) 
    = i \hbar g_0 \Theta(t) \sin (\omega_d t)(a^\dagger - a),
\end{align}
and envelope
\begin{align}
    \Theta(t) &= 
    \begin{cases}
        \Theta_+(t), & -t_{\mathrm{r}}\le t < 0,\\[3pt]
        1, & 0 \le t < t_{\mathrm{p}},\\[3pt]
        \Theta_-(t), & t_{\mathrm{p}} \le t < t_{\mathrm{p}} + t_{\mathrm{r}},
    \end{cases}\label{eq:app-ramp-disp}
\end{align}
where $t_{\mathrm{p}}$ is the flat-top duration and $t_{\mathrm{r}}$ is the ramp-up/down time. The functions $\Theta_+(t)$ and $\Theta_-(t)$ smoothly ramp from $0 \to 1$ and $1 \to 0$, respectively. For the ramp-up to implement the desired displacement, we require the corresponding propagator $U_{\mathrm{p,r}}(0,-t_{\mathrm{r}})$ to approximate the steady-state displacement operator
\begin{equation}
    U_{\mathrm{disp}}(0) = \exp\!\left(\xi(0) a^\dagger - \xi^*(0) a\right),
\end{equation}
where $\xi(t)$ is the steady-state displacement generated by a constant-amplitude drive:
\begin{equation}
    \xi(t) = -\frac{g_0}{2} \!\left(
    \frac{e^{i\omega_d t}}{\omega + \omega_d - i\kappa/2}
    - \frac{e^{-i\omega_d t}}{\omega - \omega_d - i\kappa/2}
    \right).
    \label{eq:steady_state_disp}
\end{equation}
Equivalently, the goal of the ramp is to achieve the same displacement as in the steady-state response at the flat-top amplitude, so that the steady-state DF Hamiltonian correctly describes the subsequent flat-top evolution.

Applying the steady-state displacement operator $U_{\mathrm{disp}}(t)$ to the lab-frame
Hamiltonian with a constant-amplitude drive (the flat-top portion of the pulse with
$\Theta(t)=1$) removes the linear drive term and yields a displaced-frame Hamiltonian
\begin{equation}
    \mathcal{H}_{\mathrm{disp}}
    =\hbar\omega a^\dagger a
      - \elementdepsymbol{E}{J}{}\cos_{\mathrm{nl}}\!\left[\beta(a^\dagger + a + \xi(t) + \xi^*(t))\right].
    \label{eq:DF_vs_IG_DFH}
\end{equation}
Let $\ket{E_n}$ denote
the eigenstates of the static transmon Hamiltonian
$\mathcal{H}_0 = \hbar\omega a^\dagger a - \elementdepsymbol{E}{J}{}\cos_{\mathrm{nl}}[\beta(a^\dagger + a)]$
with eigenvalues $E_n$. The displaced eigenstates
\begin{equation}
    \ket{E_{n,\mathrm{disp}}} = U_{\mathrm{disp}}(0)\ket{E_n}
\end{equation}
then form the basis used in Fig.~\hyperref[fig:DF_vs_lab]{\ref*{fig:DF_vs_lab}a}.

Now we can analyze the condition at which the ramp in Eq.~\eqref{eq:app-ramp-disp} achieves $U_{\mathrm{disp}}(0)$. We use the displacement transformation to split the full propagator during the ramp up into the two parts (the unitaries with subscript p indicate propagators)
\begin{align}
    U_{\mathrm{p, r}}(t, -t_{\mathrm{r}}) = U_{\mathrm{disp,r}}(t) U_{\mathrm{p, disp}}(t, - t_\mathrm{r}),
\end{align}
where the displacement unitary
\begin{equation}
    U_{\mathrm{disp,r}}(t) = 
    \exp\!\left[\xi_{\mathrm{r}}(t)a^\dagger - \xi_{\mathrm{r}}^*(t)a\right]
\end{equation}
generates the transformation that removes the linear drive term during the ramp, and describes the classical displacement of the linear circuit. 
The Hamiltonian under this displacement is
\begin{equation}
    \mathcal{H}_{\mathrm{disp,r}} = \hbar \omega a^\dagger a - \elementdepsymbol{E}{J}{}\cos_{\mathrm{nl}}\left[\beta(a^\dagger + a + \elementdepsymbol{\xi}{r}{}(t) + \elementdepsymbol{\xi^*}{r}{}(t)) \right].
\end{equation}
The second term, $U_{\mathrm{p, disp}}(t, - t_\mathrm{r})$, describes the parametric evolution under the displaced frame, \begin{align}
    U_{\mathrm{p,disp}}(t,-t_{\mathrm{r}})
    &= \mathcal{T}\exp\!\left[
        -\frac{i}{\hbar}\!\int_{-t_{\mathrm{r}}}^t 
            \mathcal{H}_{\mathrm{disp,r}}(t')\, dt'
    \right].
\end{align} 
The purpose of this split is to isolate two independent requirements: (i) the ramp must be slow enough that the linearized circuit follows the drive and prepares the desired steady-state displacement, i.e. $U_{\mathrm{disp,r}}(0)\approx U_{\mathrm{disp}}(0)$ and (ii) the ramp must be fast enough that the Josephson nonlinearity does not induce appreciable parametric evolution during the ramp, i.e. $U_{\mathrm{p, disp}}(0, - t_\mathrm{r})\approx\mathbb{I}$. 

We first address requirement (i), which amounts to determining the ramp-induced displacement $\xi_{\mathrm{r}}(t)$ governed by the differential equation
\begin{equation}
    \dot\xi_{\mathrm{r}}(t)
    = -i(\omega - i\kappa/2)\xi_{\mathrm{r}}(t)
      + g_0 \Theta_+(t)\sin(\omega_d t),
    \label{eq:ramp_displacement_equation}
\end{equation}
with $\xi_{\mathrm{r}}(-t_{\mathrm{r}})=0$. This initial condition reflects that the oscillator is initially undisplaced before the start of the pulse.
Solving Eq.~\eqref{eq:ramp_displacement_equation} at $t=0$, we obtain
\begin{align}
    \xi_{\mathrm{r}}(0) &= 
    \frac{g_0}{2i} \sum_{\pm}
    (\pm1) \times\Bigg[
        \frac{1}{i(\omega\pm\omega_d - i\kappa/2)} \nonumber\\
        &\qquad - \underbrace{
        \int_{-t_{\mathrm{r}}}^{0}
        \frac{e^{i(\omega\pm\omega_d - i\kappa/2)t}}
             {i(\omega\pm\omega_d - i\kappa/2)}
        \,\dot\Theta_+(t)\,dt}_{I_{\pm}}
    \Bigg]. \label{eq:ramp_displacement}
\end{align}
The first term reproduces the steady-state displacement~\eqref{eq:steady_state_disp} at $t=0$. The second term quantifies errors due to finite ramp duration. For a smooth envelope with $\dot\Theta_+(t)\sim 1/t_{\mathrm{r}}$, the magnitude of the integral is estimated as
\[
    |I_\pm|
    \sim \frac{1}
        {t_{\mathrm{r}}\,|\omega\pm\omega_d - i\kappa/2|^2}.
\]
Thus, keeping this error term small relative to the desired steady-state displacement requires
\begin{equation}
    t_{\mathrm{r}} \gg \frac{1}{|\omega\pm\omega_d - i\kappa/2|}.
\end{equation}

For requirement (ii), in practice the flat-top drive is often chosen to engineer a targeted resonant parametric interaction at a rate $\Omega_{\mathrm{param}}$ (with other terms averaged out). To prevent appreciable accumulation of this evolution during the ramp, we require
\begin{align}
    t_{\mathrm{r}} \ll 1/\Omega_{\mathrm{param}}.
\end{align}
These conditions not only ensure that the desired displacement is prepared and undone, but also justify treating the displaced eigenstates as adiabatic continuations of the lab-frame eigenstates during the ramps.

To validate these conditions, we consider the example of subharmonic Rabi driving of a weakly anharmonic oscillator~\cite{Xia2023,Sah2024}. For $\omega = 2\pi\times 3.098~\mathrm{GHz}$, $\elementdepsymbol{E}{J}{}/h = 10~\mathrm{GHz}$, and $\beta=0.359$, the qubit frequency is $(E_1-E_0)/\hbar = 3.013~\mathrm{GHz}$. Using $g_0 = 2\pi\times 5.566~\mathrm{GHz}$ and $\omega_d = 2\pi\times 0.975~\mathrm{GHz}$, the DF Hamiltonian predicts clean subharmonic Rabi oscillations (Fig.~\hyperref[fig:DF_vs_lab]{\ref*{fig:DF_vs_lab}a}). When the ramp time is set to $t_{\mathrm{r}}=5.16~\mathrm{ns}$, satisfying the conditions above, the lab-frame dynamics (Fig.~\hyperref[fig:DF_vs_lab]{\ref*{fig:DF_vs_lab}b}) closely match the DF prediction. In contrast, for ramps that are too short or too long (Fig.~\hyperref[fig:DF_vs_lab]{\ref*{fig:DF_vs_lab}c,d}), the intended displacement is not realized, and the DF and lab-frame dynamics clearly differ from each other.

\subsection{The irrotational-gauge Hamiltonian}
\label{sec:app_irg}
To derive the irrotational-gauge Hamiltonian, we apply the transformation [Eq.~\eqref{eq:irg-transform}]
\begin{align}
U_{\text{irr}} = \exp \left\{ -\frac{i}{\hbar} \int^t_{0}{dt' \sum_{i=1}^N \epsilon_i(t') n_i} \right\},
\end{align} 
to the Hamiltonian [Eq.~\eqref{eq:multimode_Hamiltonian_start}]
\begin{align}
\mathcal{H} &= \sum_{i,j = 1}^N 4 \elementdepsymbol{E}{C}{ij} n_i n_j + \sum_{i=1}^N \epsilon_i(t) n_i \nonumber \\* \nonumber 
&\quad + \frac{1}{2} \sum\limits_{\ell=1}^W \elementdepsymbol{E}{L}{\ell} {{{\left( {\sum\limits_{i=1}^N {{\elementdepsymbol{b}{L}{\ell i}}{{\varphi }_i}} } + \elementdepsymbol{\mathcal{F}}{L}{\ell}(t) \right)}^2}}\\* 
&\quad - \sum\limits_{k=1}^{M} {E_{\mathrm{J}_k}\cos {{\left( {{\sum\limits_{i=1}^N {{b_{\mathrm{J}_{ki}}}{\varphi_i}} + \mathcal{F}_{\mathrm{J}_k}(t) }} \right)}}},
\end{align}
resulting in the irrotational Hamiltonian,
\begin{align}
\mathcal{H}_{\text{irr}} &= U_{\text{irr}}^\dagger \mathcal{H} U_{\text{irr}} - i\hbar U_{\text{irr}}^\dagger \dot{U}_{\text{irr}}.
\label{eq:app-irr-transform}
\end{align}
The above transformation has the effect of shifting the $\varphi_i$ operators, easily proved with the Baker--Campbell--Hausdorff formula:
\begin{align}
U_{\text{irr}}^\dagger \varphi_i U_{\text{irr}} &= \varphi_i + \frac{i}{\hbar} \int^t_0{dt' \sum_{j=1}^N \epsilon_j(t') \left[ n_j, \varphi_i \right]}\nonumber\\
&= \varphi_i + \frac{1}{\hbar} \int^t_0 dt' \epsilon_i(t').
\label{eq:app-Uirr-shift}
\end{align}
For the remainder of this subsection, we specialize to circuits whose only retained inductive elements are Josephson junctions, so that $W=0$. Circuits containing lumped linear inductors are treated in Appendix \ref{app:lin_inductance}. The irrotational Hamiltonian is
\begin{align}
\mathcal{H}_{\text{irr}} &= \sum_{i,j=1}^N 4 \elementdepsymbol{E}{C}{ij}n_i n_j\nonumber\\
&- \sum\limits_{k=1}^M {\elementdepsymbol{E}{J}{k}\cos {{\left(  \sum\limits_{i=1}^N {\elementdepsymbol{b}{J}{ki}} \varphi_i + \elementdepsymbol{F}{J}{k}(t) \right)}}},
\label{eq:app-irr-Hamiltonian}
\end{align}
where 
\begin{align}
    \elementdepsymbol{F}{J}{k}(t) = \elementdepsymbol{\mathcal{F}}{J}{k}(t) + \frac{1}{\hbar} \int^t_0 dt' \sum_{i=1}^N \elementdepsymbol{b}{J}{ki}  \epsilon_i(t')
    \label{eq:F_J_k(t)}
\end{align}
is the effective phase modulation parameter on the $k$-th junction. Note that even though Eq.~\eqref{eq:app-Uirr-shift} assumes $\varphi_i$ to be an extended variable, the result is similar in the case of a periodic variable where $\varphi_i \rightarrow e^{i\varphi_i}$ and does not affect the rest of the derivation. 

From the irrotational-gauge Hamiltonian, Eq.~\eqref{eq:irg-Hamiltonian-ladder} of the main text is obtained by diagonalizing the linearized circuit Hamiltonian in the ladder operator basis. Here, we make an important observation that the effective phase modulation parameter $\elementdepsymbol{F}{J}{k}(t)$ is independent of the junction inductance. This can be seen from Eqs.~\eqref{eq:app-eps_i} and \eqref{eq:F_J_k(t)} which indicate that $\elementdepsymbol{F}{J}{k}(t)$ only depends on the charge and phase modulation parameters $\mathcal{V}(t)$ and $\mathcal{F}(t)$, as well as the voltage and flux participation factors $a$ and $b$, all of which are independent of the junction inductance, according to the discussion in Appendix~\ref{sec:app_describing_circuits}. This way, we can simplify simulations and the analysis by considering the junctions to be opened, as further explained in the main text.

\subsection{Numerical sensitivity of extracted phase modulation parameters in closed-junction circuits at near-resonant drive frequencies}
\label{app:IG_sensitivity}

Numerical uncertainties in the junction phase displacements $\bar{\mathbf{A}}$ obtained from microwave simulations could propagate into the junction phase modulation parameters $\elementdepsymbol{\bar{\mathbf{F}}}{J}{}$. Since these parameters are obtained by inverting the linear relation $\bar{\mathbf{A}} = \mathbf{R} \elementdepsymbol{\bar{\mathbf{F}}}{J}{}$, the sensitivity of $\elementdepsymbol{\bar{\mathbf{F}}}{J}{}$ with respect to the uncertainties in $\bar{\mathbf{A}}$ is fully determined by $\mathbf{R}^{-1}$, and therefore depends strongly on the drive frequency. This section provides a quantitative illustration, which accounts for the large deviations observed in the junction-centered example of Fig.~\ref{fig:junction-centered}.

As a concrete example, consider again the SQUID-based flux-tunable transmon in Fig.~\ref{fig:junction-centered}, and assume for simplicity that uncertainties enter only through the amplitudes of $\bar{A}_1$ and $\bar{A}_2$, and that these uncertainties are uncorrelated. In practice, the phases of simulated displacements may also have uncertainties, and inter-junction correlations may be present. 

For the symmetric SQUID considered in Fig.~\ref{fig:junction-centered} with $\elementdepsymbol{E}{J}{1}=\elementdepsymbol{E}{J}{2} = \elementdepsymbol{E}{J}{}$, the qubit mode participates in the two junctions equally. The linearized mode frequency $\omega_q$ and the participation factors are
\begin{align}
    \hbar \omega_q
    &=
    \sqrt{
    8\elementdepsymbol{E}{C}{}
    \left(
    \elementdepsymbol{E}{J}{1}
    +
    \elementdepsymbol{E}{J}{2}
    \right)
    }
    =
    \sqrt{
    16 \elementdepsymbol{E}{C}{}
    \elementdepsymbol{E}{J}{}
    },
    \label{eq:app_SQUID_mode_freq}
    \\
    \elementdepsymbol{\beta}{J}{1q}
    &=
    \elementdepsymbol{\beta}{J}{2q}
    =
    \left[
    \frac{
    2\elementdepsymbol{E}{C}{}
    }{
    \elementdepsymbol{E}{J}{1}
    +
    \elementdepsymbol{E}{J}{2}
    }
    \right]^{1/4}
    =
    (\elementdepsymbol{E}{C}{}/\elementdepsymbol{E}{J}{})^{1/4},
    \label{eq:app_squid_beta}
\end{align}
where $\elementdepsymbol{E}{C}{}$ is the charging energy of the circuit.
Substituting the participation factor expression \eqref{eq:app_squid_beta} into Eq.~\eqref{eq:irg-r-matrix} for the tunable transmon circuit,
\begin{equation}
    (\mathbf{R})_{jk}
    =
    \frac{
    2\omega_q
    \elementdepsymbol{E}{J}{k}
    \elementdepsymbol{\beta}{J}{jq}
    \elementdepsymbol{\beta}{J}{kq}
    }{
    \hbar
    (
    \omega_d^2-\omega_q^2-\kappa_q^2/4
    -i\kappa_q\omega_d
    )
    } + \delta_{jk},
\end{equation}
and eliminating the $\elementdepsymbol{E}{C}{}$-dependency using \eqref{eq:app_SQUID_mode_freq} leads to

\begin{align}
    \mathbf{R}
    &= 
    \begin{pmatrix}
    1 + b(\omega_d) & b(\omega_d)\\
    b(\omega_d) & 1+b(\omega_d)
    \end{pmatrix},
    \\
    b(\omega_d) &= \frac{\omega_q^2}{2(\omega_d^2 - \omega_q^2 - \kappa_q^2/4 - i \kappa_q \omega_d)},
\end{align}
with inverse
\begin{equation}
    \mathbf{R}^{-1} = 
    \frac{1}{1+ 2b(\omega_d)}
    \begin{pmatrix}
    1 + b(\omega_d) & - b(\omega_d)\\
    -b(\omega_d) & 1+b(\omega_d)
    \end{pmatrix}.
\end{equation}
Let the nominal junction phase displacements be $\bar{A}_k$ and the corresponding nominal parameters $\elementdepsymbol{\bar{F}}{J}{k}$ satisfy
\begin{equation}
    \elementdepsymbol{\bar{F}}{J}{k} = \sum_{j=1}^2\mathbf{R}_{kj}^{-1} \bar{A}_j.
\end{equation}
We model the phase displacement uncertainties as independent Gaussian random variables with zero mean and standard deviation $\sigma_r$ 
\begin{align}
    \delta\bar{A}_k &= \bar{A}_k r_k,\\
    \mathbb{E}[r_k] &= 0,\\
    \mathrm{Var}[r_k] &= \sigma_r^2,
\end{align}
where $r_k$ represents the relative phase displacement uncertainty, consistent with the convergence criterion used in the simulations of Fig.~\ref{fig:junction-centered}. The resulting numerical uncertainty of $\elementdepsymbol{\bar{F}}{J}{k}$ is
\begin{equation}
    \delta\elementdepsymbol{\bar{F}}{J}{k} 
    = [\mathbf{R}^{-1}]_{k1} \bar{A}_1 r_1 + [\mathbf{R}^{-1}]_{k2} \bar{A}_2 r_2.
\end{equation}
Since $r_1$ and $r_2$ are assumed to be uncorrelated, the variance of $|\delta\elementdepsymbol{\bar{F}}{J}{k}|$ is
\begin{align}
    \mathrm{Var}[|\delta\elementdepsymbol{\bar{F}}{J}{k}|] &=  \sum_{j=1}^2 \left|[\mathbf{R}^{-1}]_{kj} \bar{A}_j\right|^2 \mathrm{Var}[r_j]\\
    &= \sum_{j=1}^2 \left|[\mathbf{R}^{-1}]_{kj} \bar{A}_j\right|^2\sigma_r^2
\end{align}
This expression highlights the origin of the enhanced sensitivity near resonance: for a given junction $k$, the factor $|\mathbf{R}_{kj}^{-1} \bar{A}_j|$ determines how strongly uncertainty in the junction phase displacement of $j$ propagates into the extracted $\elementdepsymbol{F}{J}{k}$; large values of $|\mathbf{R}_{kj}^{-1} \bar{A}_j|$ therefore amplify the impact of even small uncertainties in the simulation.

\begin{figure}[t]
    \centering
    \includegraphics[width=1\linewidth]{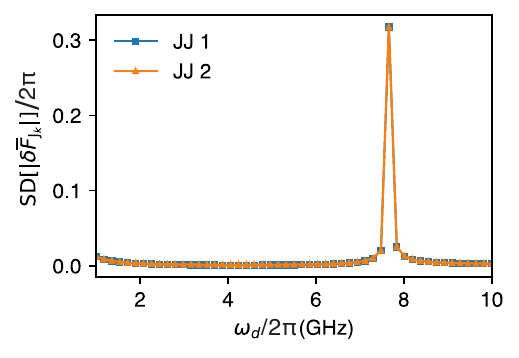}
    \caption{\textbf{Numerical sensitivity of effective phase modulation parameters.} Standard deviation of the magnitude of the numerical uncertainty $|\delta \elementdepsymbol{\bar{F}}{J}{k}|$ as a function of drive frequency based on the parameters in Fig.~\ref{fig:junction-centered}.}
    \label{fig:ig_example_sensitivity}
\end{figure}

To illustrate this effect, consider parameters of the example circuit in Fig.~\ref{fig:junction-centered}: $\omega_q/2\pi = 7.673$ GHz, $\elementdepsymbol{L}{J}{1} = \elementdepsymbol{L}{J}{2} = 15$ nH, under nominal modulations $(\elementdepsymbol{\bar{F}}{J}{1}, \elementdepsymbol{\bar{F}}{J}{2}) = (0.4, 0.6)$ for all the drive frequencies, and set $\sigma_r = 0.003$ which is the relative convergence tolerance used in the HFSS driven simulation. The resulting standard deviation of $|\delta \elementdepsymbol{\bar{F}}{J}{k}|$ is shown in Fig.~\ref{fig:ig_example_sensitivity}. When the drive frequency is near the resonance, the two terms for each junction satisfy $[\mathbf{R}^{-1}]_{k1} \bar{A}_1 \approx -[\mathbf{R}^{-1}]_{k2} \bar{A}_2$, so that each $\elementdepsymbol{\bar{F}}{J}{k}$ is obtained from the difference of two large, nearly cancelling contributions. This near-cancellation significantly amplifies the effect of relative uncertainties in $\bar{\mathbf{A}}$, which offers a plausible explanation for the observed discrepancies in Fig.~\hyperref[fig:junction-centered]{\ref*{fig:junction-centered}e}.

\subsection{Analysis of $\elementdepsymbol{L}{J}{k} \to \infty$ for opened-junction method}
\label{app:R_to_identity}

Here we prove that the matrix $(\boldsymbol{\mathbf{R}})_{jk}$ approaches $\delta_{ij}$ in the limit $\elementdepsymbol{L}{J}{k} \to \infty$ ($\elementdepsymbol{E}{J}{k} \to 0$) for all junctions, so that the expression for the effective phase modulation parameters $\elementdepsymbol{\bar{\mathbf{F}}}{J}{}$ simplifies from Eq.~(\ref{eq:f-from-Z}) to Eq.~(\ref{eq:f-from-Z-open}) by opening up junctions in the driven simulation.

For convenience, we repeat the expression for the matrix $\boldsymbol{\mathbf{R}}$ [Eq.~\eqref{eq:irg-r-matrix}]
\begin{equation}
    (\boldsymbol{\mathbf{R}})_{jk} = 
    \sum^N_{i=1} \frac{2 \omega_i \elementdepsymbol{E}{J}{k} \elementdepsymbol{\beta}{J}{ji} \elementdepsymbol{\beta}{J}{ki}}{\hbar (\omega_d^2 - \omega_i^2 - \kappa_i^2/4 - i\kappa_i \omega_d)} 
    + \delta_{jk}.
    \label{eq:app-r-matrix}
\end{equation}
As derived in \cite{Minev2021}, the normal mode participation ratio $\elementdepsymbol{\beta}{J}{ki}$ for the junction $k$ and mode $i$ is given by
\begin{equation}
    \elementdepsymbol{\beta}{J}{ki} = s_{ki} \sqrt{p_{ki} \hbar \omega_i/2\elementdepsymbol{E}{J}{k} },
    \label{eq:beta_ki_expression}
\end{equation}
where $s_{ki}$ is the EPR sign as defined in~\cite{Minev2021} and $p_{ki}$ is the inductive energy participation ratio that is bounded as $0\leq p_{ki} \leq 1$ for all $k,i$. Notice that we adopted a different index convention from the one used in \cite{Minev2021}.
To facilitate our discussion, we define a scale factor $\check{E}_J$ such that $\elementdepsymbol{E}{J}{k} = \sigma_k \elementdepsymbol{\check{E}}{J}{}$, where $\sigma_k >0$ for each junction. This guarantees that under the limit $\elementdepsymbol{\check{E}}{J}{} \to 0$, all the junctions are opened up \textit{simultaneously}. Substituting the expression for $\elementdepsymbol{\beta}{J}{ki}$ and $\elementdepsymbol{E}{J}{k}$ into the expression for a matrix element of $\boldsymbol{\mathbf{R}}$ [Eq.~\eqref{eq:app-r-matrix}] gives
\begin{equation}
    (\boldsymbol{\mathbf{R}})_{jk} = 
    \sum^N_{i=1} \frac{2 \omega_i^2 s_{ji} s_{ki}}{\omega_d^2 - \omega_i^2 - \kappa_i^2/4 - i\kappa_i \omega_d} \sqrt{p_{ji} p_{ki}}
    \sqrt{\frac{\sigma_j}{\sigma_k}}
    + \delta_{jk}.
\end{equation}
Since by construction all the circuit inductance is provided by junctions, all mode frequencies $\{\omega_i\}$ approach zero as $\check{E}_J \to 0$ as all the inductances approach infinity. This causes $(\boldsymbol{\mathbf{R}})_{jk} \to \delta_{jk}$ under finite drive frequency $\omega_d$ and proves the assertion that under such limit, Eq.~(\ref{eq:f-from-Z}) simplifies to Eq.~(\ref{eq:f-from-Z-open}).

\subsection{Applicability of the irrotational-gauge method}
\label{app:lin_inductance}

We remark on the applicability of the IG method when the circuit contains linear inductive elements. Lumped linear inductors with well-defined terminal pairs can be included by augmenting the set of branch variables, while distributed inductive elements become problematic when their effective phase modulations cannot be assigned to a finite set of localized branch variables. In Appendix \ref{sec:app_irg}, we have seen that for circuits in the absence of linear inductance, the only modulation parameters in the Hamiltonian are the effective phase modulation parameters of all the junctions, $\{\elementdepsymbol{F}{J}{k}\}$. When the modulation is sinusoidal and monochromatic, we can obtain the phasors of these modulation parameters $\elementdepsymbol{\mathbf{\bar{F}}}{J}{}$ from the phase displacement on each junction under drive, with the help of Eqs.~\eqref{eq:f-from-Z} or \eqref{eq:f-from-Z-open}. 

We now consider the case where the system has linear inductances in addition to the Josephson inductance, so that the Hamiltonian under the irrotational gauge is generalized from Eq.~(\ref{eq:app-irr-Hamiltonian}) to
\begin{align}
\mathcal{H}_{\text{irr}} &= \sum_{i,j=1}^N 4 \elementdepsymbol{E}{C}{ij}n_i n_j\nonumber\\
&+ \frac{1}{2}\sum\limits_{\ell=1}^W {\elementdepsymbol{E}{L}{\ell} {{\left(  \sum\limits_{i=1}^N {\elementdepsymbol{b}{L}{\ell i}} \varphi_i + \elementdepsymbol{F}{L}{\ell}(t) \right)}}^2}\nonumber\\
&- \sum\limits_{k=1}^M {\elementdepsymbol{E}{J}{k}\cos {{\left(  \sum\limits_{i=1}^N {\elementdepsymbol{b}{J}{ki}} \varphi_i + \elementdepsymbol{F}{J}{k}(t) \right)}}}.
\label{eq:app-irr-H-with-lin-ind}
\end{align}
In addition to the effective junction phase modulation $\elementdepsymbol{F}{J}{k}$, here we have modulation parameters $\elementdepsymbol{F}{L}{\ell}$ for linear inductive terms
\begin{align}
    \elementdepsymbol{F}{L}{\ell}(t) = \elementdepsymbol{\mathcal{F}}{L}{\ell}(t) + \frac{1}{\hbar} \int^t_0 dt' \sum_{i=1}^N \elementdepsymbol{b}{L}{\ell i}  \epsilon_i(t'),
    \label{eq: F_l(t)}
\end{align}
and we define its phasor representation as $\elementdepsymbol{\bar{F}}{L}{\ell}$.
The Hamiltonian expressed using the ladder operators is
\begin{align}
\begin{split}
\mathcal{H}_{\text{irr}} &= \sum_{i=1}^N \hbar \omega_i a_i^\dagger a_i - \frac{1}{2} \sum_{k=1}^M \elementdepsymbol{E}{J}{k} \left[ \sum_{i=1}^N \elementdepsymbol{\beta}{J}{ki} (a_i + a_i^\dagger) \right]^2\\
&- \frac{1}{2} \sum_{\ell=1}^W \elementdepsymbol{E}{L}{\ell} \left[ \sum_{i=1}^N \elementdepsymbol{\beta}{L}{\ell i} (a_i + a_i^\dagger) \right]^2 \\
&+ \frac{1}{2}\sum_{\ell=1}^W \elementdepsymbol{E}{L}{\ell} \left( \sum_{i=1}^N \elementdepsymbol{\beta}{L}{\ell i} (a_i + a_i^\dagger) + \elementdepsymbol{F}{L}{\ell}(t) \right)^2 \\
&- \sum_{k=1}^M \elementdepsymbol{E}{J}{k} \cos \left( \sum_{i=1}^N \elementdepsymbol{\beta}{J}{ki} (a_i + a_i^\dagger) + \elementdepsymbol{F}{J}{k}(t) \right),
\end{split}
\label{eq:irg-Hamiltonian-ladder-with-lin-inductance}
\end{align}
where the explicit quadratic terms subtract the linearized inductive potential already included in the harmonic normal-mode Hamiltonian, to avoid double counting. Clearly, for linear inductors well approximated by lumped inductors, such as granular aluminum nanowires, we can obtain phase displacements across these inductors $A_\ell' (t) = \langle\sum_{i=1}^N \elementdepsymbol{\beta}{L}{\ell i} (a_i + a_i^\dagger) + \elementdepsymbol{F}{L}{\ell}(t)\rangle$ (and its phasor representation $\bar A_\ell'$) with microwave simulations, in the same way as for the Josephson junctions. The relation between the phase displacements of all the inductors (including the junctions) and the phase modulation parameters is 
\begin{align}
\bar{A}^+_j = \sum_{i=1}^N \sum_{k=1}^{M+W} \frac{2\omega_i E_{\text{ind}_k} \elementdepsymbol{\beta}{\text{ind}}{ji} \elementdepsymbol{\beta}{\text{ind}}{ki} \bar{F}^+_k}{\hbar \left( \omega_d^2 - \omega_i^2 - \kappa_i^2/4 - i\omega_d \kappa_i \right)} + \bar{F}^+_j, 
\label{eq:irg-linear-equation-linear-inductance}
\end{align}
where $\bar {\mathbf A}^+ = (\bar{A}_{1}, ..., \bar{A}_M, \bar{A}'_{1}, ..., \bar{A}'_W)^\textrm{T}$
$\bar{\mathbf{F}}^+ = (\elementdepsymbol{\bar{F}}{J}{1}, ..., \elementdepsymbol{\bar{F}}{J}{M}, \elementdepsymbol{\bar{F}}{L}{1}, ..., \elementdepsymbol{\bar{F}}{L}{W})^\textrm{T}$. 
The inductive energy $\elementdepsymbol{E}{ind}{k}$ and the inductive participation factor $\elementdepsymbol{\beta}{ind}{ki}$ are defined as
\begin{align}
    \elementdepsymbol{E}{ind}{k}
    &= 
    \begin{cases}
        \elementdepsymbol{E}{J}{k} 
        \quad 
        \text{for }1 \leq k \leq M\\
        \elementdepsymbol{E}{L}{k} 
        \quad 
        \text{for }M+1 \leq k \leq M+W
    \end{cases},
    \\
    \elementdepsymbol{\beta}{ind}{ki}
    &= 
    \begin{cases}
        \elementdepsymbol{\beta}{J}{ki} 
        \quad 
        \textrm{for }1 \leq k \leq M\\
        \elementdepsymbol{\beta}{L}{ki} 
        \quad 
        \textrm{for }M+1 \leq k \leq M+W
    \end{cases}.
\end{align}

We can easily generalize Eqs.~\eqref{eq:f-from-Z} and \eqref{eq:f-from-Z-open} to invert Eq.~(\ref{eq:irg-linear-equation-linear-inductance}) and obtain both $\{\elementdepsymbol{\bar{F}}{J}{k}\}$ and $\{\elementdepsymbol{\bar{F}}{L}{\ell}\}$, thereby obtaining all the modulation parameters in the Hamiltonian. 

However, if the circuit contains distributed inductive elements, such as a 3D \(\lambda/4\) stub cavity~\cite{Reagor2013}, the corresponding phase displacement \(A'_\ell(t)\) generally cannot be obtained from a localized voltage drop as in the case of a lumped inductor. If such a distributed element acquires a non-negligible effective phase modulation, the number of unknown modulation parameters exceeds the number of branch phase-displacement equations available from the localized junction and lumped-inductor ports. In that case, Eq.~\eqref{eq:irg-linear-equation-linear-inductance} does not provide a closed inversion for all modulation parameters.

This limitation should not be interpreted as applying to all circuits containing distributed linear modes. If the effective phase modulation of the distributed inductive elements is negligible, i.e., \(\elementdepsymbol{F}{L}{\ell}\approx 0\) for these elements, the remaining modulation parameters can still be obtained from the localized branch phase displacements alone. In this case, the only unknowns are the \(M\) junction phasors \(\{\elementdepsymbol{\bar{F}}{J}{1},\ldots,\elementdepsymbol{\bar{F}}{J}{M}\}\), which are uniquely determined by the junction phase displacements \(\{\bar{A}_{1},\ldots,\bar{A}_{M}\}\) through the closed-junction IG relation. One example is a lumped-element Josephson circuit coupled to resonator modes and driven by a line that predominantly excites the Josephson circuit while only weakly coupling to the distributed modes. This represents a common class of superconducting circuit designs where the Josephson circuit serves as a nonlinear mixer among resonator modes. When this separation of coupling does not hold, the IG method no longer provides a closed extraction of all modulation parameters, and one should instead employ the displaced-frame method or overlap method to construct the driven Hamiltonian.

\subsection{Equivalence of the IG method with earlier formulations}
\label{app:You_Yao_DiVincenzo}

In this section, we establish (in appropriate limits) the equivalence between our irrotational gauge method and those of You {\it et al.} \cite{You2019} and Riwar and DiVincenzo \cite{Riwar2022}.

We first focus on the specific example of a dc SQUID and review the method of Riwar and DiVincenzo in Ref.~\cite{Riwar2022}. The electric field $\E$ and magnetic field $\B$ are expressed in terms of a scalar potential $V$ and a vector potential $\A$ as
\begin{align}
\label{eq:E}
\E &= -\nabla V - \dot{\A}, \\ 
\B &= \nabla \times \A.
\end{align}
In particular, in the presence of time-dependent flux sources, the fields must obey the Maxwell-Faraday equation 
\begin{align}
\label{eq:Maxwell_Faraday}
\oint\E\cdot\dell=-\dot{\Phi}_{\mathrm{ext}},
\end{align}
where the line integral is taken around a loop enclosing the circuit, outside of the circuit elements themselves. We will find it useful to separate the loop integral into segments for each branch of the circuit
\begin{align}
\label{eq:Kirc}
\sum_{b}\int_{b}\E\cdot\dell=-\dot{\Phi}_{\mathrm{ext}}.
\end{align}
In the time-independent case with $\frac{d}{dt}\Phie=0$ and identifying the voltage drop across each element as $V_{b}=-\int_{b}\E\cdot\dell$, we recover Kirchhoff's voltage law. In the context of circuit QED, this equation is more typically written in its integral form
\begin{align}
\sum_{b}\Phi_{b}=-\Phie,
\end{align}
where $\Phi_{b}=\int dt\int_{b}\E\cdot\dell$ is typically referred to as the branch flux across a particular element. The lower limit of integration is taken to be a time in the distant past when no flux threaded the circuit. This equation imposes a constraint on the branch fluxes and thus reduces the number of independent degrees of freedom.

Towards choosing a gauge, we decompose the electric field into $\E=\EQ+\EBd + \Eng$, where $\EQ$ is the solution to Maxwell's equations in the absence of a time-varying magnetic field or applied electric field $\Eng$. We make our gauge choice by defining the electric potential across each branch $b$ as $V_{b}=-\int_{b}(\EQ + \Eng ) \cdot\dell$. This choice corresponds to the irrotational gauge, which will become clear in the following. Note that this definition coincides with that in the electrostatic case above and thus $\EBd=-\dot{\A}$ captures the component of the electric field due to the time variation of $\B$. 

To write down the Lagrangian of the dc SQUID, we require two more pieces of information. These are the expressions for the energies of a capacitor and of a Josephson element. The energy of a capacitor with capacitance $C$ on branch $b$ is $\frac{1}{2}C V_{b}^2=\frac{1}{2}C(\int_{b}(\EQ+\Eng)\cdot\dell)^2$. The energy of a Josephson element on branch $b$ is $-\elementdepsymbol{E}{J}{b}\cos(\varphi_{b})$, where the gauge-invariant phase $\varphi_{b}$ obeys the ac Josephson relation $\dot\varphi_{b}=\frac{2\pi}{\Phi_{0}}\int_{b}\E\cdot\dell$ \cite{Devoret2021, Josephson1969}. Thus, the Lagrangian is
\begin{align}
\label{eq:L}
\mathcal{L} &= \sum_{b=\ell,r}\frac{1}{2}C_{b}\left(\int_{b}\boldsymbol(\EQ+\Eng)\cdot \dell \right)^2
\nonumber \\ 
&\quad+\sum_{b=\ell,r}\elementdepsymbol{E}{J}{b}\cos\left( \frac{2\pi}{\Phi_{0}}\int \dt\int_{b} \E\cdot \dell \right),
\end{align}
where $\ell,r$ refer to the left and right branches, respectively. To write this Lagrangian in terms of a single degree of freedom, we use the constraint Eq.~\eqref{eq:Kirc}
\begin{align}
-\dot{\Phi}_{\mathrm{ext}} &= \sum_{b}\int_{b}\E\cdot\dell \nonumber\\  &= \sum_{b=\ell,r}\left(\int_{b}[\EQ+\Eng]\cdot \dell -\int_{b}\dot{\A}\cdot\dell \right)\nonumber \\ 
&= \sum_{b=\ell,r}\int_{b}[\EQ+\Eng]\cdot \dell - \dot{\Phi}_{\mathrm{ext}},
\end{align}
where in going from the second line to the third we have used $\oint\A\cdot\dell=\Phie$.
Thus we obtain $\int_{\ell}[\EQ+\Eng]\cdot \dell = -\int_{r}[\EQ+\Eng]\cdot \dell$. The Lagrangian is now
\begin{align}
\label{eq:L_irr}
\mathcal{L} &= \frac{1}{2}C_{\Sigma}(\dot{\Phi} + V_{n_{g}})^2 \nonumber \\ 
 &\quad +\sum_{b=l,r}\elementdepsymbol{E}{J}{b}\cos\left(\frac{2\pi}{\Phi_{0}}\Phi -\frac{2\pi}{\Phi_{0}}\int_{\ell} \A\cdot\dell \right) ,
\end{align}
where $\dot\Phi=\int_{\ell}\EQ\cdot \dell$ and $V_{n_{g}}=\int_{\ell}\Eng\cdot \dell$. This result coincides with that of Ref.~\cite{Riwar2022} [c.f. Eq.~(24)], up to minus signs corresponding to different definitions of positive and negative voltage drop.

We now want to demonstrate that a different gauge choice yields terms linear in $\dot{\Phi}_{\mathrm{ext}}$ in the Lagrangian [c.f. the Methods section of Ref.~\cite{Riwar2022}]. Consider the gauge transformation $V'=V+\dot{\psi}, \A'=\A-\nabla\psi$, with $\psi=\int_{r} \A\cdot\dell$. This corresponds to the variable transformation $\Phi'=\Phi+\psi$. Ignoring any external electric field, the Lagrangian becomes
\begin{align}
\mathcal{L}& = \frac{1}{2}C_{\Sigma}\left(\dot{\Phi'} - \int_{r} \dot{\A}\cdot\dell \right)^2 
\nonumber \\ 
 &\quad +\elementdepsymbol{E}{J}{\ell}\cos\left(\frac{2\pi}{\Phi_{0}}\Phi' -\frac{2\pi}{\Phi_{0}}\Phie \right) + \elementdepsymbol{E}{J}{r}\cos\left(\frac{2\pi}{\Phi_{0}}\Phi'\right).
\end{align}
We recover the result that fully allocating the external flux to one term or another in the potential generally yields terms linear in the dynamical variable in the kinetic term. The irrotational gauge is the unique gauge where all such linear terms (proportional to $\int_r \dot{ \mathbf{A}}\cdot \mathrm{d}\boldsymbol{\ell}$)  are eliminated. 

We now show that the above results reduce to those of You {\it et al.}~\cite{You2019}, for a simple lumped-element SQUID circuit with time-dependent flux penetrating its loop. Specifically, we demonstrate that the irrotational gauge corresponds to a choice of generalized coordinate consistent with the electrostatic definition of the electric potential. To simplify the discussion, we omit any externally applied offset field $\mathbf{E}_{n_g}$, and focus solely on the total electric field
\begin{equation}
    \mathbf{E}=\mathbf{E}_Q + \EBd, 
\end{equation}
where $\mathbf{E}_Q$ arises from accumulated net charge and $\EBd$ is the total field induced by a time-varying magnetic field.
Physically, the induced field $\EBd$ has two contributions:
\begin{equation}
    \EBd = \EBd^{(a)} + \EBd^{(b)}.
\end{equation}
The first term, $\EBd^{(a)}$, is the source-free field resulting directly from Faraday’s law. The second term $\EBd^{(b)}$ is the field generated by the redistribution of charges within the circuit in response to $\EBd^{(a)}$, such that the net electric field vanishes inside the inductor. Importantly, this redistribution alters the internal charge configuration without changing the total charge on each superconducting island.

The Lagrangian of the system is given by 
\begin{equation}
    \begin{aligned}
        \mathcal{L} = & \sum_{b=l,r}\frac{1}{2}C_b\left[ \left(\int_b \EBd^{(b)} + \int_b \mathbf{E}_{Q}\right)\mathrm{d}l \right]^2 \\
        & + \sum_{b=l,r} E_{\text{J},b} \cos\frac{2\pi}{\Phi_0}\left[\int \mathrm{d}t\int_b  \left(\EBd^{(a)} + \EBd^{(b)}
        + \mathbf{E}_{Q}\right) \mathrm{d}l\right].
    \end{aligned}
\end{equation}
Here, the capacitive energy depends exclusively on fields generated by charges, both net charges and redistribution-induced charges, whereas the Josephson phase dynamics are sensitive to the total electric field.

We now define a dynamical flux variable $\Phi$, subject to 
\begin{equation}\label{eq:irr_dev}
\begin{aligned}
    \dot{\Phi} + \dot{\Phi}_l = \int_l \left(\EBd^{(a)} + \EBd^{(b)} + \mathbf{E}_{Q}\right) \mathrm{d}l, \\ 
    \dot{\Phi} + \dot{\Phi}_r = -\int_r \left(\EBd^{(a)} + \EBd^{(b)} + \mathbf{E}_{Q}\right) \mathrm{d}l,
\end{aligned}
\end{equation}
where $\dot{\Phi}_l$ and $\dot{\Phi}_r$ are auxiliary flux variables to be determined. Subtracting the two equations yields the constraint
\begin{equation}
    \dot{\Phi}_l - \dot{\Phi}_r = \oint \left(\EBd^{(a)} + \EBd^{(b)} + \mathbf{E}_{Q}\right) \mathrm{d}l  = \dot{\Phi}_\text{ext},
\end{equation}
where we have used the fact that the total electric field inside the inductor vanishes.

The irrotational gauge is defined as the choice of $\Phi$ such that the capacitive part of the Lagrangian is purely quadratic in
$\dot{\Phi}$. This condition imposes the following relation:
\begin{equation}
    C_l\left(\dot{\Phi}_l - \int_l \EBd^{(a)}\,\mathrm{d}l \right) +  C_r\left(\dot{\Phi}_r + \int_r\EBd^{(a)} \,\mathrm{d}l \right)= 0. 
\end{equation}
To solve this equation, we impose two additional physical constraints:
(1) Charge neutrality of redistribution: the redistribution process does not alter the net charge on the islands:
\begin{equation}
    C_l \int_l \EBd^{(b)} \,\mathrm{d}l = C_r \int_r \EBd^{(b)} \,\mathrm{d}l.\label{eq:C98}
\end{equation}
(2) Curl-free nature of the field from net charge:
\begin{equation}
    \int_l\mathbf{E}_{Q} \,\mathrm{d}l + \int_r\mathbf{E}_{Q}\, \mathrm{d}l = 0.
\end{equation}
Under these conditions, the auxiliary variables can be solved as
\begin{equation}
    \dot{\Phi}_{l,r} = \int_{l,r}\left(\EBd^{(a)} + \EBd^{(b)} \right) \mathrm{d}l = - \int_{l,r} \dot{A}\,\mathrm{d}l,\label{eq:C101}
\end{equation}
where $\mathbf{A}$ is the vector potential.
Substituting back into Eq.~\eqref{eq:irr_dev}, we obtain
\begin{equation}
    \dot{\Phi} = \int_l\mathbf{E}_{Q} \,\mathrm{d}l. 
\end{equation}
The above two expressions are precisely the results from the irrotational gauge as introduced in Ref.~\cite{Riwar2022}. 

Since the induced electric field $\EBd^{(a)}$ is distributed along the circuit loop, its contribution between the junctions is negligible. Thus, Eq.~\eqref{eq:C98} can be approximated as
\begin{align}
C_{\ell}\int_{\ell}\A\cdot\dell = C_{r}\int_{r}\A\cdot\dell.
\end{align}
We have also already noted that $\oint\A\cdot\dell=\int_{\ell}\A\cdot\dell + \int_{r}\A\cdot\dell=\Phie$. We may now immediately solve for both unknowns $\int_{\ell,r}\A\cdot\dell$, yielding
\begin{align}
\int_{\ell}\A\cdot\dell &= \frac{C_{r}}{C_{\Sigma}}\Phie, \quad 
\int_{r}\A\cdot\dell = \frac{C_{\ell}}{C_{\Sigma}}\Phie.
\end{align}
The Lagrangian now simplifies to
\begin{align}
\mathcal{L} &= \frac{1}{2}\left(\frac{\Phi_{0}}{2\pi}\right)^2C_{\Sigma}\dot{\varphi}^2 \nonumber \\ 
& \quad +\elementdepsymbol{E}{J}{\ell}\cos\left(\varphi - \frac{C_{r}}{C_{\Sigma}}\varphie \right) 
+ \elementdepsymbol{E}{J}{r}\cos\left(\varphi + \frac{C_{\ell}}{C_{\Sigma}}\varphie  \right),
\end{align}
c.f. Eqs.~(13, 24, 29) of Ref.~\cite{You2019}.
We have defined $\varphi=\frac{2\pi}{\Phi_{0}}\Phi$ and similarly $\varphie=\frac{2\pi}{\Phi_{0}}\Phie$. One may check that the definition of the dynamical variable $\varphi$ coincides with that in the irrotational gauge in Ref.~\cite{You2019}, c.f. Eqs.~(10, 24).

We now discuss how these results relate to our work. The Lagrangian of the dc SQUID in our irrotational gauge method is written as
\begin{align}
\mathcal{L}&=\frac{1}{2}C_{\Sigma}\dot{\Phi}^2 
+\sum_{k=1,2}\elementdepsymbol{E}{J}{k}\cos\left(\frac{\Phi}{\phi_0}+\elementdepsymbol{F}{J}{k}(t)\right). \label{eq:L_irr2}
\end{align}
We solve for the $\elementdepsymbol{F}{J}{k}(t)$ functions in the linear case following the discussion in the main text. In short, we extract the phases across each junction branch $\varphi_{1,2}=\frac{2\pi}{\Phi_{0}}\int_{-\infty}^{t}\int_{1,2}\E(t')\cdot\dell\,\mathrm d t'$, and then invert it to the modulation parameters $\elementdepsymbol{F}{J}{k}(t)$ according to the linear-response relations in Eqs.~\eqref{eq:f-from-Z} and \eqref{eq:f-from-Z-open}. The time-dependent offset charges are incorporated in $\elementdepsymbol{\mathcal{F}}{J}{k}$ through the unitary transformation defined in Eq.~\eqref{eq:irg-transform}, whereas in Ref. \cite{Riwar2022}, these remain as a coupling to the charge operator. To obtain a Lagrangian that is free of time-dependent charge coupling with Riwar and DiVincenzo's method, we can choose the gauge $V'' = V + \dot \chi$, $\mathbf{A}'' = \mathbf{A} - \nabla \chi$, with $ \chi = -\int^t_{-\infty} V_{n_g}(t') \mathrm d t'$ assuming $V_{n_g}$ vanishes at a distant past. The flux is transformed as $\Phi'' = \Phi +\chi$, thereby the Lagrangian becomes
\begin{align}
\mathcal{L} &= \frac{1}{2}C_{\Sigma}
\dot{\Phi}''^2 \nonumber \\ 
 &\quad +\sum_{k=1,2}\elementdepsymbol{E}{J}{k}\cos\left[\frac{2\pi}{\Phi_{0}}(\Phi''-\chi) -\frac{2\pi}{\Phi_{0}}\int_k \A\cdot\dell \right],\label{eq:L_irr_no_offset_charge}
\end{align}
which aligns with the Lagrangian obtained with our IG method, Eqs.~\eqref{eq:L_irr2}.

\subsection{Simulation port conventions}
\label{app:simulation_port_conventions}

\begin{figure}[t]
\centering
\includegraphics[width=\linewidth]{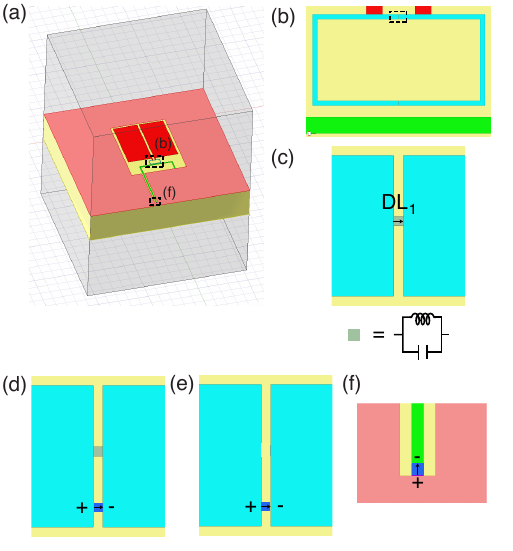}
\caption{
\textbf{HFSS terminal and port conventions used in the junction-centered calculations.}
(a) HFSS layout of the SQUID-transmon circuit in Fig.~\ref{fig:junction-centered}. The circled SQUID loop and drive-port regions are enlarged in (b) and (f).
(b) Zoomed-in view of the SQUID loop, with junctions at the top and bottom. The boxed region marks JJ1.
(c) Zoomed-in view of JJ1. The linearized junction is represented by a lumped RLC boundary across a small gap between the metal leads. The directed line $\mathrm{DL}_1$ defines the electric-field line integral used to extract the junction voltage. In the opened-junction simulation, the lumped inductive boundary is removed while the same terminal pair and directed line are retained.
(d) Closed-junction port definition. The impedance port is placed on a separate surface in parallel with the lumped RLC boundary, with the same terminal orientation as $\mathrm{DL}_1$. The port polarity is indicated by the $+$ and $-$ labels.
(e) Opened-junction port definition. The lumped RLC boundary is removed, and the port is defined directly across the same terminal pair with the same polarity convention. The resulting opened-port impedance matrix can be combined with junction shunt admittances to obtain the closed-junction response.
(f) Drive-port convention. The port connects the ground plane to the flux-bias line, with polarity defined by the indicated terminal orientation.}
\label{fig:simulation_port_conventions}
\end{figure}

In this section, we clarify the simulation conventions used for extracting phase displacements across junctions. Figure~\ref{fig:simulation_port_conventions} gives a schematic illustration for the SQUID transmon circuit in Fig.~\ref{fig:junction-centered}, showing the physical drive port, junction terminal pairs, and directed lines used for voltage/field integration. The accompanying dataset repository provides the corresponding HFSS project files and field-calculator expressions used in these examples, including the drive-port definitions and the directed junction and port line integrals used to extract the complex voltage phasors~\cite{luDatasetSystematicConstruction2026}.

In the HFSS models used here, Josephson junctions are represented by small gaps between metal leads. This representation is an effective electromagnetic model rather than a literal geometric model of the junction, since the physical dimensions of the junction are typically below the resolution practical for full-wave electromagnetic simulations. At this level of modeling, the junction is characterized by the voltage defined between the chosen terminal pair, equivalently the electric-field integral along the specified directed line, rather than by a resolved microscopic field profile inside the junction.

For both the irrotational-gauge and displaced-frame methods, the voltage phasor across the chosen inductive terminal pair is obtained from this directed electric-field integral and converted to the corresponding phase displacement. In the closed-junction calculation, the junction self-inductance, and optionally its self-capacitance, are implemented through a lumped RLC boundary condition across the gap. The integration line is drawn across this lumped boundary from one terminal to the other, as illustrated in Fig.~\hyperref[fig:simulation_port_conventions]{\ref*{fig:simulation_port_conventions}c}. In the examples shown in Fig.~\ref{fig:junction-centered}, we include only the junction inductances in the lumped RLC boundaries for simplicity.

In the opened-junction calculation, the inductive Josephson element is removed while the same terminal pair and directed integration line are retained. If a junction self-capacitance is included, it remains as a capacitive lumped boundary condition across the same terminals. In the simplified examples considered in Fig.~\ref{fig:junction-centered}, the junction capacitances are omitted, so opening the junction amounts to removing the lumped RLC boundary condition entirely.

The same directed terminal pairs are also used to define junction ports when impedance matrices are computed. In HFSS, however, a lumped RLC boundary and a port region cannot occupy the same surface. We therefore use one of two equivalent implementations. The first is to include the lumped RLC boundary between the metal leads and define a separate port region in parallel, connecting the same two terminals, to extract the closed-junction impedance response, as shown in Fig.~\hyperref[fig:simulation_port_conventions]{\ref*{fig:simulation_port_conventions}d}. This approach directly gives the closed response, but the simulation must be repeated whenever the lumped-element parameters are changed.

The second approach is to remove the lumped RLC boundary, define only the opened junction ports, and compute the opened-port impedance matrix once (see Fig.~\hyperref[fig:simulation_port_conventions]{\ref*{fig:simulation_port_conventions}e}). The shunting inductance and capacitance of each junction can then be added analytically as parallel admittance contributions,
\begin{equation}
    \mathbf{Z}_{\mathrm{closed}} =
    \left[
    \mathbf{Z}_{\mathrm{opened}}^{-1}
    + \mathbf{Y}_{\mathrm{shunt}}
    \right]^{-1}.
\end{equation}
Here $\mathbf{Y}_{\mathrm{shunt}}$ is diagonal in the junction-port subspace. For a capacitance $C_j$ and inductance $L_j$ in parallel at port $j$,
\begin{equation}
    [\mathbf{Y}_{\mathrm{shunt}}]_{jj}
    =
    i\omega C_j + \frac{1}{i\omega L_j}.
\end{equation}
This second implementation requires only one full-wave simulation for a given geometry, after which the shunting admittances can be varied without rerunning simulations.

The drive port is defined separately according to the same directed-port convention: it connects the drive line to the ground plane, with the port integration line oriented as shown in Fig.~\hyperref[fig:simulation_port_conventions]{\ref*{fig:simulation_port_conventions}f}.

\section{The overlap method}
\label{app:overlap_method}
In this section, we start by deriving the overlap method Hamiltonian with explicit charge drive terms and residual, nonlinear junction phase modulation. We then develop two parallel derivations for the electric-field overlap equation: one that maps the static and driven electric fields to voltages across the capacitive elements of the effective lumped-element circuit,  and a first-principles, field-centric derivation from Maxwell’s equations and boundary conditions, without relying on an explicit circuit model. 

\subsection{The overlap method Hamiltonian}
\label{sec:app-overlap-Hamiltonian}

Here we show how the Hamiltonian in Eq.~\eqref{eq:overlap-ov-Hamiltonian} can be obtained by performing a gauge transformation to the general Josephson circuit Hamiltonian in Eq.~\eqref{eq:multimode_Hamiltonian_start},
\begin{align}
\mathcal{H}_{\text{ov}} &= U_{\text{ov}}^\dagger \mathcal{H} U_{\text{ov}} - i\hbar U_{\text{ov}}^\dagger \dot{U}_{\text{ov}}.
\end{align}
This gauge transformation $
U_{\text{ov}} = \exp \left\{i\sum_{i=1}^N \Delta \varphi_i n_i \right\}$ shifts the phase operator $\varphi_i \rightarrow \varphi_i + \Delta \varphi_i$,
\begin{align}
\begin{split}
    \Delta \varphi_i &= -\frac{\sum_{\ell=1}^W \elementdepsymbol{E}{L}{\ell} b_{\ell i} \elementdepsymbol{\mathcal{F}}{L}{\ell} (t) 
    + \sum_{k=1}^M \elementdepsymbol{E}{J}{k} b_{k i} \elementdepsymbol{\mathcal{F}}{J}{k} (t)}{\sum_{\ell=1}^W \elementdepsymbol{E}{L}{\ell} b^2_{\ell i}
    + \sum_{k=1}^M \elementdepsymbol{E}{J}{k} b^2_{k i}}.
\end{split}
\end{align}
The transformed Hamiltonian is
\begin{align}
    \mathcal{H}_\text{ov} &= \sum_{i,j=1}^N 4 \elementdepsymbol{E}{C}{ij}n_i n_j + \sum_{i=1}^N \epsilon'_{i}(t) n_i\nonumber\\
    &+ \frac{1}{2} \Bigg\{ \sum_{\ell=1}^W \elementdepsymbol{E}{L}{\ell} \left( \sum_{i=1}^N \elementdepsymbol{b}{L}{\ell i} \varphi_i\right)^2\nonumber\\
    &+ \sum_{k=1}^M \elementdepsymbol{E}{J}{k} \left( \sum_{i=1}^N b_{k i} \varphi_i\right)^2 \Bigg\}\nonumber\\
    &- \sum_{k=1}^M \elementdepsymbol{E}{J}{k} \cos_{\mathrm{nl}} \left( \sum_{i=1}^N \elementdepsymbol{b}{J}{ki} \varphi_i + \elementdepsymbol{F}{J}{k}(t) \right), \label{app:eq:H_ov}
\end{align}
where $\epsilon'_{i}(t) = \epsilon_i(t) - \hbar\partial_t \Delta\varphi_i $, and $\elementdepsymbol{F}{J}{k}(t) = \elementdepsymbol{\mathcal{F}}{J}{k}(t) + \sum_{i=1}^N \elementdepsymbol{b}{J}{ki}\Delta\varphi_i$. Here, we assume all modes in the circuit are included in the Hamiltonian, and thus $\elementdepsymbol{F}{J}{k}(t)$ does not include contributions from the truncated modes. Mode truncation naturally leads to the modification of $\elementdepsymbol{F}{J}{k}(t)$ to $A_{\textrm{res},k}(t)$, see discussions in Appendix~\ref{app:add_nonlinearity}. Then, diagonalizing the linearized circuit part of the Hamiltonian in Eq.~\eqref{app:eq:H_ov} recovers Eq.~\eqref{eq:overlap-ov-Hamiltonian} of the main text.

\subsection{Circuit-centric derivation of the overlap method}
\label{sec:app-overlap-derivation}
\begin{figure}[h]
  \centering \includegraphics[width=0.9\columnwidth]{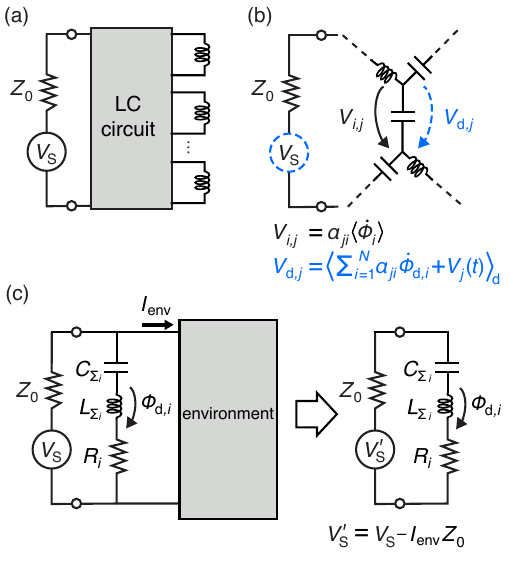}
  \caption{\textbf{Lumped-element circuit picture for the overlap method.} 
  (a) A generic electromagnetic circuit is driven from a loaded voltage source. Josephson junctions are modeled as linear inductors in the LC circuit.
(b) The LC circuit is mapped to an equivalent lumped-element network in which the $j$-th capacitor is shown here. The $i$-th normal-mode flux $\Phi_i$ has participation $a_{ji}$ in the capacitor, which results in a maximum voltage drop of $V_{i,j}$. Under drive at frequency $\omega_d$, the total voltage drop $V_{d,j}$ is the sum of the modal response contributions and the contribution from the voltage drive.
(c) To account for finite port impedance and for other ``environment'' modes drawing current through the same port, we isolate the mode of interest as an $L_{\Sigma i}$--$C_{\Sigma i}$--$R_i$ branch and model the rest of the circuit as a parallel load. The current $I_{\text{env}}$ pulled by this load reduces the device-side drive to $V_s' = V_s - I_{\text{env}} Z_0$. This picture leads directly to the frequency-dependent and lossy version of the overlap formula used in the text.
  }
\label{fig5}
\end{figure}

In this section, we derive the expression for the effective charge modulation parameter $g_i(t)$ in Eq.~\eqref{eq:gi_overlap} of the main text, from the definition of the electric overlap given in Eq.~\eqref{eq:electric_overlap}. We start with mapping the physical circuit into an abstract, lumped-element circuit network, shown in Fig.~\ref{fig5}. Here, we treat the Josephson junctions as linearized inductors for the overlap calculation of the effective charge modulation. In Appendix~\ref{app:add_nonlinearity}, we discuss how the Josephson nonlinearity can be reinserted into the Hamiltonian, along with the nonlinear phase modulation of the junction, $A_\textrm{res}(t)$.

Under drive, the electric and displacement field distribution of the real circuit are respectively $\mathbf{E}_d (\mathbf{r})$ and $\mathbf{D}_d (\mathbf{r}) = \varepsilon(\mathrm{r})\mathbf{E}_d (\mathbf{r})$, where $\varepsilon$ is the permittivity which we assume to be linear and isotropic, but may have spatial dependence. In the corresponding lumped-element model, the electric and displacement field on the $j$-th capacitor are related to the voltage across the capacitor branch
\begin{align}
E_{d,j}(t) 
&= 
\frac{V_{d,j}(t)}
{{d}_j}  
\label{eq:app-driven-E-field}\\
D_{d,j}(t) 
&= 
\varepsilon_j E_{d,j}(t) 
\label{eq:app-driven-D-field}
\end{align}
where $V_{d,j}(t)$ is the voltage across the $j$-th capacitor branch when driving at $\omega_d$, $\mathrm{d}_j$ is the distance between the parallel plates of the $j$-th capacitor, $\varepsilon_j$ is the permittivity of the $j$-th capacitor. Notice that the direction of the positive voltage difference depends on the direction of the branch in the network.

Another important ingredient of the electric overlap expression is the description of normal modes of the circuit. Suppose the total energy of the $i$-th normal mode is $\mathcal{E}_i$, the electric and displacement field mode profile of the real circuit are respectively $\mathbf{E}_i(\mathbf{r})$ and $\mathbf{D}_i(\mathbf{r}) = \varepsilon(\mathbf{r})\mathbf{E}_i(\mathbf{r})$. Likewise, for the same mode under the same total energy, the electric and displacement field profiles in the $j$-th capacitor of the mapped lumped-element circuit are given by
\begin{align}
E_{i,j} 
&= 
\frac{V_{i,j}}
{d_j} 
\label{eq:app-em-E-field}\\
D_{i,j}
&= 
\varepsilon_j E_{i,j}
\label{eq:app-em-D-field}
\end{align}
where $V_{i,j}$ is the maximum voltage drop of the $i$-th normal mode on the $j$-th capacitor. The direction of positive voltage difference for $V_{i,j}$ is the same as that for $V_{d,j}$. Mode orthogonality condition dictates that (see Ref. \cite{Glauber1991} and discussions in Appendix \ref{app:orthogonality_modes})
\begin{align}
   \int_V \mathbf{D}_i(\mathbf{r})\cdot\mathbf{E}_m(\mathbf{r}) dV = 
   \sum_j D_{i,j}E_{m,j}\mathrm{Vol}_j = 
   2\mathcal{E}_i\delta_{im}.
\end{align} 
For later convenience, we define a normalized electric field mode profile for $i$-th mode, 
$\mathbf{f}_i(\mathbf{r}) = \mathbf{E}_i(\mathbf{r})/\sqrt{2\mathcal{E}_i}$, and its counterpart in the lumped-element circuit network, $f_{i,j} = E_{i,j}/\sqrt{2 \mathcal E_i}$, such that
\begin{equation}
    \int_V \varepsilon(\mathbf{r})\mathbf{f}_i(\mathbf{r})\cdot\mathbf{f}_m(\mathbf{r}) dV 
    = \sum_{j} \varepsilon_j f_{i,j} f_{m,j}
    = \delta_{im}.
\end{equation}
With the driven displacement field and the static electric field mode function, we define the electric overlap function as
\begin{align}
O_i(t)&\equiv\int_V \mathbf{D}_d (\mathbf{r}, t)\cdot \mathbf{f}_i(\mathbf{r}) dV\nonumber\\
&= \sum_j D_{d,j}(t) f_{i,j} \mathrm{Vol}_{j} \nonumber\\
&= \sum_j \varepsilon_j\frac{V_{d,j}(t)}{d_j} \frac{V_{i,j}}{\sqrt{2\mathcal{E}_i}\text{d}_j} {S}_j {d}_j\nonumber\\
&=\sum_j \frac{C_j}{\sqrt{2\mathcal{E}_i}} V_{d,j}(t) V_{i,j}, 
\label{eq:app-overlap-discretized}
\end{align}
where $S_j$ and $\mathrm{Vol}_j=S_j d_j$ are the area and volume of the $j$-th capacitor, respectively, and in the last equality we identify $\varepsilon_j S_j/d_j$ as the capacitance of capacitor $j$, $C_j$. By construction, the lumped-element circuit faithfully represents the real circuit; therefore, the overlap function for the two circuits given by the first two lines is equivalent. In words, we see that the electric field overlap is the voltage overlap between the driven and undriven circuit, summing over all the capacitors. 

To further expand the above expression for the overlap $O_i(t)$ into a workable form, it is necessary to consider the Lagrangian description of the lumped-element network. The Lagrangian of the undriven circuit can be expressed as a sum of harmonic oscillators over all modes $i$
\begin{equation}
    \mathcal{L} = \sum_i \mathcal{L}_i = \sum_i \frac{1}{2} C_{\Sigma_i} \dot{\Phi}_i^2 - \frac{1}{2 L_{\Sigma_i}} \Phi_i^2, \label{eq:static_lagrangian}
\end{equation}
where $C_{\Sigma_i} = \sum_j C_j |\alpha_{ji}|^2$ and $L_{\Sigma_i}^{-1} = \sum_k L_k^{-1} |\beta_{ki}|^2$ are the total capacitance and inductance of the $i$-th mode, and $\alpha_{ji}$ ($\beta_{ki}$) is the participation factor of the $i$-th normal mode in the $j$-th capacitor ($k$-th inductor). In the presence of external drives, the Lagrangian of the driven circuit can be generally expressed as
\begin{align}
    \mathcal{L}_d &= \frac{1}{2}\sum\limits_j {C_j {{\left( {\sum\limits_{i=1}^N {{\alpha_{ji}}{{\dot \Phi }_{d,i}}} } + \mathcal{V}_j(t) \right)}^2}} \nonumber\\
    &- \frac{1}{2}\sum\limits_k \frac{\phi_0^2}{L_k}{{{\left( {\frac{\sum\limits_{i=1}^N{\beta_{ki}{\Phi }_{d,i}}}{\phi_0} } + \mathcal{F}_k(t) \right)}^2}},
    \label{eq:app-overlap-driven-Lagrangian}
\end{align}
where $\mathcal{V}_j(t)$ and $\mathcal{F}_k(t)$ are the external voltage and phase modulation of the corresponding circuit element. We explicitly label the flux operator with the subscript $d$ to differentiate it from the static circuit operator in Eq.~\eqref{eq:static_lagrangian}.
Now, we may perform the following gauge transformation,
\begin{equation}
    \Phi_{d,i} \rightarrow \Phi_{d,i} - \phi_0 \sum_k \frac{L_{\Sigma_i}}{L_k} \beta_{ki} \mathcal{F}_k(t),
\end{equation}
which shifts the normal mode fluxes. This effectively eliminates the linear phase modulations from the Lagrangian; however, at the level of the Josephson nonlinearity, a residual phase modulation may remain (see Appendix~\ref{app:add_nonlinearity}). As we will see in a moment, this gauge choice allows us to express the electric overlap function solely in terms of capacitive terms, eliminating the need to compute the spatial integral of $\mathbf{B}_d\cdot \mathbf{H}_i$, as well as the electric current overlap between the driven and undriven circuits summed over all the Josephson junctions/kinetic inductances. Using the orthogonality condition of normal modes
\begin{align}
    \sum_j C_j \alpha_{ji} \alpha_{jm} &= C_{\Sigma_i} \delta_{im},\label{eq:app-overlap-orthonormality-C}\\
    \sum_k L_k^{-1} \elementdepsymbol{\beta}{J}{ki} \beta_{km} &= L_{\Sigma_i}^{-1} \delta_{im}, \label{eq:app-overlap-orthonormality-L}
\end{align}
we can rewrite the Lagrangian of the driven circuit as 
\begin{align}
\begin{split}
    \mathcal{L}_d &= \sum_i \mathcal{L}_{d,i} \\&= \frac{1}{2}\sum\limits_j {C_j {{\left( {\sum\limits_{i=1}^N {{\alpha_{ji}}{{\dot \Phi }_{d,i}}} } + \mathit{V}_j(t) \right)}^2}} \\
    &- \frac{1}{2}\sum\limits_k \frac{\phi_0^2}{L_k}{{{\left( {\frac{\sum\limits_{i=1}^N{\beta_{ki}{\Phi }_{d,i}}}{\phi_0} } + \mathit{F}_k(t) \right)}^2}} 
    \\&= \sum_i \left(\frac{1}{2} C_{\Sigma_i} \dot{\Phi}_{d,i}^2 - \frac{\Phi_{d,i}^2}{2 L_{\Sigma_i}}
    + \sum_j C_j \alpha_{ji} V_j(t) \dot{\Phi}_{d,i}\right),
    \label{eq:app-overlap-driven-L}
\end{split}
\end{align}
where $V_j(t) = \mathcal{V}_j(t) - \phi_0 \sum_i \alpha_{ji} \sum_k \frac{L_{\Sigma_i}}{L_k} \beta_{ki} \mathcal{\dot{F}}_{k}(t)$, and $F_k(t)=\mathcal{F}_k(t)-\sum_{i}\beta_{ki}\sum_h \frac{L_{\Sigma_i}}{L_h} \beta_{hi} \mathcal{F}_h(t)$.

We now obtain an expression for the effective charge modulation strength $g_i(t)$ of the $i$-th mode. Performing a Legendre transformation on the Lagrangian of the driven $i$-th mode $\mathcal{L}_{d,i}$, we have
\begin{align}
\begin{split}
\mathcal{H}_{d,i} &= \dot{\Phi}_{d,i}\frac{\partial \mathcal{L}_{d,i}}{\partial \dot{\Phi}_{d,i}}-\mathcal{L}_{d,i}\\
&= \frac{Q_{d,i}^2}{2C_{\Sigma_i}}+\frac{\Phi_{d,i}^2}{2L_{\Sigma_i}} - \sum_j \frac{C_j}{C_{\Sigma_i}} \alpha_{ji} V_j(t) Q_{d,i}\\
&=\hbar\omega_i a_i^{\dagger} a_i + ig_i(t) (a_i^{\dagger}-a_i),
\end{split}
\label{eq:app-overlap-driven-H}
\end{align}
where 
\begin{align}
Q_{d,i} = \frac{\partial \mathcal{L}}{\partial \dot{\Phi}_{d,i}} &= C_{\Sigma_i} \dot{\Phi}_{d,i} + \sum_j C_j \alpha_{ji} V_j(t),\\
g_i(t) &= -\frac{Q_{\mathrm{zpf}_i}}{C_{\Sigma_i}} \sum_j C_j \alpha_{ji} V_j(t).\label{app:eq:g(t)}
\end{align}
Here, $\omega_i = 1/\sqrt{L_{\Sigma_i} C_{\Sigma_i}}$ and $Q_{\mathrm{zpf}_i} = \sqrt{\hbar \omega_i C_{\Sigma_i}/2}$ correspond to the frequency and zero-point charge fluctuation of the $i$-th normal mode, respectively. 

From Eq.~\eqref{eq:static_lagrangian} and Eq.~\eqref{eq:app-overlap-driven-L}, we relate the $j$-th capacitor voltage for the driven circuit and the static circuit, to the flux operators in the respective Lagrangian, through   \begin{align}
    V_{d,j}(t) 
    &= 
    \left\langle
    \sum^N_{i=1} \alpha_{ji} \dot{\Phi}_{d,i} + V_j(t)
    \right\rangle_{d},
    \\
    V_{i,j} &= \alpha_{ji} \braket{\dot{\Phi}_i}.
\end{align}
Here, $\braket{\cdot}_\text{d}$ represents taking the expectation value with respect to the steady state (displaced vacuum) of the driven system, and $\braket{\cdot}$ denotes the expectation value with respect to the coherent state of mode $i$ with energy $\mathcal E_i$. Plugging these in Eq.~\eqref{eq:app-overlap-discretized}, we obtain the electric overlap function as
\begin{align}
\begin{split}
    O_i(t) 
    &= \sum_j \frac{C_j}{\sqrt{2\mathcal{E}_i}} \left\langle\sum_{n=1}^N \alpha_{jn} \dot{\Phi}_{d,n} + V_j(t)\right\rangle_{d} \alpha_{ji} \braket{\dot{\Phi}_i}\\
    &= \left\langle C_{\Sigma_i} \dot{\Phi}_{d,i} + \sum_j C_j \alpha_{ji} V_j(t) \right\rangle_{d} \frac{\braket{\dot{\Phi}_i}}{\sqrt{2\mathcal{E}_i}}\\
    &= \left\langle \dot{\Phi}_{d,i} - Q_{\mathrm{zpf}_i}^{-1} g_i(t) \right\rangle_{d} \frac{\braket{Q_i}}{\sqrt{2\mathcal{E}_i}},
\end{split}
\label{eq:app-overlap-product}
\end{align}
where the second line made use of the orthogonality condition in Eq.~\eqref{eq:app-overlap-orthonormality-C}, and the third line made use of Eq.~\eqref{app:eq:g(t)}. The driven circuit expectation value can be found from the equation of motion derived from $\mathcal{L}_{d,i}$ in Eq.~\eqref{eq:app-overlap-driven-L},
\begin{equation}
    \ddot{\Phi}_{d,i} + \omega_i^2 \Phi_{d,i} - Q_{\mathrm{zpf}_i}^{-1}\dot{g}_i(t)=0,
    \label{eq:app-overlap-EOM}
\end{equation}
having solution
\begin{align}
    \left\langle \dot{\Phi}_{d,i} - Q_{\mathrm{zpf}_i}^{-1} g_i(t) \right\rangle_{d} = \frac{\omega_i^2}{\omega_d^2 - \omega_i^2} \frac{g_i(t)}{Q_{\mathrm{zpf}_i}}.
    \label{eq:app-overlap-EOM-solution}
\end{align}
The undriven circuit expectation value $\braket{Q_i}$ is obtained by taking mode $i$ in a coherent state $\ket{\alpha}$ such that
\begin{align}
    \braket{Q_i} &= i Q_{\mathrm{zpf}_i} \left( \braket{a^\dagger} - \braket{a} \right)\nonumber\\
    &= 2 Q_{\mathrm{zpf}_i} \Imag(\alpha)\nonumber\\
    &= 2 Q_{\mathrm{zpf}_i} \sqrt{\frac{\mathcal{E}_i}{\hbar \omega_i}},
    \label{eq:app-overlap-Qi-expect}
\end{align}
where in the last line we have used the fact that we are interested in computing the peak expectation value that occurs for Im$(\alpha) = |\alpha|$, and $\mathcal{E}_i = \hbar \omega_i |\alpha|^2$ is the energy of the $i$-th mode in the undriven circuit. As we are ultimately using classical electromagnetic simulations to obtain the electric overlap, the assumption of a coherent state is valid. Combining Eqs.~\eqref{eq:app-overlap-product}, \eqref{eq:app-overlap-EOM-solution}, and \eqref{eq:app-overlap-Qi-expect} we obtain Eq.~\eqref{eq:gi_overlap} of the main text
\begin{align}
    O_i(t) =\frac{\omega_i^2 }{\omega_d^2-\omega_i^2}\sqrt{\frac{2}{\hbar\omega_i}}g_i(t).
    \label{eq:app-overlap-in-terms-of-g}
\end{align}

This result can be generalized to incorporate weak dissipation from the lossy boundary conditions, which, in general, disturb the electric field profile of the normal modes and break the orthogonality condition between them. Thus, we limit ourselves to consider high-Q modes which are only weakly perturbed by the lossy boundaries, and treat other modes as the environment under the appropriate condition (see Appendix~\ref{app:mode_truncation}). In the circuit picture of Fig.~\ref{fig5}, the $i$-th mode of interest is represented by a single series $R_i$--$L_{\Sigma_i}$--$C_{\Sigma_i}$ branch, while the rest of the structure is modeled as a parallel black box that draws a current $I_\textrm{env}(t)$. This treatment is valid when mode frequency spacings and Q factors are large compared to the small dissipative couplings, so that the driven response of each kept mode is well captured by a single Foster branch, and the environment can be represented by the net current it pulls at the port node. The environment current reduces the device side voltage seen by the branches:
\begin{align}
    V'_\textrm{S}(t) = V_\textrm{S}(t) - Z_0 I_\textrm{env}(t).
\end{align}
In the overlap gauge, the drive modulation of mode $i$ is set by this available device voltage, $g_i(t) = Q_{\textrm{zpf}_i}(t) V'_\textrm{S}(t)$. After the $i$-th branch is connected, its own response current $I_i(t)$ produces an additional voltage drop $Z_0 I_i(t)$ across the port resistance $Z_0$. Therefore, the voltage across the $i$-th mode capacitor $C_{\Sigma_i}$ is
\begin{align}
    V_{C,i} = \left\langle \dot{\Phi}_{d,i} - Q_{\mathrm{zpf}_i}^{-1} g_i(t) + \kappa_i^\textrm{port}\Phi_{d,i}\right\rangle_{d}\,,
\end{align}
with $\kappa_i^\textrm{port} = Z_0/L_{\Sigma_i}$. Projecting this voltage onto the static mode profile exactly as in Eq.~\eqref{eq:app-overlap-product} yields the lossy overlap identity 
\begin{equation}
    O_i(t) = \left\langle \dot{\Phi}_{d,i} - Q_{\mathrm{zpf}_i}^{-1} g_i(t) + \kappa_i^\textrm{port} \Phi_{d,i} \right\rangle_{d} \frac{\braket{Q_i}}{\sqrt{2\mathcal{E}_i}}.
    \label{eq:app-overlap-product-lossy}
\end{equation}

The corresponding equation of motion for $\Phi_{d,i}$ is given by the Kirchhoff's law,
as
\begin{align}
    \frac{\Phi_{d,i}}{L_{\Sigma,i}} = C_{\Sigma_i}\frac{d}{dt}\left[V'_\textrm{S}(t)-(Z_0+R_i)\frac{\Phi_{d,i}}{L_{\Sigma,i}}-\dot{\Phi}_{d,i}\right].
\end{align}
Using $\omega_i^2 = (L_{\Sigma_i}C_{\Sigma_i})^{-1/2}$, $\kappa^\textrm{int} = R_i/L_{\Sigma_i}$, $\kappa_i^\textrm{tot} = \kappa_i^\textrm{int}+\kappa_i^\textrm{port}$, and $g_i(t) = Q_{\textrm{zpf}_i}(t) V'_\textrm{S}(t)$,
we obtain
\begin{equation}
    \ddot{\Phi}_{d,i} + (\omega_i^2 +(\kappa_i^\textrm{tot})^2/4)\Phi_{d,i} - Q_{\mathrm{zpf}_i}^{-1}\dot{g}_i(t) + \kappa_i^\textrm{tot} \dot{\Phi}_{d,i} = 0,
    \label{eq:app-overlap-EOM-lossy}
\end{equation}
where we renormalize $\omega_i^2 \to \omega_i^2 +(\kappa_i^\textrm{tot})^2/4$ such that the real part of the eigenmode frequency remains $\omega_i$.

Following the same procedure to obtain Eq.~(\ref{eq:app-overlap-in-terms-of-g}), we arrive at the weak dissipation version of the electric overlap function 
\begin{align}
    O_i(t) 
    &= \Real[\bar{O}_i e^{i\omega_d t}], \\
    \bar{O}_i 
    &= 
    \frac{\omega_i^2 + (\kappa_i^\textrm{tot})^2/4+i\omega_d\kappa_i^\textrm{int}}{\omega_d^2 - \omega_i^2 -(\kappa_i^\textrm{tot})^2/4 - i\kappa_i^\textrm{tot} \omega_d } \sqrt{\frac{2}{\hbar \omega_i}} \bar{g}_i(\omega_d).
    \label{eq:gbar_fieldcentric}
\end{align}
Under the assumption of no intrinsic loss, i.e. $\kappa_i^\textrm{int} = 0$, this formula reduces to Eq.~\eqref{eq:gi_overlap} in the main text.

\subsection{Field-centric derivation of the overlap method}
\label{app:overlap_EM}
In this section, we present an alternative field-centric derivation of Eq.~\eqref{eq:gi_overlap} without reference to lumped-circuit elements, instead modeling the device as an electromagnetic cavity with perfectly conducting walls and an interior described by the inhomogeneous dielectric function $\varepsilon(\mathbf{r})$. For simplicity, we will assume this dielectric function to be isotropic, though we note that extension of our approach to the case of an anisotropic (but diagonal) dielectric is straightforward, with the final relation Eq.~\eqref{eq:gbar_fieldcentric} unchanged between the two. We comment on this extension after the main steps of our derivation below Eq~\eqref{eq:gauss_law_no_charge}.

The modes of a dielectric cavity correspond to the independent solutions to the sourceless wave equation,
\begin{equation}
\nabla\times\nabla\times\mathbf{A}(\mathbf{r},t) +\mu_0\varepsilon(\mathbf{r}) \ddot{\mathbf{A}}(\mathbf{r},t) = 0,
\label{eq:app-waveeqn}
\end{equation}
where $\mathbf{A}(\mathbf{r},t)$ is the vector potential and $\mu_0$ is permeability of free space. In writing the above equation, we have adopted the generalized Coulomb gauge $\nabla\cdot(\varepsilon(\mathbf{r})\mathbf{A}(\mathbf{r},t)) = 0$, which allows us to choose the scalar potential to vanish in the absence of free charges and external drives \cite{Glauber1991}. Therefore, the electric and magnetic fields can be expressed entirely in terms of the vector potential as 
\begin{align}
    \mathbf{E}(\mathbf{r},t) &= -\dot{\mathbf{A}}(\mathbf{r},t), \\
    \mathbf{B}(\mathbf{r},t) &= \nabla\times\mathbf{A}(\mathbf{r},t).
\end{align}

We now search for separable solutions to Eq.~\eqref{eq:app-waveeqn} of the form $\mathbf{A}_i(\mathbf{r},t) = -q_i(t)\mathbf{f}_i(\mathbf{r})$, where $q_i(t)$ is a time-dependent amplitude and $\mathbf{f}_i(\mathbf{r})$ the corresponding mode function that characterizes the spatial profile of mode $i$. Inserting this form into Eq.~\eqref{eq:app-waveeqn}, we find that $q_i(t)$ and $\mathbf{f}_i(\mathbf{r})$ must independently obey
\begin{align}
     \ddot{q}_i +\omega_i^2q_i &=0, \label{eq:app-separated_eqs-a}\\
     \nabla\times\nabla\times \mathbf{f}_i(\mathbf{r}) - \omega_i^2\mu_0\varepsilon(\mathbf{r}) \mathbf{f}_i(\mathbf{r}) &= 0, \label{eq:app-separated_eqs-b}
\end{align}
where $\omega_i^2$ is a (positive) constant resulting from the separation of variables. Eq.~\eqref{eq:app-separated_eqs-a} is the equation of motion for the dynamical coordinate $q_i$, equivalent to that of a harmonic oscillator with natural frequency $\omega_i$. Furthermore, we see that $\mathbf{f}_i(\mathbf{r})$ must obey the generalized Helmholtz equation, Eq.~\eqref{eq:app-separated_eqs-b}.

\subsubsection{Orthogonality between modes}
\label{app:orthogonality_modes}
We now derive the relevant orthogonality condition between the mode functions $\mathbf{f}_i(\mathbf{r})$, following the arguments of Ref.~\cite{Glauber1991}. To simplify, let us define the rescaled mode function $\mathbf{g}_i(\mathbf{r}) = \sqrt{\varepsilon(\mathbf{r})/\varepsilon_0}\mathbf{f}_i(\mathbf{r})$, such that
\begin{equation}
\left[\sqrt{\frac{\varepsilon_0}{\varepsilon(\mathbf{r})}}\nabla\times\nabla\times\sqrt{\frac{\varepsilon_0}{\varepsilon(\mathbf{r})}}\right]\mathbf{g}_i(\mathbf{r}) = -\frac{\omega_i^2}{c^2}\mathbf{g}_i(\mathbf{r}).
\end{equation}
Thus, $\mathbf{g}_i(\mathbf{r})$ is an eigenfunction of the operator in brackets with corresponding eigenvalue $-\omega_i^2/c^2$. Noting that the operator in brackets is Hermitian, this implies the orthogonality condition
\begin{equation}
\int_V d^3r\, \mathbf{g}_i(\mathbf{r})\cdot\mathbf{g}_j(\mathbf{r}) = \delta_{ij}/\varepsilon_0,
\label{eq:gorthogonality}
\end{equation}
where we have chosen the normalization $1/\varepsilon_0$ without loss of generality, and have denoted the space enclosed by the cavity by $V$. Rewriting Eq.~\eqref{eq:gorthogonality} in terms of the mode functions $\mathbf{f}_i(\mathbf{r})$, we find that the mode functions $\{\mathbf{f}_i(\mathbf{r})\}$ satisfy the orthogonality condition
\begin{equation}
\int_V d^3r \varepsilon(\mathbf{r})\mathbf{f}_i(\mathbf{r})\cdot\mathbf{f}_j(\mathbf{r}) = \delta_{ij}.
\label{eq:app-normalization}
\end{equation}
As a side remark, while we have here chosen a particularly simple normalization condition, more generally, this can be freely chosen at the expense of rescaling the time-dependent coordinates $q_i(t)$. At the level of the mode quantization and/or dynamics, this is analogous to rescaling the mass of an oscillator or, equivalently, performing a single-mode squeezing transformation.

\subsubsection{Computing the overlap integral}
Next, we leverage this formalism to analyze the overlap integral in Eq.~\eqref{eq:electric_overlap}. Using the fact that the mode functions $\{\mathbf{f}_i(\mathbf{r})\}$ form a complete orthonormal basis set over the space of vector functions $\mathbf{F}(\mathbf{r})$ that obey $\nabla\cdot(\varepsilon~(\mathbf{r})\mathbf{F}(\mathbf{r}))=0$, the driven vector potential can be entirely expressed in terms of the mode decomposition
\begin{equation}
\mathbf{A}_d(\mathbf{r},t) =  -\sum_m q_m(t)\mathbf{f}_m(\mathbf{r}).
\end{equation}

The core of the overlap method is to compute the quantity
\begin{equation}
    O_i\left( t\right)=\int_V d^3r\, \varepsilon(\mathbf{r})\mathbf{E}_d(\mathbf{r},t) \cdot \mathbf{f}_i(\mathbf{r}),
\end{equation}
where $\mathbf{E}_d(\mathbf{r},t)$ is the total electric field comprising both an external voltage source (the drive) and the response of the modes. Taking into account the input voltage at the port, we can write the total driven field as 
\begin{align}
        \mathbf{E}_d(\mathbf{r},t) &= -\dot{\mathbf{A}}_d(\mathbf{r},t) - \nabla V_{\textrm{ext}}(\mathbf{r},t) \nonumber\\
        &= \sum_m \dot{q}_m(t)\mathbf{f}_m(\mathbf{r}) - \nabla V_{\textrm{ext}}(\mathbf{r},t).
\end{align}
Using this expansion, the overlap becomes
\begin{align}
    O_i(t) =& \int_V d^3r\, \varepsilon(\mathbf{r})\mathbf{E}_d(\mathbf{r},t) \cdot \mathbf{f}_i(\mathbf{r}) \nonumber\\
    =& \sum_m \dot{q}_m(t) \int_V d^3 r \varepsilon(\mathbf{r})\mathbf{f}_m(\mathbf{r})\cdot\mathbf{f}_i(\mathbf{r}) \nonumber\\ &- \int_V d^3r \varepsilon(\mathbf{r})\nabla V_{\textrm{ext}}(\mathbf{r},t)\cdot\mathbf{f}_i(\mathbf{r}).
\label{eq:app-overlap_steady}
\end{align}
The first integral corresponds to the generalized orthogonality condition in Eq.~\eqref{eq:app-normalization}, collapsing the sum over $m$ to the case $m=i$. The second integral can be simplified by applying integration by parts; only the boundary term contributes as $\nabla\cdot(\varepsilon(\mathbf{r})\mathbf{f}_i(\mathbf{r})) = 0$ due to the modified Coulomb gauge condition. Altogether, we find
\begin{equation}
O_i(t) = \dot{q}_i(t) - \alpha_i(t),
\end{equation}
where we adopted a shorthand
\begin{equation}
    \alpha_i(t) = \oint_{\partial V} d\mathbf{a}\cdot \varepsilon(\mathbf{r}) V_{\textrm{ext}}(\mathbf{r},t) \mathbf{f}_i(\mathbf{r}).
    \label{eq:app-boundary}
\end{equation}

\subsubsection{Deriving the field-centric Lagrangian}
Next, we aim to express the $\dot{q}_i(t)$ in terms of system parameters such as the mode frequency $\omega_i$ and drive strength. In the following, we closely follow a simplified variant of the formalism developed in Refs.~\cite{smith2020active, smith2025} in the context of interacting dielectric cavities. We begin by writing the electromagnetic Lagrangian of the driven system,

\begin{equation}
\mathcal{L}_d = \frac{1}{2}\int_V d^3 r \,\mathbf{D}_d\cdot\mathbf{E}_d - \frac{1}{2}\int_V d^3 r\, \mathbf{H}_d\cdot \mathbf{B}_d.
\end{equation}
As above, we simplify to the scenario of non-magnetic media (i.e., $\mathbf{H}_d = \mathbf{B}_d/\mu_0$). Re-expressing the Lagrangian in terms of our mode expansion, we find 
\begin{align}
\mathcal{L}_d
&= \frac{1}{2} \int_V d^3 r \,
\Bigl[
  \varepsilon(\mathbf r)\bigl(-\dot{\mathbf A}_d - \nabla V_{\textrm{ext}}(\mathbf r,t)\bigr)^2
  \nonumber\\
&\qquad\qquad
  - \varepsilon_0 c^2 (\nabla \times \mathbf A_d)^2
\Bigr] \nonumber
\\
&= \frac{1}{2}\sum_{ij} \bigl(\dot q_i \dot q_j I_1 - q_i q_j I_2\bigr)
   - \sum_i \dot q_i I_3 + I_4 ,
\label{eq:app-LagrangianEM}
\end{align}
where $I_1$--$I_4$ denote four distinct integrals,
\begin{align}
    I_1 &= \int_V d^3r\, \varepsilon(\mathbf{r})\mathbf{f}_i(\mathbf{r})\cdot\mathbf{f}_j(\mathbf{r}),\\
    I_2 &= \varepsilon_0c^2\int_V d^3r \, (\nabla\times\mathbf{f}_i(\mathbf{r})\cdot(\nabla\times\mathbf{f}_j(\mathbf{r})),\\
    I_3 & = \int_V d^3r\, \varepsilon(\mathbf{r})\mathbf{f}_i(\mathbf{r})\cdot\nabla V_{\textrm{ext}}(\mathbf{r},t),\\
    I_4 &= \frac{1}{2}\int_V d^3r \varepsilon(\mathbf{r})\nabla V_{\textrm{ext}}(\mathbf{r},t)\cdot \nabla V_{\textrm{ext}}(\mathbf{r},t).
\end{align}

We now evaluate the integrals in sequence. Via the orthogonality condition Eq.~\eqref{eq:app-normalization}, the first
integral simplifies as,
\begin{equation}
\begin{split}
I_1 = \delta_{ij}.
\end{split}
\end{equation}
Likewise, the second integral can be simplified using integration by parts:
\begin{align}
I_2 &= \varepsilon_0c^2\int_V d^3r\, \mathbf{f}_i(\mathbf{r})\cdot(\nabla\times\nabla\times \mathbf{f}_j(\mathbf{r})) \nonumber\\ &- \oint_{\partial V} d\mathbf{a}\cdot(\mathbf{f}_i(\mathbf{r})\times\nabla\times \mathbf{f}_j(\mathbf{r})) \nonumber\\
&=\varepsilon_0\mu_0c^2\omega_i^2\int_V d^3r\,\varepsilon(\mathbf{r})\mathbf{f}_i(\mathbf{r})\cdot \mathbf{f}_j(\mathbf{r}) \nonumber\\
&=\omega_i^2\delta_{ij},
\end{align}
where we have applied the generalized Helmholtz equation in Eq.~\eqref{eq:app-separated_eqs-b} and again used mode orthogonality. Note that the boundary term arising from integration by parts vanishes due to the boundary condition at the perfect conductor interface $\hat{\mathbf{n}}\times\mathbf{E}|_{\partial V} = 0$.

The third integral can also be simplified using integration by parts:
\begin{equation}
    \begin{split}
        I_3 = \oint_{\partial V} d\mathbf{a}\cdot \varepsilon(\mathbf{r}) V_{\textrm{ext}}(\mathbf{r},t) \mathbf{f}_i(\mathbf{r}) 
        = \alpha_i(t),
    \end{split}
\end{equation}
where the bulk term vanishes due to the generalized Coulomb gauge condition $\nabla\cdot(\varepsilon(\mathbf{r})\mathbf{f}_i(\mathbf{r}))=0$ and $\alpha_i(t)$ is defined in Eq.~\eqref{eq:app-boundary}.

The fourth and final term of Eq.~\eqref{eq:app-LagrangianEM} contributes only at the boundary. Again, using integration by parts,
\begin{equation}
    \begin{split}
       I_4 =\frac{1}{2}\oint_{\partial V} d\mathbf{a}\cdot V_{\textrm{ext}}(\mathbf{r},t)\nabla V_{\textrm{ext}}(\mathbf{r},t),
   \end{split}
\end{equation}
where the bulk term vanishes due to Gauss's law in the absence of free charge,
\begin{equation}
\nabla\cdot(\varepsilon(\mathbf{r})\nabla V_{\textrm{ext}}(\mathbf{r},t))=0.
\label{eq:gauss_law_no_charge}
\end{equation}
Because $V_{\textrm{ext}}(\mathbf{r},t)$ is not a dynamical degree of freedom, the nonvanishing boundary term contributes an energy offset which we can discard.

Piecing all simplified terms together, the driven system Lagrangian can be written as
\begin{equation}
\mathcal{L}_d = \sum_i \left[\frac{1}{2}\dot{q}_i^2 - \frac{1}{2} \omega_i^2q_i^2\right] - \sum_i \alpha_i(t)\dot{q}_i.
\end{equation}
Thus, we see that the driven Lagrangian is exactly equivalent to a set of forced, independent harmonic oscillators with natural frequency $\omega_i$.

For completeness, we also derive the corresponding Hamiltonian. Solving for the momentum conjugate to $q_i$, we find
\begin{equation}
 p_i = \frac{\partial \mathcal{L}}{\partial \dot{q}_i} = \dot{q}_i - \alpha_i(t)
\end{equation}
and, correspondingly, the Hamiltonian
\begin{equation}
\mathcal{H}_d = \sum_i \left[\frac{1}{2}p_i^2 + \frac{1}{2}\omega_i^2 q_i^2\right] + \sum_i \alpha_i(t) p_i,
\end{equation}
where we have dropped a non-dynamical time-dependent factor proportional to $\alpha_i(t)^2$. We can further express the Hamiltonian in terms of raising and lowering operators, defined via the canonical relations,
\begin{align}
    q_i &= \sqrt{\frac{\hbar}{2\omega_i}}(a^\dagger+a) \\
    p_i &=i\sqrt{\frac{\hbar\omega_i}{2}}(a^\dagger-a).
\end{align}
Plugging this into the Hamiltonian,
\begin{align}
\mathcal{H}_d &= \sum_i \hbar \omega_i a_i^\dagger a_i + i\sum_i \alpha_i(t) \sqrt{\frac{\hbar\omega_i}{2}}(a^\dagger-a) \nonumber\\
& = \sum_i \hbar \omega_i a_i^\dagger a_i +i\sum_i g_i(t)(a^\dagger - a), 
\label{eq:app-Hamiltonian_aadag}
\end{align}
where we have dropped the vacuum energy and have defined $g_i(t) = \alpha_i(t)\sqrt{\hbar\omega_i/2}$, adopting the notation of Eq.~\eqref{eq:overlap-ov-Hamiltonian} for the drive term.

Recalling that the overlap for the $i$-th mode is given by 
\begin{equation}
O_i(t) = \dot{q}_i - \alpha_i(t) = p_i,
\label{eq:overlap_coord_mom}
\end{equation}
we see that computation of the overlap corresponds to the determination of the steady-state solution for $\dot{q}_i$ or, equivalently, $p_i$. Below, we derive the steady-state solution $\dot{q}_i$ for the situation where the system is driven at a single frequency $\omega_d$, i.e., $V_{\textrm{ext}}(\mathbf{r},t)=
\Real[\bar{V}_{\textrm{ext}}(\mathbf{r})e^{i\omega_d t}]$. This, in turn, enables us to determine a relationship between $O_i(t)$ and the drive strength $g_i(t)$.

Before proceeding, we reiterate that the derivations above are naturally extendable to the case of a diagonal anisotropic dielectric. The only role played by $\varepsilon(\mathbf r)$ is to define the weighted metric $\int_V d^3r\,\mathbf f_i\cdot\varepsilon\cdot\mathbf f_j$ on the mode functions, and the rescaling $\mathbf g_i=\sqrt{\varepsilon/\varepsilon_0}\,\mathbf f_i$ used above is precisely the transformation that flattens this metric to the trivial inner product $\int_V d^3r\,\mathbf g_i\cdot\mathbf g_j$, rendering the eigenvalue operator Hermitian. For a dielectric tensor $\boldsymbol{\varepsilon}(\mathbf r)$ the same flattening is accomplished by the matrix square root, $\mathbf g_i=\boldsymbol{\varepsilon}^{1/2}\mathbf f_i/\sqrt{\varepsilon_0}$, which exists and is symmetric whenever $\boldsymbol{\varepsilon}$ is positive definite; in the diagonal case this is simply a component-wise rescaling $f_{i,a}\to\sqrt{\varepsilon_a}\, f_{i,a}$. With $\varepsilon$ promoted to $\boldsymbol{\varepsilon}$ and the gauge condition to $\nabla\cdot(\boldsymbol{\varepsilon}\,\mathbf f_i)=0$, the simplifications of $I_1$--$I_4$ then carry through unchanged, leaving Eq.~\eqref{eq:gbar_fieldcentric} the same as in the isotropic case.

\subsubsection{Introducing loss and deriving Eq.~\eqref{eq:gi_overlap}}
From the Lagrangian, the equation of motion of the $i$-th mode is
\begin{equation}
\ddot{q}_i + \kappa^{\textrm{tot}}_i \dot{q}_i + (\omega_i^2 + \kappa_i^2/4) q_i = \dot{\alpha}_i(t),
\label{eq:eom_w_damping}
\end{equation}
where we have inserted a damping term proportional to $\kappa^{\textrm{tot}}_i = \kappa^{\textrm{int}}_i + \kappa^{\textrm{port}}_i$ that captures both the intrinsic loss of the high-Q modes ($\kappa^{\textrm{int}}_i$) as well as that inherited from the 50 $\Omega$ resistor at the port ($\kappa^{\textrm{port}}_i$). Likewise, the expression for the overlap $O_i(t)$ is modified as
\begin{equation}
    O_i(t) = \dot{q}_i - \alpha_i(t) + \kappa_i^{\textrm{port}} q_i,
\end{equation}
where the final term accounts for the voltage drop across the resistor at the port, effectively reducing the drive strength that mode $m$ is subject to; see discussion below Eq.~\eqref{eq:app-overlap-product-lossy}.

We note that there will be additional perturbing effects due to the presence of this resistor, such as a weak induced coupling between high-Q modes and a modification of the external voltage ``seen'' by the high-Q modes. Here, we consider sufficiently high-Q modes such that these effects are negligible. We note, however, that this is not the case for low-Q modes \cite{Solgun2014}, and their inclusion is an interesting direction for future work. In the present field-centric setting, this would involve generalizing the current approach to incorporate quasi-normal mode theory for lossy resonators \cite{kristensen2020modeling, sauvan2022normalization}. However, such a treatment is beyond the scope of this work, and we instead treat the perturbing effects of loss in a phenomenological manner consistent with numerical simulations.

We assume the case of a harmonic external potential $V_{\textrm{ext}}(\mathbf{r},t)=\Real[\bar{V}_{\textrm{ext}}(\mathbf{r})e^{i\omega_d t}]$. From Eq.~\eqref{eq:app-boundary}, this in turn implies $\alpha_i(t)=\Real[\bar{\alpha}_i e^{i\omega_d t}]$ and $g_i(t)=\Real[\bar{g}_i e^{i\omega_d t}]$. The steady-state solution to Eq.~\eqref{eq:eom_w_damping} can then be expressed as $q_i(t)=\Real[\bar{q}_i(\omega_d)e^{i\omega_d t}]$ where
\begin{equation}
 \bar{q}_i(\omega_d) = \frac{i\omega_d}
 {-\omega_d^2 + \omega_i^2 +
 (\kappa_i^{\textrm{tot}})^2/4
 +i\omega_d \kappa_i^{\textrm{tot}}}\bar{\alpha}_i.
\end{equation}
Expressing the overlap in Eq.~\eqref{eq:overlap_coord_mom} in the frequency domain, i.e., $O_i(t) = \Real[\bar{O}_i(\omega_d) e^{i\omega_d t}]$, we find
\begin{equation}
\begin{split}
\bar{O}_i(\omega_d) &= (i\omega_d +\kappa_i^{\textrm{port}})\bar{q}_i - \bar{\alpha}_i  \\
& = \left[
\frac{-\omega_d^2 + i\omega_d\kappa_i^{\textrm{port}}}
{-\omega_d^2 + \omega_i^2 +
 (\kappa_i^{\textrm{tot}})^2/4
 +i\omega_d \kappa_i^{\textrm{tot}}}
- 1
\right]\bar{\alpha}_i \\
& = \left[
\frac{\omega_i^2 + (\kappa_i^{\textrm{tot}})^2/4 +i\omega_d\kappa_i^{\textrm{int}}}
{\omega_d^2 - \omega_i^2-(\kappa_i^{\textrm{tot}})^2/4 -i\omega_d \kappa_i^{\textrm{tot}}}
\right]\bar{\alpha}_i. 
\end{split}
\end{equation}
Casting this relation in terms of the drive strength $g_i$, we arrive at the desired expression:
\begin{align}
    \bar{g}_i 
    &= \left[
    \frac{\omega_d^2 - \omega_i^2-(\kappa_i^{\textrm{tot}})^2/4 -i\omega_d \kappa_i^{\textrm{tot}}}
    {\omega_i^2 + (\kappa_i^{\textrm{tot}})^2/4 +i\omega_d\kappa_i^{\textrm{int}}}
    \right]\nonumber\\
    &\times
    \sqrt{\frac{\hbar\omega_i}{2}}\bar{O}_i(\omega_d),
\end{align}
which reduces to Eq.~\eqref{eq:gi_overlap} of the main text when $\kappa_i^\textrm{int}=0$. In practice, $\kappa_i^{\textrm{port}}$ can be obtained by including the port load in the microwave simulation and extracting the corresponding external quality factor or linewidth, or, in the weak-loss regime, from a participation estimate of the mode coupling to the port load~\cite{Minev2021}. The intrinsic linewidth $\kappa_i^{\textrm{int}}$ can be obtained from material-loss models in the electromagnetic simulation, or participation estimates using independently characterized loss tangents or surface losses. When both mechanisms are present, the two contributions can be separated by evaluating the corresponding loss channels independently.

\subsection{Adding in the Josephson nonlinearity}
\label{app:add_nonlinearity}
Both preceding derivations of the overlap method were performed for a linearized circuit without explicit Josephson junctions. By the earlier argument that the drive parameters are independent of the Josephson nonlinearity (see Appendix~\ref{sec:app_describing_circuits}), the effective charge modulation $g_n(t)$ and the junction phase modulation $\elementdepsymbol{F}{J}{k}(t)$ obtained from the linear circuit are unaffected when the lumped inductors are replaced by Josephson junctions. While all phase modulations are eliminated on the quadratic order, they are preserved for the Josephson nonlinearities when they are reintroduced into the Hamiltonian, 
\begin{align}
\label{app:eq:overlap-Hamiltonian}
    \mathcal{H}_{\text{ov}} &= \sum_{i=1}^N \hbar \omega_i a_i^\dagger a_i + i \sum_{i=1}^N g_i(t) (a_i^\dagger - a_i) \nonumber \\ 
    &- \sum_{k=1}^M \elementdepsymbol{E}{J}{k} \cos_\mathrm{nl} \left( \sum_{i=1}^N \elementdepsymbol{\beta}{J}{ki} (a_i + a_i^\dagger) +  \elementdepsymbol{F}{J}{k} (t) \right).
\end{align}
Here, we still assume this Hamiltonian model includes all modes in the circuit, and $\elementdepsymbol{F}{J}{k}(t)$ is given by Eq.~\eqref{app:eq:H_ov}. Under finite mode truncation, extra junction phase contributed by the truncated modes needs to be added to the phase modulation parameters, naturally giving rise to the residual junction displacements $A_{\textrm{res},k}$ in Eq.~\eqref{eq:overlap-Fk} of the main text.

\subsection{Simulation details for electric-field overlap calculations}

The electric-field overlap calculation combines eigenmode and driven-field solutions of the same linearized circuit. 
In this appendix, we describe our HFSS implementation of the overlap integral, and discuss the associated computational cost. A step-by-step HFSS workflow is provided in Ref.~\cite{luDatasetSystematicConstruction2026}.

In our HFSS workflow, we first perform an eigenmode simulation to obtain the electric-field mode profiles of the linearized structure. We denote the normalized electric-field profile of mode $i$ by $\mathbf{f}_i(\mathbf{r})$, with the normalization chosen according to Eq.~\eqref{eq:normalization_normal_mode}.

The driven displacement field is then obtained from a driven frequency-domain simulation of the same structure. For a monochromatic drive at frequency $\omega_d$, the solver returns the complex driven electric-field phasor $\bar{\mathbf{E}}_{d}(\mathbf{r},\omega_d)$, from which the displacement-field phasor $\bar{\mathbf{D}}_{d}(\mathbf{r},\omega_d)$ is formed using the local permittivity. The time-domain field for the monochromatic drive is obtained by taking the real part of $\bar{\mathbf{D}}_{d}(\mathbf{r},\omega_d)e^{i\omega_d t}$.

In practice, the driven-field solution and the eigenmode profiles must be evaluated on a common integration representation. In our HFSS implementation, after completing the eigenmode simulation, we import the resulting mesh into the driven frequency-domain simulation so that the driven field and eigenmode profiles are evaluated on compatible finite-element grid points. After the driven simulation is completed, we import the eigenmode field profiles exported from the eigenmode simulation into the field calculator of the driven simulation, where the overlap integrals are evaluated. This avoids ambiguities associated with separately exporting and interpolating both fields between different meshes. More generally, neither using the same mesh nor evaluating the integral inside HFSS is a formal requirement of the method; the overlap integral can also be evaluated outside the solver, provided that the driven field and eigenmode profiles are extracted with sufficient accuracy and interpolated consistently over the same physical domain.

The computational cost of this procedure is mainly set by the underlying full-wave field solves and by the storage and processing of volumetric field data. For a fixed geometry and mesh, the eigenmode part has a cost comparable to a standard eigenmode-based EPR calculation. The overlap method additionally requires driven field solutions for the relevant drive ports and drive frequencies. Unlike impedance- or admittance-based extraction, which only retains scalar port-response data, the overlap method requires that the full three-dimensional driven-field and eigenmode profiles be stored or exported.

Once the field solutions are available, the additional overlap post-processing consists of volume integration over the simulation domain. This cost scales with the number of finite-element degrees of freedom and linearly with the number of retained modes, sampled drive frequencies, and independently excited drive ports. The number of finite-element degrees of freedom is set by the mesh and basis choice, and therefore depends on the required accuracy, convergence criteria, device geometry, material parameters, and frequency range. For larger devices, the cost will generally grow with the size and complexity of the meshed layout, as in other full-wave simulation workflows.

This additional computational and memory cost is the price for retaining spatially resolved field information. In return, the overlap method provides the mode-resolved decomposition of the driven response and yields the driven Hamiltonian for distributed circuits, information that is not directly available from an EPR calculation or from an impedance-only drive extraction. A possible future direction for improving scalability is to develop reduced-region or mode-targeted versions of the overlap analysis, where only selected field data relevant to the design question is retained. Establishing the accuracy and validity range of such reduced workflows is left for future work.

\section{Effect of the finite mode truncation on the circuit Hamiltonian model}
\label{app:mode_truncation}
Any real Josephson circuit device is inherently distributed and, in principle, supports an unbounded spectrum of electromagnetic modes. However, to keep the model tractable, we retain only a subset of these modes as explicit quantum degrees of freedom in our Hamiltonian models, treating the remainder as the environment (or ``filter" modes) that still shape the classical fields and noise seen by the system. Such a mode truncation must be done with care. In practice, a sufficient (but not necessary) criterion for keeping a mode explicit in the model is that it actively participates in the desired driven processes. Whether the remaining modes can be relegated to the environment is case-dependent, especially when they are driven strongly and near resonantly. Here, we briefly discuss the implications of such truncation for the three methods introduced in this work.

In the displaced-frame (DF) method, the junction phase displacement $A(t)$ is obtained from RF simulations treating the full linearized circuit. On the surface level, it is insensitive to the number of quantum modes retained in the Hamiltonian model. However, this is assuming that the drive does not induce nonlinear effects that renormalize the linear susceptibility underlying $A(t)$, nor does it create non-negligible, parasitic interactions between the filter modes and the system modes.  When these conditions are violated, the DF Hamiltonian that omits the involved filter modes will fail to describe the system dynamics accurately. Operationally, one can test whether the omission of a particular mode would violate these assumptions under the drive conditions (by, for example, a convergence test) to determine the appropriate truncation.

The irrotational-gauge (IG) method presumes lumped-element inductors and cannot explicitly represent driven distributed modes (e.g., CPW sections, cables, 3D-cavity resonances). Because of this limitation, it is inherently incapable of judging whether such a distributed mode can be treated as the environment or must be kept explicitly. In practice, the IG method is often applied to a lumped-element subsystem that is well separated from its distributed environment, a typical example being the low-energy subspace of a transmon (explicitly kept in the IG Hamiltonian) weakly coupled to drive lines (treated as the external linear network). Even in this favorable regime, truncation must be checked to avoid omitting drive-activated parametric processes between omitted and retained modes. Furthermore, omitting a mode can also bias the extracted effective phase modulation parameter, $\elementdepsymbol{F}{J}{k}(t)$. In the closed-JJ formulation, the drive is reconstructed from the junction phase displacement via the matrix $\mathbf{R}$ [Eq.~\eqref{eq:irg-r-matrix}], which is a function of retained modes' frequencies, damping rates, and their junction participation ratios. If a driven mode with appreciable junction participation is left out, the modal reconstruction is incomplete and the resulting $\elementdepsymbol{F}{J}{k}(t)$ is systematically biased. This sensitivity to truncation is avoided by the opened-JJ approach, where the modulation parameter is computed directly from the circuit impedance functions [Eq.~\eqref{eq:f-from-Z-open}], closely analogous to how $A_k(t)$ is obtained in the displaced-frame method. 

In the overlap method, retained modes receive explicit charge drives from the overlap calculation, which uses the full linear circuit information; therefore, these drive amplitudes are not affected by the Hamiltonian truncation. The remaining linear response appears as a residual phase at the junctions, obtained by subtracting the junction-phase contribution of the retained modes from the total junction phase (rather than summing over omitted modes). This preserves the full linear information of the junction response. Since filter modes enter only through their classical contribution to the junction phase displacement, the same truncation conditions as in the DF method apply.

\section{Numerically calculating decoherence rates}
\label{app:pvnr_details}
\subsection{Relating the drive parameter noise PSD to input voltage noise PSD}
\label{sec:app-Sxy-to-SVN}
Suppose $\delta s_i(t)$ and $\delta s_j(t)$ are noises in the circuit, generated by the same voltage noise source $\delta V(t)$ with power spectral density $S_{\delta V}(\omega)$. Their cross power spectral density is given by
\begin{equation}
    S_{\delta s_i\delta s_j}(\omega) = \int_{-\infty}^{\infty}{\left\langle \delta s_i(t) \delta s_j(0) \right\rangle e^{i \omega t} dt}. 
    \label{eq:app-Sij_time_correlation}
\end{equation}
The linear susceptibility function $\chi_{\delta s_i}$, obtainable from microwave simulations (Eq.~\eqref{eq:susceptibility}), connects $\delta s_{\delta_{i\left(j\right)}}$ to $\delta V(t)$ via
\begin{equation}
    \delta s_{i\left(j\right)}(t) = \int_{-\infty}^t{\chi_{i\left(j\right)}\left(t - t'\right)\delta V\left(t'\right)dt'},
    \label{eq:app-del_ij_convolution}
\end{equation}
same as Eq.~\eqref{eq:del_i_convolution} of the main text. Using Eqs.~\eqref{eq:app-Sij_time_correlation} and \eqref{eq:app-del_ij_convolution}, we can express the drive parameter noise PSD $S_{\delta s_i\delta s_j}(\omega)$ in terms of the input voltage noise PSD $S_{\delta V}(\omega)$, as
\begin{align}
    S_{\delta s_i\delta s_j}(\omega) &= \int_{-\infty}^{\infty}{dt\ e^{i \omega t}} \int_{-\infty}^t {dt'\ \chi_{s_i}\left(t - t' \right)} \nonumber\\
    \times \int_{-\infty}^0 &{dt''\ \chi_{s_j}\left(- t'' \right) {\left\langle \delta V\left(t'\right) \delta V\left(t''\right) \right\rangle}}\nonumber\\
    &= \frac{1}{2\pi} \int_{-\infty}^{\infty} d\omega' S_{\delta V}\left(\omega'\right) \int_{-\infty}^{\infty} e^{i\left(\omega - \omega'\right)t} dt\nonumber\\
    \times \int_0^\infty &\chi_{s_i}\left(\tau_1\right) e^{i \omega' \tau_1} d\tau_1 \int_0^\infty \chi_{s_j}\left(\tau_2\right) e^{-i \omega' \tau_2} d\tau_2 \nonumber\\ 
    & =\bar{\chi}_{s_i}(\omega) \bar{\chi}_{s_j}^*(\omega) S_{\delta V}(\omega).
    \label{eq:app-Sdidj_chi_SVN}
\end{align}
Here, we have used the Wiener–Khinchin theorem in the second equation~\cite{Clerk2010}. For the special case of $\delta s_i(t) = \delta s_j(t)$, we obtain Eq.~\eqref{eq:SVV_to_SFF} of the main text.

\subsection{Linear susceptibility relations between quantum noises}
\label{sec:app-quantum-noise}
When the system is subject to quantum voltage fluctuations, the classical perturbations $\delta s_i$ and $\delta V$ are promoted to quantum operators $\widehat{\delta s_i}$ and $\widehat{\delta V}$ acting on the Hilbert space of the bath. Within the PVNR framework, the response of $\delta s_i$ to $\delta V$ is characterized by the linear response function $\chi_{s_i}$. Owing to the linearity of the system, the same relation holds for the operators:
\begin{equation}
    \widehat{\delta s_i}(t) 
    = \int_{-\infty}^{t} \chi_{s_i}(t-t') \, \widehat{\delta V}(t') \, dt' .
\end{equation}

Following the standard treatments in Refs.~\cite{Schoelkopf2003,Clerk2010}, and under the Born--Markov approximation, the total density matrix approximately factorizes as $\rho \simeq \rho_S \otimes \rho_B$, where $\rho_S$ and $\rho_B$ denote the density matrices of the system and the bath, respectively. Assuming the bath remains in thermal equilibrium, the transition rates between states $\ket{m}$ and $\ket{n}$ depend on the generalized (quantum) noise spectral densities, defined as
\begin{align}
    S_{\widehat{\delta s_i} \widehat{\delta s_j}} (\omega) 
    &= \int_{-\infty}^{\infty} \! dt \;
    \mathrm{Tr}\!\left[\rho_B \, 
    \widehat{\delta s_i}(t) \, \widehat{\delta s_j}(0)\right] e^{i \omega t},\\
    S_{\widehat{\delta V}} (\omega) 
    &= \int_{-\infty}^{\infty} \! dt \;
    \mathrm{Tr}\!\left[\rho_B \, 
    \widehat{\delta V}(t) \, \widehat{\delta V}(0)\right] e^{i \omega t}.
\end{align}
With these definitions, the correspondence expressed in Eq.~\eqref{eq:app-Sdidj_chi_SVN} remains valid in the quantum noise case:
\begin{align}
\begin{split}
    S_{\widehat{\delta s_i}\widehat{\delta s_j}}(\omega) 
    & =\bar{\chi}_{s_i}(\omega) \bar{\chi}_{s_j}^*(\omega) S_{\widehat{\delta V}}(\omega).
\end{split}
\end{align}

\subsection{FGR calculation under gauge and displacement transformation}
\label{sec:app-gauge}

This section demonstrates that the FGR transition rate remains invariant under a continuous gauge transformation transforming the external noise between charge and phase. It also explains why a displaced frame centered on the coherent trajectory hides the leading Purcell channel yet still supports higher-order parametric transitions. 

Let $\mathcal{H}_0(\hat\theta,\hat n)$ be the unperturbed Hamiltonian of a circuit, where $\hat\theta$ and $\hat n$ are the superconducting phase and the Cooper-pair number operators, and they satisfy~\cite{Koch2007}
\begin{align}
\hat{n}=-i\partial/\partial\hat{\theta}\label{n-theta}    
\end{align}
from the canonical commutation relation between the two operators. We start from the gauge where noise $\delta\epsilon(t)$ is coupled to the Cooper-pair number operator $\hat{n}$, 
\begin{equation}
\mathcal{H}(t)=\mathcal{H}_0+\delta\epsilon(t)\hat n.
\end{equation}
Consider the continuous gauge transformation
\begin{equation}
U_\zeta(t)=\exp\!\big[-i\zeta \delta F(t)\hat n\big],
\end{equation}
where $\zeta$ is a real number, and $\delta F(t)$ satisfies 
\begin{align}
\delta \dot{F}(t) =  \delta\epsilon(t)/\hbar.\label{F-epsilon}
\end{align}
This transforms the original Hamiltonian to
\begin{align}
\mathcal{H}_\zeta(t)&=U_\zeta^\dagger \mathcal{H} U_\zeta - i\hbar U_\zeta^\dagger\dot U_\zeta
\\&= \mathcal{H}_0\big(\hat\theta+\zeta \delta F(t),\hat n\big) + (1-\zeta)\delta\epsilon(t)\hat n.
\end{align}
Apparently, $\zeta=1$ corresponds to the gauge where the noise is coupled to the phase operator. 
In the weak-noise limit, $\mathcal{H}_0(\hat\theta+\zeta \delta F(t))$ can be expanded to the first order:
\begin{equation}
\mathcal{H}_\zeta(t)\approx \mathcal{H}_0+(1-\zeta)\delta\epsilon(t)\hat n + \zeta\delta F(t)\frac{\partial \mathcal{H}_0}{\partial\theta}.\label{Hzeta}
\end{equation}

The transition rate between two arbitrary eigenstates, $\lvert i\rangle$ and $\lvert j\rangle$, can be calculated by applying FGR to Eq.~\eqref{Hzeta},
\begin{align}
\Gamma_{i\to j}
&=\frac{1}{\hbar^2}\Big\{
(1-\zeta)^2 \lvert n_{ij}\rvert^2 S_{\delta\epsilon}(\omega_{ij})\nonumber\\ 
&+\zeta^2 \Big\lvert \Big\langle i \Big\lvert \frac{\partial \mathcal{H}_0}{\partial \theta} \Big\rvert j \Big\rangle \Big\rvert^2 S_{\delta F}(\omega_{ij})\nonumber
\\  &+2\zeta(1-\zeta)\textrm{Re}\Big[n_{ij} \Big\langle i \Big\lvert \frac{\partial \mathcal{H}_0}{\partial \theta} \Big\rvert j \Big\rangle S_{\delta \epsilon \delta F}(\omega_{ij})\Big]
\Big\},\label{FGR rate}
\end{align}
where $\omega_{ij}$ is the transition frequency, and $n_{ij} = \langle i\lvert\hat n\rvert j\rangle$. From Eq.~\eqref{n-theta}, we obtain
\begin{equation}
\Big\langle i\Big\lvert\frac{\partial \mathcal{H}_0}{\partial\theta}\Big\rvert j\Big\rangle
=i\langle i\lvert[\hat n,\mathcal{H}_0]\rvert j\rangle=-i\hbar\omega_{ij} n_{ij}.\label{matrix element}
\end{equation}
From Eq.~\eqref{F-epsilon}, assuming both noises are stationary, it can be proved that~\cite{Koch2009}
\begin{equation}
S_{\delta\epsilon}(\omega)=\hbar^2\omega^2 S_{\delta F}(\omega),\quad
S_{\delta\epsilon \delta F}(\omega)= i\hbar\omega S_{\delta F}(\omega).\label{F-epsilon PSD}
\end{equation}
Substituting
 Eqs.~\eqref{matrix element} and \eqref{F-epsilon PSD} into Eq.~\eqref{FGR rate} gives
\begin{equation}
\Gamma_{i\to j}
=\omega_{ij}^2\lvert n_{ij}\rvert^2 S_{\delta F}(\omega_{ij}) = \frac{\lvert n_{ij}\rvert^2}{\hbar^2}S_{\delta\epsilon}(\omega_{ij}),
\end{equation}
which is independent of $\zeta$. Thus, the FGR rate is invariant under gauge transformations when noise and system operators are transformed consistently.

In contrast to the gauge transformation of the lab frame Hamiltonian, the displaced frame transformation is defined by the classical response of the linearized driven system. In strongly driven circuits with significant nonlinearity, the susceptibility generally depends on the drive amplitude, which can deviate from the linear susceptibility due to effects such as ac Stark shifts and Kerr corrections. Consequently, the transformed noise in the displaced frame is reliable only when drive-induced renormalization is small over the operating range, which is often the case in practice for off-resonantly driven, weakly anharmonic circuits. If the displacement follows the full trajectory (coherent plus noise), all terms linear in $a$ and $a^\dagger$ (linearized mode operator) disappear, so the static single-photon Purcell channel does not show up as an explicit noise term. If, instead, the transformation is applied only to the coherent drive and the noise is left in the lab frame, FGR calculation of the linear Purcell rate still applies; however, this requires the explicit displaced frame transformation, which is not available in the DF method introduced in this paper. On the other hand, even when the leading one-photon loss is hidden in the displaced frame, higher-order terms from the cosine expansion remain and allow for the calculation of parametric, multi-photon transitions, whose rates follow from the main-text formula once the derivative operator and the susceptibility functions are identified. In the presence of coherent drives, the same expansion acquires time-periodic coefficients, and the displaced-frame formulation naturally generalizes to driven decoherence from these parametric processes, which can be formally treated within a Floquet-Markov approach (see main text).

\subsection{Calculating voltage-noise-induced transition rates}
\label{sec:app-decoherence-rates}
Let $\mathcal{H}_0$ denote the static unperturbed Hamiltonian, with eigenstates $\ket{m}$ and eigenenergies $E_m$. We consider weak stationary noise perturbations of the form $\mathcal{H}(t)=\mathcal{H}_0+\sum_i \delta s_i(t)\partial\mathcal{H}/\partial\delta s_i$. To leading order in the perturbations, the interaction-picture state evolves as~\cite{Schoelkopf2003, Clerk2010}
\begin{equation}
    \ket{\psi_I(t)} = \ket{m} - \frac{i}{\hbar} \int_0^t{d\tau \sum_i{\delta s_i\left(\tau\right) \frac{\partial \mathcal{H}(\tau)}{\partial \delta s_i}} \ket{m}}.
    \label{eq:app-decay-rate-initial-state}
\end{equation}
With the transition amplitude to another eigenstate $\ket{n}$ computed as
\begin{equation}
    \alpha_{m\rightarrow n}(t) = \braket{n|\psi_I(t)},
\end{equation}
and the interaction-picture operator expressed in terms of Schr\"odinger-picture operators
\begin{equation}
    \frac{\partial \mathcal{H}(\tau)}{\partial \delta s_i} = e^{i \mathcal{H}_0\tau/\hbar}
    \frac{\partial \mathcal{H}}{\partial \delta s_i}e^{-i \mathcal H_0 \tau/\hbar},
\end{equation}
one finds
\begin{equation}
    \alpha_{m\rightarrow n} = -\frac{i}{\hbar} \int_0^t d\tau \sum_i \delta s_i\left(\tau\right) D_{\delta s_i,nm} e^{i \omega_{nm} \tau},
\end{equation}
with $\omega_{nm} = (E_n - E_m)/\hbar$ the frequency difference between the states, and $D_{\delta s_i,nm}$ defined as
\begin{align}
D_{\delta s_i,nm}=\Big\langle n\Big\lvert\frac{\partial \mathcal{H}}{\partial \delta s_i}\Big\rvert m\Big\rangle\Big|_{\delta s_i=0}.
\end{align}
The transition probability, after ensemble-averaging over the noise realizations, is 
\begin{align}
\begin{split}
    \bar{p}_{m \rightarrow n} &= \left\langle \left\lvert \alpha_{m\rightarrow n}\right\rvert^2 \right\rangle = \frac{1}{\hbar^2} \sum_{i,j} D_{\delta s_i,nm} D^*_{\delta s_j,nm}\\
    &\times \int_0^t d\tau_2 \int_0^t d\tau_1 \left\langle \delta s_i\left( \tau_1 \right) \delta s_j\left( \tau_2 \right) \right\rangle e^{i\omega_{nm}\left(\tau_1 - \tau_2\right)} \\
    &= \frac{1}{\hbar^2} \sum_{i,j} D_{\delta s_i,nm} D^*_{\delta s_j,nm}\\
    &\times \int_0^t d\tau_2 \int_{-\tau_2}^{t - \tau_2} d\tau_1 e^{i \omega_{nm} \tau_1} \left\langle \delta s_i\left(\tau_1 + \tau_2\right)\delta s_j\left(\tau_2\right)\right\rangle,
\end{split}
\end{align}
where, assuming time invariance $\left\langle \delta s_i\left(\tau_1 + \tau_2\right)\delta s_j\left(\tau_2\right)\right\rangle = \left\langle \delta s_i\left(\tau_1\right)\delta s_j(0)\right\rangle$ and short correlation time (compared with the noise-induced transition time constant) such that we can extend the bounds of the integral of $\tau_1$ to infinity, we obtain
\begin{equation}
    \bar{p}_{m \rightarrow n} = \frac{t}{\hbar^2} \sum_{i,j} D_{\delta s_i,nm} D^*_{\delta s_j,nm} S_{\delta s_i\delta s_j}\left(\omega_{nm}\right)\label{eq:app-pmn}
\end{equation}
using Eq.~\eqref{eq:app-Sij_time_correlation}. We can then use Eq.~\eqref{eq:app-Sdidj_chi_SVN} and the fact that the transition rate is the time derivative of the transition probability to obtain
\begin{align}
    \Gamma_{m \rightarrow n} &= \frac{1}{\hbar^2} \sum_{i,j} D_{\delta s_i,nm} D^*_{\delta s_j,nm} \nonumber\\
    &\times \bar{\chi}_{s_i}\left(\omega_{nm}\right) \bar{\chi}_{s_j}^*\left(\omega_{nm}\right) S_{\delta V}\left(\omega_{nm}\right)\nonumber\\
    &=\frac{1}{\hbar^2}\left\lvert\sum_iD_{\delta s_i,nm}\bar{\chi}_{s_i}\left(\omega_{nm}\right)\right\rvert^2S_{\delta V}\left(\omega_{nm}\right),
    \label{eq:app-Gamma-m-to-n}
\end{align}
retrieving Eq.~\eqref{eq:Gamma_m_to_n} of the main text. 

Similarly, we can obtain the dephasing rate between the states $\ket{n}$ and $\ket{m}$ by considering the transition between the superposition states $\ket{m} \pm \ket{n}$ and $\ket{m} \mp \ket{n}$. That is, replacing $\ket{m}$ in Eq.~\eqref{eq:app-decay-rate-initial-state} with $(\ket{m} \pm \ket{n})/\sqrt{2}$, one finds
\begin{align}
\begin{split}
    \alpha_{\pm\rightarrow\mp} &= \frac{1}{\sqrt{2}} (\bra{m} \mp \bra{n})\ket{\psi_I(t)}\\
    &=-\frac{i}{2\hbar} \int_0^t d\tau \sum_i \delta s_i\left(\tau\right) \left(D_{\delta s_i,mm} - D_{\delta s_i,nn} \right),
\end{split}
\end{align}
where there are no cross terms since pure dephasing relates to a system-bath coupling operator diagonal in the states $\ket{m}$ and $\ket{n}$. The derivation follows as previously, and the pure dephasing rate is then
\begin{align}
\begin{split}
    \Gamma^{\phi}_{mn} &\equiv \frac{d}{dt}\bigg(\bar{p}_{+ \rightarrow -} + \bar{p}_{- \rightarrow +}\bigg)\\
    = &\frac{1}{2\hbar^2} \sum_{i,j} \left(D_{\delta s_i,mm} - D_{\delta s_i,nn} \right) \left(D_{\delta s_j,mm} - D_{\delta s_j,nn} \right) \\
    &\times \bar{\chi}_{s_i}(0) \bar{\chi}_{s_j}^*(0) S_{\delta V}(0)\\
    &= \frac{1}{2\hbar^2}\left\lvert\sum_i\left(D_{\delta s_i,mm}-D_{\delta s_i,nn}\right)\bar{\chi}_{s_i}(0)\right\rvert^2S_{\delta V}(0),\label{eq:app-Gamma-phi}
\end{split}
\end{align}
retrieving Eq.~\eqref{eq:Gamma_phi_mn} of the main text.

\subsection{Generalization to multi-port with correlated voltage noise}
\label{sec:app-decoherence-rates-multiports}
We now allow the circuit to be coupled to multiple ($P$) drive ports with possibly correlated voltage noise. In this case, the $i$-th noise variable $\delta s_i$ in the circuit Hamiltonian is a combination of linear responses to voltage fluctuations from different ports, with susceptibility $\bar{\chi}_{s_i,p}(\omega)$ to the $p$-th port noise, $\{\delta V_p\}$. Following a similar derivation to Eq.~\eqref{eq:app-Sdidj_chi_SVN}, it is straightforward to obtain the relationship between the cross PSD of $\delta s_i$ and $\delta s_j$, and that of $\delta V_p$ and $\delta V_q$, as
\begin{align}
    S_{\delta s_i\delta s_j}(\omega) = \sum_{p,q}\bar{\chi}_{s_i,p}(\omega)\bar{\chi}_{s_j,q}^*(\omega)S_{\delta V_p\delta V_q}(\omega).
\end{align}
Plugging this result into Eq.~\eqref{eq:app-pmn} and taking the time derivative, we may write the decay rate in a compact, matrix form as 
\begin{equation}
    \Gamma_{m \rightarrow n} = \frac{1}{\hbar^2} \mathbf{X}_{nm}^\dagger\mathbf{S}_{\delta V}\left(\omega_{nm}\right)\mathbf{X}_{nm},\label{eq:app-gamma_mn-matrix}
\end{equation}
where $\mathbf{X}_{nm}$ is a susceptibility vector with its $p$-th element being $\left[\mathbf{X}_{nm}\right]_p = \sum_{i}D_{\delta s_i,nm}\bar{\chi}_{s_i,p}(\omega_{nm})$,
and $\mathbf{S}_{\delta V}$ is a $P\times P$ matrix, with entry $\left[\mathbf{S}_{\delta V}(\omega)\right]_{pq}$ being the cross PSD $S_{\delta V_p\delta V_q}(\omega)$. 

Through a similar derivation, we obtain the generalization of the dephasing rate in Eq.~\eqref{eq:app-Gamma-phi} to the multi-port case, as
\begin{equation}
    \Gamma_{mn}^\phi = \frac{1}{2\hbar^2} \mathbf{X}_{mn}^{\left(\Delta\right)\dagger}\mathbf{S}_{\delta V}\left(0\right)\mathbf{X}_{mn}^{\left(\Delta\right)},\label{eq:app-gamma_phi-matrix}
\end{equation}
with $\mathbf{X}_{mn}^{\left(\Delta\right)} = \mathbf{X}_{mm}-\mathbf{X}_{nn}$.

Finally, we generalize the transition rate and dephasing rate between the Floquet modes in Eqs.~\eqref{eq:Floquet_decay} and \eqref{eq:Floquet_dephasing}, to multi-port driven circuits, as 
\begin{equation}
    \Gamma_{\alpha\rightarrow\beta} = \frac{1}{\hbar^2}\sum_{n=-\infty}^{+\infty}\mathbf{P}^\dagger_{\alpha\beta n}\mathbf{S}_{\delta V}(\Delta_{\alpha\beta n})\mathbf{P}_{\alpha\beta n},
\label{eq:app-Floquet_decay}
\end{equation}
\begin{equation}
    \Gamma^{\phi}_{\alpha\beta} = \frac{1}{2\hbar^2}\sum_{n=-\infty}^{+\infty}\mathbf{P}^{(\Delta)\dagger}_{\alpha\beta n}\mathbf{S}_{\delta V}(n\omega_d)\mathbf{P}^{(\Delta)}_{\alpha\beta n},
\label{eq:app-Floquet_dephasing}
\end{equation}
where $\mathbf{P}_{\alpha\beta n}$ and $\mathbf{P}^{(\Delta)}_{\alpha\beta n}$ are driven susceptibility vectors,
\begin{align}
    \left[\mathbf{P}_{\alpha\beta n}\right]_p &= i\sum_j\frac{\bar{\chi}_{s_j,p}\left(\Delta_{\alpha\beta n}\right)}{T}\nonumber\\
    &\times\int^T_0 e^{-in\omega_d t} \bra{\Phi_\alpha} \mathcal{V}_j(t) \ket{\Phi_\beta}dt ,
    \\
    \mathbf{P}^{(\Delta)}_{\alpha\beta n} &= \mathbf{P}_{\alpha\alpha n}-\mathbf{P}_{\beta\beta n}.
\end{align}

\subsection{Single-JJ and SQUID-based transmon decay rates}
\label{sec:app-transmon-SQUID-decay}

\begin{figure}
    \centering
    \includegraphics[width=\linewidth]{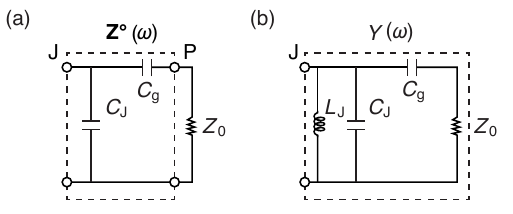}
    \caption{\textbf{Lumped-element circuit model for computing drive-port induced decay rates.} (a) A two-port circuit model for extracting the susceptibility function of the circuit's effective phase modulation by applying the opened-junction IG method. (b) A one-port circuit model for calculating the decay rate using admittance of the junction port.}
    \label{fig:transmon_opened_JJ}
\end{figure}
Here, we derive the Purcell decay rate formulae [Eqs.~\eqref{eq:transmon_decay_LR} and \eqref{eq:decay_SQUID}], for the static, single-JJ transmon, and the SQUID-based transmon composed of two JJs.
The irrotational-gauge Hamiltonian of the transmon in Fig. \hyperref[fig:transmon-decay-rate]{\ref*{fig:transmon-decay-rate}b} is given by
\begin{align}
    \mathcal{H} &= \hbar \omega_{ge} a^\dagger a - \frac{\elementdepsymbol{E}{J}{}}{2}\elementdepsymbol{\beta}{J}{}^2(a + a^\dagger)^2 \nonumber\\ &-  \elementdepsymbol{E}{J}{} \cos\left[\elementdepsymbol{\beta}{J}{}(a + a^\dagger)+ \delta \elementdepsymbol{F}{J}{}(t)\right] ,
    \label{eq:app-H-transmon}
\end{align}
where $\elementdepsymbol{E}{J}{} = \phi_0^2/\elementdepsymbol{L}{J}{}$ is the Josephson energy, $a$ ($a^\dagger$) is the annihilation (creation) operator of the (linearized) transmon mode at frequency $\omega_{ge}$, $\elementdepsymbol{\beta}{J}{}$ is the participation factor of transmon phase in the JJ, and $\delta \elementdepsymbol{F}{J}{}(t)$ is the JJ phase modulation noise as the result of the noisy drive port. Here, we have effectively traced out the resonator modes, treating the latter as the environment that modifies the phase modulation noise acting on the transmon. The susceptibility function is obtained from Eqs.~\eqref{eq:Vs-vs-P-and-Vp}, \eqref{eq:f-from-Z-open} and \eqref{eq:susceptibility}, as
\begin{equation}
    \bar{\chi}_{\elementdepsymbol{F}{J}{}}(\omega) = \frac{\elementdepsymbol{\bar{F}}{J}{}}{\elementdepsymbol{\bar{V}}{S}{}} = \frac{Z^\circ_{\mathrm{J}\mathrm{P}}(\omega)}{i\phi_0 \omega \left(Z_0 + Z^\circ_{\mathrm{P}\mathrm{P}}(\omega)\right)},
    \label{eq:160}
\end{equation}
where $Z^\circ_{\mathrm{J}\mathrm{P}}$ $Z^\circ_{\mathrm{P}\mathrm{P}}$ are the impedance parameters for the opened-junction circuit (see Fig.~\ref{fig:transmon_opened_JJ}). From Eq.~\eqref{eq:linear_noise_coupling} and Eq.~\eqref{eq:app-H-transmon}, we obtain the leading-order noise coupling,  
\begin{align}
    D_{\delta \elementdepsymbol{F}{J}{},ge}&=\Big\langle g\Big\lvert\frac{\partial \mathcal{H}}{\partial \delta \elementdepsymbol{F}{J}{}}\Big\rvert e \Big\rangle\Big\vert_{\delta \elementdepsymbol{F}{J}{} = 0} \nonumber \\&= \bra{g}\elementdepsymbol{E}{J}{} \sin\left[\elementdepsymbol{\beta}{J}{}(a + a^\dagger)\right]\ket{e}.\label{eq:161}
\end{align}
Feeding Eqs.~\eqref{eq:160}--\eqref{eq:161} into Eq.~\eqref{eq:Gamma_m_to_n}, and assuming Johnson-Nyquist voltage noise at the port (Eq.~\eqref{eq:Johnson-Nyquist-spectral-density} with $S_\textrm{in}(\omega)=0$), we obtain the Purcell decay rate as 
\begin{align}
    &\Gamma_{\downarrow} = \frac{1}{\hbar^2}\lvert D_{\delta \elementdepsymbol{F}{J}{},ge}\bar{\chi}_{\elementdepsymbol{F}{J}{}}\left(\omega_{ge}'\right)\rvert^2S_{\delta V}\left(\omega_{ge}'\right)\nonumber\\
    &=2\frac{\lvert D_{\delta \elementdepsymbol{F}{J}{},ge}\rvert^2}{\hbar\omega_{ge}'\phi_0^2}\Big\lvert \frac{Z^\circ_{\mathrm{J}\mathrm{P}}(\omega_{ge}')}{ Z_0 + Z^\circ_{\mathrm{P}\mathrm{P}}(\omega_{ge}')}\Big\rvert^2  Z_0 \left[1+n_\mathrm{B}(\omega_{ge}')\right],\label{app:eq-transmon_full_decay}
\end{align}
where $\omega_{ge}'$ represents transmon's $\ket{g}\rightarrow\ket{e}$ transition frequency, dressed by the Josephson nonlinearity. 

While Eq.~\eqref{app:eq-transmon_full_decay} supports the comprehensive decay rate calculation that accounts for the full nonlinearity, we can also consider its simplification to a linear transmon circuit, allowing us to compare our result with other methods established for linear circuits, such as the HFSS eigenmode Q or the admittance matrix method. Thus, we rewrite the transmon Hamiltonian in Eq.~\eqref{eq:app-H-transmon}, keeping only the quadratic components: 
\begin{align}
    \mathcal{H}_\textrm{lin} &= \hbar \omega_{ge} a^\dagger a + \elementdepsymbol{E}{J}{}\varphi_\mathrm{zpf}\delta \elementdepsymbol{F}{J}{}(t)(a + a^\dagger),
    \label{eq:app-Hlin-transmon}
\end{align}
where $\varphi_{\mathrm{zpf}} = \sqrt{\hbar \omega_{ge}/2 \elementdepsymbol{E}{J}{}}$ is the zero-point phase fluctuation. Under this approximation, Eq.~\eqref{eq:161} becomes
\begin{align}
    D_{\delta \elementdepsymbol{F}{J}{},ge} = \elementdepsymbol{E}{J}{}\varphi_\mathrm{zpf},
\end{align}
and the decay rate of the linearized transmon mode reduces to
\begin{align}
    \Gamma_{\downarrow} =\elementdepsymbol{L}{J}{}^{-1}\left\lvert \frac{Z^\circ_{\mathrm{J}\mathrm{P}}(\omega_{ge})}{ Z_0 + Z^\circ_{\mathrm{P}\mathrm{P}}(\omega_{ge})}\right\rvert^2  Z_0 \left[1+n_\mathrm{B}(\omega_{ge})\right],
\end{align}
which corresponds to Eq.~\eqref{eq:transmon_decay_LR} in the main text. 

As an important special case, let us consider a transmon capacitively coupled to a 50\,$\Omega$ port (see Fig.~\ref{fig:transmon_opened_JJ}). We may express the decay rate in terms of circuit parameters using $\omega_{ge} = \sqrt{1/\elementdepsymbol{L}{J}{}C_{\text{eff}}}$ and the impedance relations (with $j = -i$)
\begin{align}
    Z_{\mathrm{J}\mathrm{P}}^\circ(\omega_{ge}) &= \frac{1}{j \omega_{ge} \elementdepsymbol{C}{J}{}},\\
    Z_{\mathrm{P}\mathrm{P}}^\circ(\omega_{ge}) &= \frac{\elementdepsymbol{C}{g}{} + \elementdepsymbol{C}{J}{}}{j \omega_{ge} \elementdepsymbol{C}{g}{} \elementdepsymbol{C}{J}{}},
\end{align}
such that we obtain 
\begin{equation}
\Gamma_{\downarrow} \approx C_{\text{eff}}^{-1} \left( \frac{C_{\text{eff}}}{C_\Sigma} \right)^2\left[\frac{1}{Z_0\omega_{ge}^2\elementdepsymbol{C}{g}{}^2}+Z_0 \left(\frac{\elementdepsymbol{C}{J}{}}{C_\Sigma}\right)^2\right]^{-1}, 
\label{eq:app-transmon_decay_LR_circuit_params}
\end{equation}
where $C_\Sigma = \elementdepsymbol{C}{J}{} + \elementdepsymbol{C}{g}{}$ is the total capacitance of the transmon. The effective capacitance of the transmon mode, $C_{\text{eff}} = (1/2)\Imag Y'(\omega_{ge})$, includes the loading from the port. Meanwhile, expressing Eq.~\eqref{eq:transmon_decay_BBQ} in terms of circuit parameters, we obtain a similar albeit different expression
\begin{align}
\Gamma_{\downarrow} \approx C_{\text{eff}}^{-1}\left[\frac{1}{Z_0\omega_{ge}^2\elementdepsymbol{C}{g}{}^2}+Z_0\right]^{-1}. 
\label{eq:app-transmon_decay_BBQ_circuit_params}
\end{align}
Eqs.~\eqref{eq:app-transmon_decay_LR_circuit_params} and ~\eqref{eq:app-transmon_decay_BBQ_circuit_params} agree closely in the weak-coupling regime $\elementdepsymbol{C}{g}{} \ll \elementdepsymbol{C}{J}{}$, whereas they are both invalid in the strong-coupling regime $\elementdepsymbol{C}{g}{} \gtrsim \elementdepsymbol{C}{J}{}$.

For the SQUID-based transmon in Fig.~\ref{fig:junction-centered}, we can similarly write down its Hamiltonian in the irrotational gauge, as
\begin{align}
    \mathcal{H} &= \hbar \omega_{ge} a^\dagger a - \sum_{k=1,2}\frac{\elementdepsymbol{E}{J}{k}}{2}\elementdepsymbol{\beta}{J}{k}^2(a + a^\dagger)^2 \nonumber\\ &-  \sum_{k=1,2}\elementdepsymbol{E}{J}{k} \cos\left[\elementdepsymbol{\beta}{J}{k}(a + a^\dagger)+ \delta \elementdepsymbol{F}{J}{k}(t)\right].
    \label{eq:app-H-SQUID}
\end{align}
The linearized Hamiltonian is given by 
\begin{equation}
    \mathcal{H}_\textrm{lin} = \hbar \omega_{ge} a^\dagger a +  \sum_{k=1,2} \elementdepsymbol{E}{J}{k} \varphi_{\mathrm{zpf}} (a + a^\dagger) \elementdepsymbol{\delta F}{J}{k}(t),
    \label{eq:app-H-SQUID-lin}
\end{equation}
where $\varphi_{\mathrm{zpf}}$ is the zero-point fluctuation of the phase,
\begin{equation}
     \varphi_{\mathrm{zpf}} =  \phi_0^{-1} \sqrt{\hbar \omega_{ge} L_\Sigma/2}.
\end{equation}
and ${L_\Sigma} = ( \elementdepsymbol{L}{J}{1}^{-1}+ \elementdepsymbol{L}{J}{2}^{-1})^{-1}$ is the total inductance of the two junctions. Eq.~\eqref{eq:decay_SQUID} can be readily obtained from Eq.~\eqref{eq:app-H-SQUID-lin} and Eq.~\eqref{eq:Gamma_m_to_n}, which captures the interference of $\elementdepsymbol{\delta F}{J}{1}(t)$ and $\elementdepsymbol{\delta F}{J}{2}(t)$ which originate from the same source. For the special case of a differentially-coupled ($Z^\circ_{\mathrm{J}_1\mathrm{P}}=-Z^\circ_{\mathrm{J}_2\mathrm{P}}$), symmetric SQUID ($\elementdepsymbol{L}{J}{1}=\elementdepsymbol{L}{J}{2}$) transmon, the noise terms destructively interfere with each other, resulting in the reduction of the Purcell decay rate (see Fig.~\hyperref[fig:SQUID_Q]{\ref*{fig:SQUID_Q}d}).

\subsection{First-order PVNR estimates of transmon decoherence}
\label{app:dephasing_susceptibility}
In this section, we show how several key decoherence processes in a Josephson circuit can be understood from the viewpoint of a noise-perturbed Josephson nonlinearity, and we use the PVNR expansion to obtain leading-order analytic rates. These results are intended for intuition and quick estimates. For quantitatively exact rates, we refer to the Floquet-Markov treatment presented in the main text. We consider a weakly-coupled transmon-resonator circuit, where the resonator is capacitively coupled to a lossy drive port. The lab-frame Hamiltonian of the driven circuit is
\begin{align}
\label{eq:transmon_filter}
&\mathcal{H}_{\rm lab}  = \hbar \omega_q q^\dag q + \hbar \omega_a a^\dag a + i \left[g_a(t)+\delta g_a(t)\right]\left(a^\dag-a\right)\nonumber \\ 
&- E_\mathrm{J}\cos_\textrm{nl}\left[\beta_q \left(q + q^\dag\right)+ \beta_a \left(a + a^\dag\right)\right].
\end{align} 
Here, $q, a$ are the annihilation operators of the transmon and the resonator, respectively. $g_a(t)$ is the linear drive (effective charge modulation) strength on the resonator, and $\delta g_a(t)$ represents fluctuations in the linear drive due to the voltage noise at the drive port. Here, we have ignored the direct drive and its noise of the transmon inherited from the resonator, which can lead to important physical effects such as the static Purcell loss. The linear susceptibility function that relates the source voltage to the resonator drive $\bar{g}_a(\omega)$ is
\begin{align}
\label{eq:chi_gv}
\bar{\chi}_{g_a}(\omega) =  \left(8PZ_0 \right)^{-1/2} \bar{g}_a(\omega),
\end{align}
where $P$ is the drive power, and $Z_0$ is the load impedance of the port, see Eq.~\eqref{eq:Vs-vs-P-and-Vp}. Assuming the lossy port being the only loss channel in the system, Fermi's golden rule dictates that the relaxation spectrum of the resonator is given by~\cite{Clerk2010}
\begin{align}
\label{relaxtion_spec}
    K_a(\omega) &= \lvert\bar{\chi}_{g_a}(\omega)\rvert^2 \left[S_{\delta V}(\omega)-S_{\delta V}(-\omega)\right] \\ \nonumber  &=\lvert \bar{g}_a(\omega)\rvert^2 \hbar\omega \left(4P\right)^{-1},
\end{align}
which gives the relaxation rate $\kappa_a = K_a\left(\omega_a\right)$ of the resonator at its resonance. Under the linear drive at $\omega_d$, the steady-state displacement of the resonator field $a+a^\dagger$ is 
\begin{align}
    \xi_a (t) = \Real\left[ \frac{2{\omega _d}\bar{g}_a\left( \omega _d \right)e^{i\left( \omega _d t + \varphi _a \right)}}{\sqrt {{\left( {\omega _d^2 - \omega_a^2} \right)}^2 + K_a ^2\left( \omega_d \right)\omega_d^2}} \right],
\end{align}
where $\varphi_{a}$ is the phase delay due to the damping. It follows that the susceptibility function connecting this displacement to the source voltage is
\begin{align}
    \bar{\chi}_{\xi_a}(\omega) = \frac{\bar{\xi}_a(\omega)}{\elementdepsymbol{\bar{V}}{S}{}} = \frac{ \sqrt{\frac{2\omega K_a(\omega)}{\hbar Z_0}}e^{i\varphi_a}}{\sqrt {{\left( \omega ^2 - \omega_a^2 \right)}^2 + K_a^2(\omega){\omega ^2}} }.\label{chi_xiv}
\end{align}
The fluctuation in the displacement, $\delta\xi_a(t)$, is transformed from the voltage noise $\delta V(t)$ through the same susceptibility function. Considering the quantum extension of Johnson-Nyquist noise at the drive port, the resulting PSD of $\delta\xi_a$ is
\begin{align}
S_{\delta\xi_a}(\omega) &= \lvert \bar{\chi}_{\xi_a}(\omega) \rvert^2 S_{\delta V}(\omega) \nonumber \\ 
&= \frac{\left[ 1 + {n_\mathrm{B}}(\omega) \right]K_a(\omega)}{\left( \omega  - \omega_a \right)^2\left( \omega  + \omega_a \right)^2\left( {\frac{1}{{2\omega }}} \right)^2 + \left( \frac{K_a(\omega)}{2} \right)^2}.
\label{eq:S_xixi}
\end{align}

\begin{figure}
    \centering
    \includegraphics[width=\linewidth]{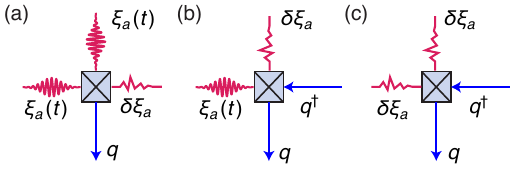}
    \caption{\textbf{Leading-order incoherent parametric interactions.}
    Diagrams illustrating four-wave mixing processes involving incoherent drive photons, which give rise to various decoherence processes.
    (a) Representation of $\mathcal{H}_{\text{int,1}}$ [Eq.~\eqref{eq:hint1}], which models drive-induced Purcell decay. The outgoing arrow denotes the term proportional to the annihilation operator $q$. The reverse process involving $q^\dagger$ models drive-induced heating. 
    (b) Representation of $\mathcal{H}_{\text{int,2}}$  [Eq.~\eqref{eq:hint2}] for drive-induced dephasing. 
    (c) Representation of $\mathcal{H}_{\text{int,3}}$ [Eq.~\eqref{eq:hint3}] for thermal-photon dephasing.}
    \label{fig:incoherent_4wm_processes}
\end{figure}

Now, we are ready to move to the displaced-frame Hamiltonian, containing the displacement $\xi_a(t)$ and its noise $\delta\xi_a(t)$,
\begin{align}
\label{eq:app-qubit-cavity-dephasing}
&\mathcal{H}_{\rm disp}  = \hbar \omega_q q^\dag q + \hbar \omega_a a^\dag a \nonumber \\ 
&- E_\mathrm{J}\cos_\textrm{nl}\left[\beta_q \left(q + q^\dag\right)+ \beta_a \left(a + a^\dag + \xi_a (t) + \delta\xi_a(t)\right)\right].
\end{align}
Among various coherent and incoherent parametric interactions induced by the stimulated nonlinearity, we focus on the following three terms from the leading-order expansion:
\begin{align}
    \mathcal{H}_{\rm int,1} &=- \frac{1}{2}E_\mathrm{J}\beta_q\beta_a^3\xi_a^2 (t) \delta\xi_a(t)\left(q+q^\dag\right), \label{eq:hint1} \\
    \mathcal{H}_{\rm int,2} &=-E_\mathrm{J}\beta_q^2\beta_a^2\xi_a(t)\delta\xi_a(t)q^\dag q, \label{eq:hint2} \\
    \mathcal{H}_{\rm int,3} &=-\frac{1}{2}E_\mathrm{J}\beta_q^2\beta_a^2\left(\delta\xi_a(t)\right)^2q^\dag q. \label{eq:hint3}
\end{align}
These terms are visually represented in Fig.~\ref{fig:incoherent_4wm_processes}.
Next, we will show how these incoherent interactions can lead to drive-induced Purcell decay, coherent-photon-induced dephasing, and thermal-photon dephasing of the transmon, respectively.

For $\mathcal{H}_{\rm int,1}$ in Eq.~\eqref{eq:hint1}, we treat $\xi^2_a(t)\delta\xi_a(t)$ as a noise convoluted by the coherent modulation, rewriting it as
\begin{align}
    \xi^2_a(t)\delta\xi_a(t) = 2\bar{n}_a^\textrm{coh}\left[1-\cos(2\omega_dt+2\varphi_a)\right]\delta\xi_a(t),
\end{align}
where $\bar{n}_a^\textrm{coh}=\lvert\xi_a(t)\rvert^2/4$ is the coherent photon population in the resonator mode. The PSD of this convoluted noise is
\begin{align}
S_{\xi^2_a*\delta\xi_a}&(\omega)=\left(\bar{n}_a^\textrm{coh}\right)^2\times
\nonumber\\ &\left[4S_{\delta\xi_a}\left(\omega\right)
+S_{\delta\xi_a}\left(\omega-2\omega_d\right)+S_{\delta\xi_a}\left(\omega+2\omega_d\right)\right].
\end{align}
Thus, $\mathcal{H}_\textrm{int,1}$ results in a Purcell decay rate of 
\begin{align}
    \gamma_{\downarrow} = \frac{1}{4}(\chi^0_{qa})^2\beta_a^2\beta_q^{-2}S_{\xi_a^2*\delta\xi_a}(\omega_q'),\label{eq:gamma_decay_first}
\end{align}
where $\chi^0_{qa}=-E_\mathrm{J}\beta_q^2\beta_a^2/\hbar$ is the leading-order transmon-resonator cross-Kerr from the expansion of $\cos_\textrm{nl}$, and $\omega_q'$ is the transmon frequency renormalized by the nonlinearity. Consider the practical regime where the sideband frequency is near-resonant with the resonator frequency, i.e., $\omega_q'-2\omega_d\approx\omega_a$, the drive-induced Purcell decay rate is approximated by 
\begin{align}
    \gamma_{\downarrow} &\approx \frac{1}{4}(\chi^0_{qa})^2\beta_a^2\beta_q^{-2}\left(\bar{n}_a^\textrm{coh}\right)^2S_{\delta\xi_a}\left(\omega_q'-2\omega_d\right)\nonumber\\
    &\approx  \frac{g_\textrm{sb}^2\kappa_a}{\Delta_d^2+ \left( \frac{\kappa_a}{2} \right)^2},\label{eq:gamma_decay_second}
\end{align}
where $\Delta_d = \omega_q'-2\omega_d-\omega_a$, and we assumed a negligible thermal population of the resonator. Noticing that the prefactor $g_\textrm{sb}=\frac{1}{2}\chi^0_{qa}\beta_a\beta_q^{-1}\bar{n}_a^\textrm{coh} $ represents the transmon-resonator sideband coupling rate, Eq.~\eqref{eq:gamma_decay_second} is consistent with the well-known drive-induced Purcell rate formula~\cite{Sete2014}.

Now we move to $\mathcal
{H}_\textrm{int,2}$ in Eq.~\eqref{eq:hint2} and see how it leads to transmon dephasing due to the fluctuation in the coherent field of the resonator. Treating $\xi_a(t)\delta\xi_a(t)$ as the convoluted noise, we obtain its spectral density as  
\begin{align}
    S_{\xi_a*\delta\xi_a}(\omega) = \bar{n}_a^\textrm{coh}\left[S_{\delta\xi_a}(\omega-\omega_d)+S_{\delta\xi_a}(\omega+\omega_d)\right],
\end{align}
whose zero-frequency component sets the dephasing rate of the transmon in the long-time limit (Eq.~\eqref{eq:Gamma_phi_mn}),
\begin{align}
    \gamma_{\phi,1} &= \frac{(\chi^0_{qa})^2}{2}S_{\xi_a*\delta\xi_a}(0)\nonumber \\ &=\frac{\frac{(\chi^0_{qa})^2}{2}\bar{n}_a^\text{coh}\left[ 1 + 2n_\mathrm{B}(\omega_d) \right]K_a(\omega_d)}{\left( \omega_d  - {\omega _a} \right)^2\left( {\omega_d  + {\omega _a}} \right)^2\left( {\frac{1}{{2\omega_d }}} \right)^2 + \left( \frac{K_a(\omega_d)}{2} \right)^2}.\label{eq:gamma_dephase_coh}
\end{align}
The above result requires the dephasing rate $\gamma_{\phi,1}$ to be much smaller than the bandwidth of the noise $\delta\xi_a$, which is set by the resonator relaxation rate $\kappa_a$. In the low-temperature limit, and when the drive frequency is close to the resonator frequency, Eq.~\eqref{eq:gamma_dephase_coh} reduces to 
\begin{align}
\gamma _{\phi,1} \approx \frac{\frac{(\chi^0_{qa})^2}{2}\bar{n}_a^\text{coh}\kappa_a}{\left( \omega_d  - {\omega _a} \right)^2 + \left( \frac{\kappa_a}{2} \right)^2},\label{eq:gamma_phi_one_reduced}
\end{align}
which matches the standard coherent-photon (measurement-induced) dephasing expression \cite{Gambetta2006, Yan_dephasing, Clerk_Utami} in the weak-$\chi$ limit, up to the coefficient $\chi^0_{qa}$ used here. 

Finally, for the third interaction $\mathcal{H}_\textrm{int,3}$ in Eq.~\eqref{eq:hint3}, we need to calculate the spectral density of the ``quadratic noise" $\left(\delta\xi_a(t)\right)^2$. According to Wick's theorem, the correlation function of $\left(\delta\xi_a(t)\right)^2$ can be expressed as 
\begin{align}
C_{\left(\delta\xi_a\right)^2}(t) = &\, \big\langle\left(\delta\xi_a(t)\right)^2 \left(\delta\xi_a(0)\right)^2\big\rangle\nonumber\\
=&\,\big\langle\delta\xi^2_a(0)\rangle^2+2\langle \delta\xi_a(t)\delta\xi_a(0)\big\rangle^2,\end{align}
which is exact when $\delta\xi_a(t)$ is Gaussian noise. The first term of the right-hand side is an unimportant offset that is ultimately absorbed by the qubit frequency. The Fourier transform of the second term is a convolution of two $S_{\delta\xi_a}(\omega)$,
\begin{align}
    S_{\left(\delta\xi_a\right)^2}(\omega) =&\,2 (S_{\delta\xi_a}*S_{\delta\xi_a})(\omega)\nonumber\\=&\, \frac{1}{\pi}\int_{-\infty}^{\infty} S_{\delta\xi_a}(\omega')S_{\delta\xi_a}(\omega-\omega').
\end{align}

In the weak-$\chi$ limit ($\chi^0_{qa}\ll\kappa_a$), this yields
\begin{align}
\gamma_{\phi,2} = \frac{1}{8}\left(\chi^{0}_{qa}\right)^2S_{\left(\delta\xi_a\right)^2}(0) = \frac{\left(1+\bar{n}_a^{\text{th}}\right)\bar{n}_a^{\text{th}}\left(\chi^0_{qa}\right)^2}{\kappa_a},
\label{eq:gamma_phi_two}
\end{align}
which is consistent with the thermal-photon dephasing \cite{Yan_dephasing}, despite the different coefficient $\chi^0_{qa}$. We remark that the standard result with the renormalized $\chi_{qa}$ can be recovered by directly applying PVNR to the full dispersive interaction, $-\chi_{qa}q^\dagger q a^\dagger a$. We omit the short derivation here.

The above analysis for the transmon-resonator system can be extended to a more general circuit, where a transmon is weakly coupled to a linear network terminated by noisy drive ports. Eliminating the network modes within a Markovian treatment, we keep the collective contribution of the driven linear network to the coherent and incoherent junction phase displacement as $A(t)$ and $\delta A(t)$,
\begin{align}
&\mathcal{H}_{\rm disp}  = \hbar \omega_q q^\dag q - \elementdepsymbol{E}{J}{} \cos_{\mathrm{nl}}\left[\elementdepsymbol{\beta}{J}{} (q +q^\dagger)+A(t)+\delta A(t) \right].
\end{align}
In the main text, we highlighted the case of two nearly degenerate resonators, which hybridize into two normal modes that both participate in the JJ phase. Because both modes are driven from the same port, their contributions are correlated and can interfere either constructively or destructively, leading to suppression or enhancement of the drive-induced decoherence rate, as shown in Fig.~\ref{fig:two_filter_modes}. This is made explicit by writing the phase susceptibility as a coherent sum of the modal susceptibilities,
\begin{align}
 \bar\chi_A(\omega)= \beta_a \bar\chi_{\xi_a}(\omega)+\beta_b \bar\chi_{\xi_b}(\omega),
\end{align}
so that the noise PSD becomes
\begin{align}
S_{\delta A}(\omega)&=\lvert\bar\chi_A(\omega)\rvert^2S_{\delta V}(\omega)\nonumber\\
  &=\left\{
      \beta_a^2 S_{\delta\xi_a}(\omega)
      + \beta_b^2 S_{\delta\xi_b}(\omega) \right. \nonumber \\
   &\left. + 2 \beta_a \beta_b
      \Real\big[ \bar\chi_{\xi_a}(\omega)\bar\chi^*_{\xi_b}(\omega) \big]
      \right\}S_{\delta V}(\omega).
\end{align}
The last term is the interference term that captures whether the two network paths reinforce or cancel each other at a given frequency.

Finally, we consider the generalization to a multi-JJ transmon circuit, with phase displacement of the $k$-th JJ denoted as $A_k(t)$ and its fluctuation $\delta A_k(t)$. From Eq.~\eqref{eq:app-Sdidj_chi_SVN}, we obtain the cross PSD between $\delta A_k$ and $\delta A_l$, as
\begin{align}
    S_{\delta A_k \delta A_l}(\omega) = \bar{\chi}_{A_k}(\omega)\bar{\chi}^*_{A_l}(\omega)S_{\delta V}(\omega).
\end{align}
Then, we can readily write down the driven Purcell decay rates, the coherent-photon and thermal-photon dephasing rates, as
\begin{align}
    \gamma_\downarrow = \frac{1}{64\hbar^2}&\sum_{k,l}\elementdepsymbol{E}{J}{k} \elementdepsymbol{E}{J}{l} \elementdepsymbol{\beta}{J}{k} \elementdepsymbol{\beta}{J}{l} \bar{A}_k^2(\omega_d)\bar{A}_l^2(\omega_d)\nonumber\\
    &\times S_{\delta A_k \delta A_l}(\omega_q'-2\omega_d),\label{eq:F251}\\
    \gamma_\phi^\textrm{coh}=\frac{1}{8\hbar^2}&\sum_{k,l}\elementdepsymbol{E}{J}{k} \elementdepsymbol{E}{J}{l} \elementdepsymbol{\beta^2}{J}{k} \elementdepsymbol{\beta^2}{J}{l} \bar{A}_k(\omega_d)\bar{A}_l(\omega_d)\nonumber\\
    &\times \left[S_{\delta A_k \delta A_l}(\omega_d)+S_{\delta A_k \delta A_l}(-\omega_d)\right],\label{eq:F252}\\
    \gamma_\phi^\textrm{th}=\frac{1}{8\pi\hbar^2}&\sum_{k,l}\elementdepsymbol{E}{J}{k} \elementdepsymbol{E}{J}{l} \elementdepsymbol{\beta^2}{J}{k} \elementdepsymbol{\beta^2}{J}{l}\nonumber\\
    &\times \int_{-\infty}^\infty S_{\delta A_k \delta A_l}(\omega)S_{\delta A_l \delta A_k}(-\omega)d\omega.\label{eq:F253}
\end{align}
These expressions recover the single-junction results derived earlier when $M=1$.

For an outlook, our PVNR framework can be generalized to address short and non-periodic pulses. Our current calculation of the decoherence rates assumes sinusoidal drives with a constant amplitude, and the FGR calculation samples the power of noise only at one or a few discrete frequencies (aside from higher-order processes in Eq.~\eqref{eq:F253}). For short and non-periodic drive pulses, a spectral-averaged sampling following the framework of the pseudo-Lindblad master equation or Keldysh map becomes necessary, where $\gamma_{\downarrow}$ and $\gamma_\phi$ take on the format $\int_{-\infty}^{\infty}d\omega f(\omega)S_{\delta A_k\delta A_l}(\omega)$ that involves a filter function $f(\omega)$. A systematic calculation of these filter functions entails a numerical Fourier transform of the time-dependent system propagator, as derived and explained in detail in~\cite{Green2013, Huang2023, Groszkowski2023}. In the special case of no drive or a constant amplitude tone, $f(\omega)$ collapses to one or a sum of Dirac delta functions, recovering Eqs.~\eqref{eq:F251} and \eqref{eq:F252}.

\section{Master equation for the drive-induced decay and dephasing example}\label{app:master}
In this section, we present the construction of the Lindblad master equation used to compute drive-induced dephasing and Purcell decay for the example circuit studied in Section~\ref{sec:drive-induced-decay-dephasing} (Fig.~\hyperref[fig:two_filter_modes]{\ref*{fig:two_filter_modes}a}). The form of the master equation, specifically the choice of relevant dissipators, is guided by the HFSS microwave simulation results that are summarized in Table~\ref{tab:drive_induced_decoherence_parameters}. Appendices~\ref{app:tmon_filter_static_circuit} and \ref{app:tmon_filter_bath} introduce the Hamiltonian and the dissipative part of the master equation and the approximations involved. Appendices~\ref{app:tmon_filter_dephasing} and \ref{app:tmon_filter_decay} describe the procedure for extracting the drive-induced dephasing and decay rates. Details of the parameter-extraction workflow for the master equation are provided in Appendix~\ref{app:tmon_filter_parameter_extraction}.

\begin{figure}
    \centering
    \includegraphics[width=1\linewidth]{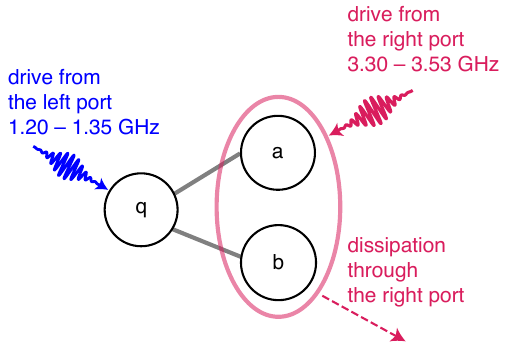}
    \caption{\textbf{Schematic of the example system used to study drive-induced Purcell decay and dephasing.} A transmon mode $q$ is dispersively coupled to two filter modes $a$ and $b$. The right port is strongly coupled to the filter modes, setting the dominant decay and dephasing rates in the following studies (the noise in the weakly-coupled left port is neglected). For the drive-induced dephasing study, the filter modes are driven through the right port near their resonance frequencies. For the drive-induced Purcell decay study, the transmon is driven through the left port at frequencies near half of the transmon–filter detuning frequency.
    }    \label{fig:drive_induced_decoherence_lumped_element}
\end{figure}

\begin{table}[]
\raggedright
\begin{tabular}{@{}ll@{}}
\toprule
\textbf{Parameter} & \textbf{Value} \\
\midrule
\multicolumn{2}{l}{{Linearized circuit eigenmode information} (from HFSS)}\\
$\elementdepsymbol{E}{J}{}/h$ (input parameter) & 34.948 GHz \\
$\omega_a^0/2\pi$ & 3.3331 GHz \\
$\omega_b^0/2\pi$ & 3.4823 GHz \\
$\omega_q^0/2\pi$ & 6.0863 GHz \\
$\kappa_a^0/2\pi$ & 30.627 MHz \\
$\kappa_b^0/2\pi$ & 45.686 MHz \\
$\kappa_q^0/2\pi$ & 38.547 kHz \\
\midrule
\multicolumn{2}{l}{{Participation factors in the transmon JJ} (from EPR)}\\
$\beta_a$ & 0.011158\\
$\beta_b$ & -0.012171\\
$\beta_q$ & -0.29064\\
\midrule
\multicolumn{2}{l}{{Lindblad master equation parameters} (from fitting)}\\
$\tilde{\omega}_a/2\pi$ & 3.3308 GHz\\
$\tilde{\omega}_b/2\pi$ & 3.4846 GHz\\
$\tilde{\omega}_q/2\pi$ & 6.0863 GHz\\
$c_a/\sqrt{2\pi}$ & 0.44029 $\sqrt{\mathrm{GHz}}$\\
$c_b/\sqrt{2\pi}$ & 0.53444 $\sqrt{\mathrm{GHz}}$\\
$\tilde\kappa_q/\sqrt{2\pi}$ & 38.574 kHz \\
\midrule
\multicolumn{2}{l}{{Cross-Kerr frequencies} (from diagonalizing the Hamiltonian)}\\
$\chi_{aq}/2\pi$ & -452.87 kHz\\
$\chi_{bq}/2\pi$ & -536.87 kHz\\
$\chi_{ab}/2\pi$ & -0.85639 kHz\\
\bottomrule
\end{tabular}
\caption{Parameters of the drive-induced decay and dephasing example}
\label{tab:drive_induced_decoherence_parameters}
\end{table}

\subsection{Modeling the static circuit}
\label{app:tmon_filter_static_circuit}
The example system is made of a transmon and two filter modes. The filter modes are near-degenerate, and one of them is strongly coupled to a lossy drive port. 

Writing in the normal-mode basis of the linearized circuit, the static Hamiltonian is
\begin{align}
    \mathcal{H} &= \mathcal{H}_{\mathrm{lin}}
    + \mathcal{H}_{\mathrm{nl}},
    \label{eq:tmon_filter_hamiltonian}\\
    \mathcal{H}_{\mathrm{lin}} &= \hbar \tilde{\omega}_{q} q^\dagger q 
    + \hbar \tilde{\omega}_a a^\dagger a 
    + \hbar \tilde{\omega}_b b^\dagger b ,
    \\
    \mathcal{H}_{\mathrm{nl}} & = - \elementdepsymbol{E}{J}{}\cos_{\mathrm{nl}}[\beta_{a}(a^\dagger + a)
    + \beta_b(b^\dagger + b) \nonumber \\
    &\quad + \beta_q (q^\dagger + q)].
\end{align}
The operators $a$ and $b$ are the annihilation operators of the two filter normal modes, and $q$ is the annihilation operator of the transmon mode. A schematic diagram for the normal-mode system is shown in Fig.~\ref{fig:drive_induced_decoherence_lumped_element}. 
For convenience, the parameters are compiled in Table \ref{tab:drive_induced_decoherence_parameters}; the procedure of extracting them from HFSS simulations is detailed in Appendix ~\ref{app:tmon_filter_parameter_extraction}.

We diagonalize the Hamiltonian in \eqref{eq:tmon_filter_hamiltonian} and denote the low-lying eigenstates of the system as $\ket{\overline{n_a, n_b, n_q}}$,  where $n_a$, $n_b$ and $n_q$ are the photon numbers of the linear modes $a$, $b$ and $q$ in the bare Fock states $\{\ket{n_a, n_b, n_q}\}$ closest to the eigenstates. The corresponding eigenenergies are denoted as $E_{\overline{n_a, n_b, n_q}}$. The cross-Kerr frequency between modes $a$ and $q$ is defined as 
\begin{align}
    \chi_{aq} = \left(E_{\overline{1, 0, 1}}
    - E_{\overline{1, 0, 0}}
    -
    E_{\overline{0, 0, 1}}
    + E_{\overline{0, 0, 0}}\right)/\hbar.
\end{align}
We define the cross-Kerr frequencies for the other mode pairs analogously. The quantities $\chi_{aq}$, $\chi_{bq}$, and $\chi_{ab}$ are obtained by numerically diagonalizing the Hamiltonian in Eq.~\eqref{eq:tmon_filter_hamiltonian}, and the resulting values are listed in Table~\ref{tab:drive_induced_decoherence_parameters}. The small magnitudes of these cross-Kerr shifts, compared to the frequency separations among the modes, indicate that these modes are dispersively coupled.

\subsection{Modeling the bath}
\label{app:tmon_filter_bath}

Because one of the bare filter modes is strongly coupled to the right port, the dissipation in the dressed-mode basis is well captured by a single joint bath channel. We therefore write the master equation as
\begin{align}
\dot \rho &= \frac{1}{i\hbar}[\mathcal{H} + \mathcal{H}_d(t), \rho] + \mathscr{L}[\rho],
\label{eq:tmon_filter_master_eq}
\\
\mathscr{L}[\rho] &=[1+n_\mathrm{B}(\bar{\omega})]\mathcal{D}[c_a a
+ c_b b] \rho \nonumber \\
&+ n_\mathrm{B}(\bar{\omega})\mathcal{D}[c_a a^\dagger
+ c_b b^\dagger]\rho\,. 
\label{eq:tmon_filter_master_eq_dissipator}
\end{align}
Here, $\mathcal{H}_d(t)$ is the time-dependent part of the Hamiltonian describing the drive, $\bar{\omega}$ is the average of the two filter mode frequencies $\tilde{\omega}_a$ and $\tilde{\omega}_b$, and $c_a$ and $c_b$ are two real, positive-valued parameters. Equivalently, the common dissipator in Eq.~\eqref{eq:tmon_filter_master_eq_dissipator} can be obtained by starting from the physical port-loss channel associated with the port-coupled bare filter coordinate and expressing that channel in the filter normal-mode basis. The same common-channel structure can also be justified from the Bloch--Redfield equation using a partial secular approximation~\cite{Hofer_2017, Trushechkin_2021, Rohan_2026}; here the partial secular approximation is used to retain the correlated dissipator in the near-degenerate normal-mode basis, not as the only route to obtaining the coherent sum. It is important to note that $\mathcal{D}[c_a a + c_b b]\neq\mathcal{D}[c_a a] + \mathcal{D}[c_b b]$, where the latter expression treats the normal-mode loss channels as independent and therefore ignores interference effects. Since the linewidths of the two filter modes $\tilde{\kappa}_a$ and $\tilde{\kappa}_b$ are comparable in magnitude to their detuning $|\tilde{\omega}_a - \tilde{\omega}_b|$, such effects cannot be ignored. The choice of signs for $c_a$ and $c_b$ should be understood as relative to the JJ participation ratios $\beta_a$ and $\beta_b$, which we find from the EPR simulation to have opposite signs. For two strongly hybridized, near-degenerate filter modes, the normal modes are approximately symmetric and antisymmetric combinations, so their fields at the left and right ends can acquire an approximately $\pi$ relative phase. This phase assignment is an approximation motivated by the lumped-element circuit model, whereas in the PVNR approach, the corresponding relative phases are contained directly in the complex port-to-junction susceptibility and do not require an explicit sign choice. In the drive-induced decay example, we explicitly compare the difference between the two ways of modeling decoherence (Fig.~\hyperref[fig:two_filter_modes]{\ref*{fig:two_filter_modes}c}). By contrast, the PVNR method automatically accounts for this interference through the susceptibility function. This provides a convenient way to include such interference effects, especially when extending the analysis to systems with larger filter networks.

\subsection{Drive-induced dephasing rate calculation}
\label{app:tmon_filter_dephasing}
In the drive-induced dephasing case, we consider the filter modes being driven through the right port. The form of the drive is 
\begin{equation}
    \mathcal{H}_{d}(t) = i\cos(\omega_d t)[g_{a} (a^\dagger - a) + g_{b} (b^\dagger - b)]\,.
\end{equation}
To compare the LME results with the PVNR calculation, we must relate the drive power and frequency to the effective drive strengths $g_a$ and $g_b$.
Since the drive is nearly resonant with the two filter modes, by considering input--output theory \cite{Girvin2014, Ganjam2024}, the drive strengths $g_a$ and $g_b$ under input power $\elementdepsymbol{P}{in}{}$ is proportional to the coefficients $c_a$ and $c_b$ in the jump operators
\begin{equation}
    g_\mu = 2 c_\mu \left( \frac{\elementdepsymbol{\hbar P}{in}{}}{ \omega_d}\right)^{1/2}\,,
    \label{eq:app_drive_induced_dephasing_drive_strength}
\end{equation}
for $\mu = a, b$. The dephasing rate is extracted by simulating a Ramsey-type experiment, initializing the system in the state $(\ket{\overline{0,0,0}} + \ket{\overline{0,0,1}})/\sqrt{2}$. We implemented a ramp-up of the drive amplitude for approximately 80 ns, rounding up to the nearest integer multiple of drive period, sufficient for adiabatically displacing the filter modes to their steady states. An elaborate discussion on the adiabatic ramp can be found in Appendix \ref{app:df_vs_lab}. The direct displacement of the transmon mode due to the drive is negligible due to its weak coupling to the filter modes. 

The dephasing rate of the system is estimated by computing the expectation value of
\begin{equation}
    M_{01} = \sum_{j,k} \ket{\overline{j,k,0}}\bra{\overline{j,k,1}}\,,
\end{equation}
which mimics the action of tracing out the filter modes and measures the (0,1) component of the reduced density matrix of the transmon mode. The Ramsey decay rate $\gamma_2$ is estimated as the negative slope of $P_{01}(t) = \mathrm{Tr}[\rho(t) M_{01}]$ as a function of flat-pulse time $t$ rather than performing a fit to an exponential decay model, since the time scale for the decay of $P_{01}(t)$ is much longer than our measurement time. This rate, however, also contains the Ramsey decay rate of the static system $\gamma_{2,\mathrm{static}}$, which is obtained by repeating the extraction procedure under no drive. The drive-induced pure dephasing rate is then obtained from subtracting $\gamma_{2, \mathrm{static}}$ from $\gamma_2$. We verified that this rate is a good representation of the drive-induced pure dephasing rate, because the contribution from drive-induced depolarization is negligible. 

The master-equation simulation is computationally demanding because the system contains three modes. In addition, the two filter modes require large Hilbert space sizes to reach numerical convergence due to their drive-induced displacements and thermal populations. To accelerate the computation, we run the simulations on a graphical processing unit (GPU) using the \texttt{dynamiqs} package \cite{guilmin2025dynamiqs}.
In this example, tracing out the two filter modes and incorporating their interference through the susceptibility functions allows the PVNR method to estimate the drive-induced dephasing rate much more efficiently.

\subsection{Drive-induced decay rate calculation}
\label{app:tmon_filter_decay}
In the drive-induced decay case, we consider a drive near half of the transmon-filter detuning frequency at around 1.25 GHz, which is far detuned from both the filter modes and the transmon mode. A sufficiently large junction phase displacement is achieved by driving through the left port, which couples primarily to the transmon mode and negligibly to the filter modes. Achieving the same junction phase displacement by displacing the filter modes through driving from the right port would require large coherent occupations of the filter modes, making the LME calculation prohibitively expensive. 

The form of the drive is
\begin{equation}
    \mathcal{H}_d(t) = i g_q\cos(\omega_d t) (q^\dagger - q)\,,
\end{equation}
where $g_q$ is the drive strength. Because this drive is off-resonant, the near-resonant input–output relation used in \eqref{eq:app_drive_induced_dephasing_drive_strength} does not apply. Instead, we calibrate $g_q$ as a function of input power and frequency by matching the steady-state junction phase displacement produced by this term to that obtained from the HFSS driven-system simulation. 

In the simulation, we first initialize the system in the state $\ket{\overline{0,0,1}}$, then ramp up the drive pulse in approximately 100 ns to displace the transmon mode to its steady state, followed by a flat-top pulse of time $t$. Instead of explicitly simulating the dynamics during the ramp-down pulse, which is computationally expensive, we approximate its effect by applying the inverse steady-state displacement $\rho'(t) = U_{\mathrm{disp}}^\dagger(t) \rho(t) U_{\mathrm{disp}}(t)$. Similar to the drive-induced dephasing rate calculation, the leakage rate from the transmon's first excited state is obtained from the negative slope of $P_{11}(t) = \mathrm{Tr}[\rho'(t) M_{11}]$, where the measurement operator $M_{11}$ is
\begin{equation}
    M_{11} = \sum_{j,k} \ket{\overline{j,k,1}}\bra{\overline{j,k,1}}\,.
\end{equation}
This mimics the action of measuring the (1,1) component of the reduced density matrix of the transmon mode. 
Similarly, we leveraged the \texttt{dynamiqs} package and GPU calculation to speed up the computation.

\subsection{Parameter extraction}
\label{app:tmon_filter_parameter_extraction}
Because the correlated dissipators in Eq.~\eqref{eq:tmon_filter_master_eq_dissipator} renormalize the complex normal-mode frequencies, the effective LME parameters that govern the dynamics, including the dressed frequencies $\tilde\omega_a$, $\tilde\omega_b$, and $\tilde\omega_q$ and the dissipator weights $c_a$ and $c_b$, are not given directly by the bare HFSS eigenmode outputs, $(\omega_a^0, \omega_b^0 , \omega_q^0)$ and $(\kappa_a^0, \kappa_b^0, \kappa_q^0)$. Instead, the HFSS results must be mapped onto a consistent set of LME parameters that reproduce the same complex spectrum and port-induced dissipation, and we detail that mapping procedure below.

Consider a modified circuit for which we drop the Josephson nonlinearity and the drive. Such circuit corresponds to the linearized circuit simulated with HFSS eigenmode solver, and the corresponding Lindblad master equation is
\begin{align}
\dot \rho = \frac{1}{i\hbar}[\mathcal{H}_{\textrm{lin}}, \rho] +
    \mathscr{L}[\rho] 
     + (1+\elementdepsymbol{n}{B}{})\tilde{\kappa}_q \mathcal{D}[q] + \elementdepsymbol{n}{B}{}\tilde{\kappa}_q \mathcal{D}[q^\dagger].
    \label{eq:tmon_filter_master_eq_lin} 
\end{align}
We note that in comparison with Eq.~\eqref{eq:tmon_filter_master_eq}, here we include static transmon decay to match microwave simulation results. The corresponding dissipator is treated separately, since the detuning between the transmon and the filter modes is large compared to the linewidths of all three modes.

To obtain the normal mode parameters of the system, we consider the Heisenberg operator $\elementdepsymbol{O}{H}{}(t) = A(t) a^\dagger + B(t) b^\dagger + C(t) q^\dagger$ with 
\begin{equation}
    \begin{pmatrix}
        A(t)\\B(t)\\C(t)
    \end{pmatrix}
    = 
    \begin{pmatrix}
        A_0\\B_0\\C_0
    \end{pmatrix} e^{i(\omega + i\kappa/2) t},
\end{equation}
that describes a creation operator of a normal mode of the LME with a complex frequency $\omega + i\kappa/2$. Its corresponding equation of motion is
\begin{align}
    \elementdepsymbol{\dot{O}}{H}{}(t) &= \frac{i}{\hbar}[\elementdepsymbol{\mathcal{H}}{lin}{}, \elementdepsymbol{O}{H}{}(t)] + \mathscr{L}^{\dagger} [\elementdepsymbol{O}{H}{}(t)] \\
    &+ (1+\elementdepsymbol{n}{B}{})\kappa_q \mathcal{D}^{\dagger}[q]\elementdepsymbol{O}{H}{}(t) + \elementdepsymbol{n}{B}{}\kappa_q \mathcal{D}^\dagger[q^\dagger] \elementdepsymbol{O}{H}{}(t), \nonumber
\end{align}
where
\begin{align}
    \mathscr{L}^\dagger[\elementdepsymbol{O}{H}{}(t)] &= (1+\elementdepsymbol{n}{B}{})\mathcal{D}^\dagger[c_a a + c_b b] \elementdepsymbol{O}{H}{}(t) \nonumber \\
    &+ \elementdepsymbol{n}{B}{} \mathcal{D}^\dagger[c_a a^\dagger + c_b b^\dagger] \elementdepsymbol{O}{H}{}(t)\,,
\end{align}
and the action of the adjoint dissipator is~\cite{Breuer_Petruccione_2007}
\begin{equation}
    \mathcal{D}^\dagger[M] \elementdepsymbol{O}{H}{}(t) = M^\dagger \elementdepsymbol{O}{H}{}(t) M - \{M^\dagger M, \elementdepsymbol{O}{H}{}(t)\}/2.
\end{equation}
From the Heisenberg equation of motion, the coefficients in $\elementdepsymbol{O}{H}{}(t)$ satisfy
\begin{align}
    &[\mathbf{M} - (i\omega - \kappa/2) \mathbf{I}]
    \begin{pmatrix}
        A_0 \\
        B_0 \\
        C_0
    \end{pmatrix}
    =
    \begin{pmatrix}
        0\\0\\0
    \end{pmatrix}
    \label{eq:heisenberg_eom_for_normal_modes},\\
    &\mathbf{M} = 
    \begin{pmatrix}
        i \tilde{\omega}_a -  c_{a}^2 /2  & -c_{a} c_{b}/2 & 0 \\
        -c_{a} c_{b}/2 & i\tilde{\omega}_b  - c_{b}^2/2 & 0\\
        0 & 0 & i \tilde{\omega}_q - \tilde{\kappa}_q/2
    \end{pmatrix}.
\end{align}
Solving this set of equations yields three normal modes of the system, with complex frequencies $i {\omega}_n - {\kappa}_n/2$ ($n \in \{a,b,q\}$). These normal modes correspond to the eigenmodes obtained from HFSS simulations, $({\omega}_n, {\kappa}_n) = (\omega_n^0, \kappa_n^0)$. We treat the transmon normal mode as decoupled from the rest of the system, leading to $A_0 = B_0 = 0$, $\tilde{\omega}_q = \omega_q^0$ and $\tilde{\kappa}_q = \kappa_q^0$. The filter normal mode solutions are obtained by diagonalizing the submatrix
\begin{equation}
    \begin{pmatrix}
        i \tilde{\omega}_a -  c_{a}^2 /2  & -c_{a} c_{b}/2 \\
        -c_{a} c_{b}/2 & i\tilde{\omega}_b  - c_{b}^2/2 
    \end{pmatrix}.
\end{equation}
Their normal mode amplitudes satisfy $C_0 = 0$ but both $A_0$ and $B_0$ are nonzero because of the hybridization induced by the off-diagonal elements of $\mathbf{M}$, which originates from the correlated jump operators in the dissipator $\mathscr{L}[\rho]$. 

We numerically determine $(\tilde{\omega}_a, \tilde{\omega}_b, c_a, c_b)$ by minimizing the sum of squared relative differences between the LME normal-mode frequencies and those extracted from HFSS,
\begin{align}
f_\mathrm{obj} &= 
\left(\frac{{\omega}_a - \omega_a^0}{\omega_a^0}\right)^2 
+
\left(\frac{{\omega}_b - \omega_b^0}{\omega_b^0}\right)^2 \nonumber\\
&+\left(\frac{{\kappa}_a - \kappa_a^0}{\kappa_a^0}\right)^2 
+\left(\frac{{\kappa}_b - \kappa_b^0}{\kappa_b^0}\right)^2\,.
\end{align}
The resulting mean squared difference is $1.2\times 10^{-6}$, indicating the good quality of the fit. The resulting parameters are summarized in Table \ref{tab:drive_induced_decoherence_parameters}.
\nocite{hfsshelp}
\bibliography{aps-bibliography}

\end{document}